\documentclass[11pt,a4paper]{article}

\usepackage{cite}
\usepackage{stackrel}
\usepackage{graphicx}
\usepackage{amsmath}
\usepackage{amssymb,MnSymbol}
\usepackage{amsfonts}
\usepackage{dsfont}
\usepackage{mathtools}
\usepackage{array}
\usepackage{rotating}
\usepackage{bbold,amsfonts}
\allowdisplaybreaks

\usepackage[utf8]{inputenc}
\usepackage{bm}
\usepackage{xcolor}
\usepackage{float}
\usepackage{braket}
\usepackage{placeins}
\usepackage[height=8.8in,width=6.45in]{geometry}
\usepackage[font=small,labelfont=bf]{caption}
\usepackage[hidelinks]{hyperref}
\usepackage{booktabs,float,slashed}
\usepackage{standalone}
\usepackage{caption}
\usepackage{subcaption}
\usepackage{makecell}

\usepackage{stackengine,scalerel}
\newcommand\gl{\mathrel{\hstretch{1.5}{%
  \stackanchor[1pt]{\scriptscriptstyle>}{\scriptscriptstyle<}}}}
\stackMath

\numberwithin{equation}{section}
\newcommand{\vev}[1]{{\left\langle #1 \right\rangle}}

\newcommand{\beq}{\begin{align}}
\newcommand{\eeq}{\end{align}}

\newcommand{\e}{\epsilon}

\DeclareMathOperator{\Tr}{Tr}
\DeclareMathOperator{\tr}{tr}

\newcommand{\ii}{\mathrm{i}}

\makeatletter
\newcommand*{\letterdef@}{}
\newcommand*{\letterdef}[3]{%
	\def\letterdef@##1{\expandafter\newcommand\csname #1\endcsname{#2{##1}}}%
	\@tfor\@tempa :=#3\do{\expandafter\letterdef@\expandafter{\@tempa}}}
\makeatother
\letterdef{c#1} {\mathcal}{ABCDEFGHIJKLMNOPQRSTUVWXYZ} 
\letterdef{rm#1}{\mathrm} {dDeimM} 

\newcommand{\gym}{g_{_{\rm YM}}}

\def\Schur#1{{\mathcal{O}^{{\rm Schur}}_{#1}}}

\usepackage{bbm}
\DeclareMathAlphabet{\mathbb}{U}{msb}{m}{n} 

\begin{document}

\begin{titlepage}

\begin{flushright}
{\small QMUL-PH-25-17}
\end{flushright}

\vspace*{10mm}
\begin{center}
{\LARGE \bf 
Universality of giant graviton correlators}

\vspace*{15mm}

{\Large Augustus Brown$^{(a)}$, Daniele Dorigoni$^{(b)}$, Francesco Galvagno$^{(a)}$,  and Congkao Wen$^{(a)}$ }

\vspace*{8mm}

$(a)$ Centre for Theoretical Physics, Department of Physics and Astronomy, \\
Queen Mary University of London, London E1 4NS, UK \\[0.2cm]
$(b)$ Centre for Particle Theory \& Department of Mathematical Sciences Durham University, \\
Lower Mountjoy, Stockton Road, Durham DH1 3LE, UK
    
\vspace*{0.8cm}

email: a.a.x.brown@qmul.ac.uk, daniele.dorigoni@durham.ac.uk, f.galvagno@qmul.ac.uk, c.wen@qmul.ac.uk 
\end{center}

\begin{abstract}

We study a class of heavy-heavy-light-light (HHLL) integrated correlators of superconformal primary operators in $SU(N)$ $\mathcal{N}=4$ super Yang-Mills theory involving two light operators from the stress-tensor multiplet and two heavy operators whose conformal dimensions are proportional to the number of colours $N$. In the large-$N$ limit these heavy operators are dual to sphere and AdS giant gravitons, realised holographically as D3-branes wrapping an $S^3$ inside either the $S^5$ or the $AdS_5$ factor of the $AdS_5 \times S^5$ background geometry. These HHLL correlators thus describe the scattering of two gravitons off D3-branes.
In the planar limit we derive exact expressions for the HHLL integrated correlators as functions of both the ’t Hooft coupling and the giant graviton dimension. Remarkably, despite exhibiting distinct perturbative expansions at weak coupling, these integrated correlators share the same universal asymptotic series at strong coupling. We also demonstrate that a seemingly unrelated integrated correlator in a $USp(2N)$ $\mathcal{N}=2$ gauge theory, holographically dual to gluon-graviton scattering off D7-branes, exhibits precisely the same strong coupling asymptotic series. This reveals a striking universality of D-brane scattering processes. Furthermore, we compute the exponentially suppressed corrections at strong coupling for all these observables, showing that they are precisely the non-perturbative effects that account for the differences between these integrated correlators beyond the universal asymptotic series. Finally, we comment on the resurgent properties and the holographic interpretation of these exponentially suppressed terms.
\vspace*{0.3cm}

\end{abstract}
\end{titlepage}
\setcounter{tocdepth}{2}
\tableofcontents

\section{Introduction}
\label{sec:intro}

One of the main challenges in theoretical physics resides in
understanding the strong coupling regime of gauge theories. A natural playground to explore non-perturbative techniques is
$\mathcal{N}=4$ super Yang-Mills theory (SYM), the maximally supersymmetric gauge theory in four dimensions, where the high degree of symmetry allows for many observables to be determined in an exact analytical way. A major example is represented by the study of half-BPS superconformal primary operators. Such operators have conformal dimensions that do not depend on the gauge coupling constant, and their two- and three-point correlation functions are fully protected and tree-level exact \cite{Lee:1998bxa}. Four and higher-point correlation functions exhibit rich dynamical structures and have been extensively studied in both the weak-coupling, e.g.~\cite{Eden:2012tu,Drummond:2013nda,Chicherin:2015edu, Bourjaily:2016evz, Coronado:2018cxj, Caron-Huot:2021usw, Bargheer:2022sfd, He:2024cej, Bourjaily:2025iad}, and strong-coupling regimes, e.g.~\cite{Goncalves:2014ffa,Rastelli:2016nze,Alday:2017xua,Caron-Huot:2018kta,Alday:2019nin,Goncalves:2019znr, Chester:2019pvm,Drummond:2020dwr,Abl:2020dbx, Aprile:2020mus,Green:2020eyj,Huang:2021xws,Drummond:2022dxw, Goncalves:2023oyx,Alday:2023mvu,Aprile:2024lwy,Fernandes:2025eqe}, see also the reviews~\cite{Heslop:2022xgp,Bissi:2022mrs} for further references. According to the AdS/CFT dictionary, half-BPS four-point functions in $SU(N)$ $\cN=4$ SYM represent boundary observables dual to type IIB string scattering processes on an $AdS_5\times S^5$ background. 

Most efforts have been focused on correlation functions of superconformal primary operators $\cO_p$ with fixed conformal dimension $p \ll N$. 
In type IIB superstring theory, these operators are holographically dual to the graviton (for $p=2$) and higher Kaluza-Klein modes (for $p>2$). 
The main focus in this paper is on Heavy-Heavy-Light-Light (HHLL) four-point correlation functions of the type $\langle \mathcal{O}_2 \mathcal{O}_2 \mathcal{H} \mathcal{H} \rangle$, where the heavy operators $\cH$ have conformal dimensions proportional to the number of colours,~i.e. $\Delta \propto N$, thus becoming heavy in the large-$N$ limit. 
In particular, we concentrate on two classes of heavy operators: the subdeterminant operators with dimension $\Delta=\alpha N$ (where $0<\alpha \leq 1$), and the symmetric Schur polynomial operators with dimension $\Delta=\beta N$ (where $\beta>0$).\footnote{The cases where the dimension of heavy operators $\Delta_{\mathcal{H}} \gg N^2$ have been studied in various set ups in $\cN=4$ SYM \cite{Paul:2023rka, Brown:2023why, Caetano:2023zwe, Brown:2024yvt, Grassi:2024bwl,  Brown:2025cbz}; such cases are in some sense simpler because $\Delta_{\mathcal{H}}$ is the only large parameter, and indeed one may determine these HHLL correlators completely \cite{Caetano:2023zwe, Brown:2024yvt, Brown:2025cbz} in the so-called large-charge 't Hooft limit where $\Delta_{\mathcal{H}} \to \infty$ with $\Delta_{\mathcal{H}}\, g_{_{\rm YM}}^2$ being fixed \cite{Bourget:2018obm}. }

Since these operators become heavy in the large-$N$ limit, they play a major role in the holographic correspondence as they correspond to extended objects,~i.e. D-branes, in the dual type IIB superstring theory on an $AdS_5 \times S^5$ background \cite{Witten:1998xy,McGreevy:2000cw,Corley:2001zk, Balasubramanian:2001nh,Berenstein:2002ke,Balasubramanian:2002sa,Berenstein:2003ah, Lin:2004nb}.
Via the AdS/CFT dictionary, subdeterminant operators are dual to \textit{sphere giant gravitons}, which are D3-brane solutions that wrap an $S^3$ inside the $S^5$, and are thus heavy states that travel along one-dimensional geodesics in $AdS$. On the other hand, the symmetric Schur polynomial operators are holographically dual to D3-branes wrapping an $S^3$ inside the $AdS_5$ part of the geometry while moving along $S^5$ geodesics. For this reason they are denoted as \textit{AdS} (or \textit{dual}) \textit{giant gravitons}.

Evaluating correlation functions in the presence of these heavy operators in the planar limit is a challenging task since their dimension $\Delta\sim N$ affects the large-$N$ expansion. Several results have been obtained for Heavy-Heavy-Light (HHL) three-point functions $\langle \mathcal{H} \mathcal{H}  \mathcal{O}_L \rangle$, where $\mathcal{O}_L$ is a light half-BPS chiral primary operator, or a non-BPS single-trace operator, and $\cH$ stands for either type of giant gravitons. The planar results have been obtained by using various powerful methods such as integrability, holography, and  semi-classical analysis \cite{Jiang:2019xdz,Jiang:2019zig,Yang:2021kot,Holguin:2022zii}.

On the contrary, the HHLL four-point correlators of two light operators in the presence of two giant gravitons have been much less explored. In fact, we are not aware of any explicit results regarding HHLL correlators with AdS giant gravitons. For the case of sphere giant gravitons of dimension exactly equal to $N$ (namely the maximal determinant operator with $\alpha=1$), the HHLL correlator has been determined in the weak 't Hooft coupling regime at one and two loops in \cite{Jiang:2019xdz,Jiang:2019zig}, and studied at three loops in \cite{Jiang:2023uut}. The leading strong-coupling regime holographically corresponding to the supergravity approximation was recently obtained in \cite{Chen:2025yxg} by interpreting the heavy insertions as a zero-dimensional defect. 

A novel method for constraining the dynamics of four-point correlation functions in $\mathcal{N}=4$ SYM has been introduced in \cite{Binder:2019jwn, Chester:2020dja}; such an approach consists in integrating the spacetime dependence of the inserted operators over a supersymmetric invariant measure. The resulting integrated correlators can be computed exactly via supersymmetric localisation \cite{Pestun:2007rz}, thus mapping the problem of evaluating such integrated four-point functions to the calculation of matrix model integrals. This powerful tool allows us to derive results which are exact both in the gauge coupling constant $g_{_{\rm YM}}^2$ as well as in the number of colours $N$, thus allowing us to investigate the non-perturbative properties and S-duality of $\mathcal{N}=4$ SYM \cite{Chester:2019jas, Chester:2020vyz, Dorigoni:2021bvj, Dorigoni:2021guq, 
Chester:2021aun, Alday:2021vfb, Dorigoni:2022zcr, Dorigoni:2022iem,
Collier:2022emf,
Dorigoni:2022cua, Dorigoni:2023ezg,  Chester:2023ehi,  Alday:2023pet, Caron-Huot:2024tzr}. 

In this paper, we consider an integrated version of the HHLL correlator $\langle \mathcal{O}_2 \mathcal{O}_2 \mathcal{H} \mathcal{H} \rangle$ where the heavy operator $\mathcal{H}$ is a giant graviton. We study such observables in the 't Hooft planar limit where $N\to\infty$ while the 't Hooft coupling $\lambda= g_{_{\rm YM}}^2 N$ is kept fixed. At leading order in the $1/N$ expansion, we show that the integrated correlators can be determined exactly as functions of $\lambda$ and of the conformal dimension of the giant graviton operators considered. 
 We compute these HHLL integrated correlators following the analysis of \cite{Brown:2024tru}, which relies on matrix model techniques coming from supersymmetric localisation combined with the inputs of HHL three-point functions computed via semi-classical techniques. Our work generalises the results of \cite{Brown:2024tru} which used precisely this approach to determine the planar HHLL integrated correlator for the sphere giant graviton of dimension $\Delta =N$. In this work, we extend that analysis to both sphere and AdS giant gravitons of generic dimensions, and uncover universal structures in the strong-coupling expansion of such HHLL integrated correlators.

At large $N$, we denote by $N \, \mathcal{I}_S(\lambda;\alpha)$ the planar HHLL integrated correlator with two sphere giant gravitons of equal dimension $\Delta = \alpha N$, and similarly by $N\, \mathcal{I}_{AdS}(\lambda;\beta)$ the planar  HHLL integrated correlator with two AdS giant gravitons of equal dimension $\Delta= \beta N$. In the weak 't Hooft coupling regime $\lambda \to 0$, we find the following all-loop expressions,
\begin{align}
    \mathcal{I}_{S}( \lambda ;\alpha) &=  4\sum_{\ell=1}^\infty (-1)^{\ell+1} \,\zeta(2\ell{+}1)\Big(\frac{\lambda}{16\pi^2}\Big)^\ell \left[ \binom{2\ell{+}1}{\ell}^2 - \alpha^{2(\ell+1)} \binom{2\ell {+} 1}{\ell}  \, _2 F_1 \big(\ell{+}1,\ell{+}2;1;1{-}\alpha\big) \right] \, , \\ 
    \mathcal{I}_{AdS}(\lambda;\beta) &=  4\sum_{\ell=1}^\infty (-1)^{\ell} \,\zeta(2\ell{+}1)\Big(\frac{\lambda}{16\pi^2}\Big)^\ell \left[ \binom{2\ell{+}1}{\ell}^2 - \beta^{2(\ell+1)} \binom{2\ell {+} 1}{\ell}  \, _2 F_1 \big(\ell {+} 1,\ell {+} 2;1;1{+}\beta\big) \right] \, ,
\end{align}
where $\zeta(s)$ denotes the standard Riemann zeta function while $_2F_1(a,b;c;z)$ is the hypergeometric function. 
We will show that both finite coupling expansions have finite radius of convergence in $\lambda$.
We note that the two expansions are remarkably similar and that $\cI_{AdS}(\lambda;\beta)$ can be obtained from $\mathcal{I}_{S}(\lambda;\alpha)$ by simply substituting $\alpha = -\beta$, i.e. $\cI_{AdS}(\lambda;\beta)=-\mathcal{I}_{S}(\lambda; -\beta)$. Given that the dimensions of the giant graviton operators are both proportional to $N$, we see that the two HHLL integrated correlators are simply related by the exchange $N\leftrightarrow -N$.

Conversely,  at strong coupling $\lambda \gg 1$ we find, 
\begin{align}
  \mathcal{I}_{S}(\lambda;\alpha) &\sim 
  2 \alpha - \frac{4 \pi ^2}{3 \lambda }  - \sum_{n=1}^{\infty}   \frac{16 n \zeta (2 n+1) \Gamma
   \left(n-\frac{1}{2}\right)^2 \Gamma \left(n+\frac{1}{2}\right)}{  \lambda ^{  n+\frac{1}{2}}\, \pi ^{3/2}\, \Gamma (n)} \, ,  \\
       \cI_{AdS}(\lambda;\beta)  &\sim 2 \log(1+\beta)- \frac{4 \pi ^2}{3 \lambda }  +  \sum_{n=1}^{\infty}   \frac{16 n \zeta (2 n+1) \Gamma
   \left(n-\frac{1}{2}\right)^2 \Gamma \left(n+\frac{1}{2}\right)}{  \lambda ^{  n+\frac{1}{2}}\, \pi ^{3/2}\, \Gamma (n)} \, ,
\end{align}
where the notation $\sim$ denotes that, contrary to the weak coupling regime, both strong coupling perturbative series are only formal asymptotic expansions in the sense of Poincar\'e.
At strong coupling, the leading order terms are holographically dual to the supergravity contributions and we see that these depend explicitly on the giant graviton parameters $\alpha$ and $\beta$. However, higher-order terms, which correspond to stringy corrections, are completely independent of $\alpha$ and $\beta$. Up to an overall minus sign we furthermore note that  $ \mathcal{I}_{S}(\lambda;\alpha)$ and $\mathcal{I}_{AdS}(\lambda;\beta)$ have exactly the same asymptotic series at large $\lambda$, despite having different weak coupling expansions. This shows the remarkable universality of these observables in the strong coupling perturbative regime.

The remaining difference between  $ \mathcal{I}_{S}(\lambda;\alpha)$ and  $\mathcal{I}_{AdS} (\lambda;\beta)$ at strong coupling then lies entirely in their exponentially suppressed non-perturbative contributions. In particular, for $ \mathcal{I}_{S}(\lambda;\alpha)$ we find that there are three different types of non-perturbative corrections classified in terms of their exponential suppression scales as follows:
\begin{align}
    \exp [-2n \sqrt{\lambda}] \, , \quad \exp\left[ -n\sqrt{\lambda}(1 - \sqrt{1-\alpha}) \right] \, , \quad 
    \exp \left[ -n\sqrt{\lambda}(1 + \sqrt{1-\alpha}) \right] \, ,
\end{align}
where $n$ runs over the positive integers.
On the other hand, for $\mathcal{I}_{AdS}(\lambda;\beta)$ we find that the non-perturbative sectors are weighted by the following exponential suppression scales,
\begin{align}
    \exp [-2n \sqrt{\lambda}] \, , \quad \exp \left[ -{n\sqrt{\lambda}(\sqrt{\beta+1} - 1)}  \right] \, , \quad 
    \exp \left[ -{n\sqrt{\lambda}(\sqrt{\beta+1} + 1)}  \right] \, ,
\end{align}
where again $n$ runs over the positive integers.
Note that in both $ \mathcal{I}_{S}(\lambda;\alpha)$ and $\mathcal{I}_{AdS}(\lambda;\beta)$, each non-perturbative scale is accompanied by its own perturbative series in powers of $1/\sqrt{\lambda}$.

Furthermore, we also consider the AdS giant graviton integrated correlators with dimension $\Delta = \beta N$ in the limit where $\beta \gg 1$. As discussed in \cite{Ivanovskiy:2024vel}, the large-$\beta$ behaviour of the AdS giant graviton maps $\cN=4$ SYM to its Coulomb branch physics.  Here we provide an exact expression \eqref{eq:large_beta_weak} for the HHLL integrated correlator in the presence of such `very heavy' AdS giant graviton insertions. In particular, we discuss the connection between the large-$\beta$ expansion of the AdS giant graviton HHLL integrated correlator and a semi-classical approach to $\mathcal{N}=4$ SYM Coulomb branch physics, finding a surprisingly similarity between the present case and the results of \cite{Caetano:2023zwe, Brown:2024yvt,Brown:2025cbz} obtained in the so-called large charge 't Hooft limit \cite{Bourget:2018obm}.

Our final results concerns a different observable in a different theory. Recently, \cite{Chester:2025ssu} studied the integrated correlator of moment map operators in an $\mathcal{N}=2$ SCFT with $USp(2N)$ gauge group and matter content given by four fundamental and one antisymmetric hypermultiplets. This set-up is holographically dual to the four-point scattering amplitude of two gluons and two gravitons in the presence of D7-branes on an $AdS_5 \times S^5/\mathbb{Z}_2$ background. Even though the weak-coupling expansion of this observable is very different from that of $\mathcal{I}_{S}(\lambda;\alpha)$ and $\mathcal{I}_{AdS}(\lambda;\beta)$, surprisingly its strong-coupling expansion, displayed in \eqref{eq:F1-strong}, is once again given by the same universal asymptotic series as for the giant graviton correlators. This fact demonstrates a remarkable universal behaviour of D-brane scattering processes.
Lastly, we show that it is again the non-perturbative, exponentially suppressed corrections which distinguish this integrated correlator from the giant graviton integrated correlators at strong coupling. 

\subsection*{Outline}

The rest of the paper is structured as follows. In Section \ref{BPS}, we introduce the sphere and AdS giant gravitons operators and review some of their key properties. In particular, we discuss HHL three-point functions of two giant gravitons and an additional single-trace operator in the large-$N$ limit.  In Section \ref{integrated correlators}, we review the matrix model computation of integrated Heavy-Heavy-Light-Light correlators via supersymmetric localisation. In Section \ref{exact}, we derive exact expressions for the HHLL integrated correlators of both sphere and AdS giant gravitons at the leading order in the large-$N$ limit. We use these results to show that these two different integrated correlators share remarkably the same asymptotic perturbative series at strong 't Hooft coupling. Furthermore, we also consider the AdS giant gravitons correlator in the limit when the parameter $\beta \gg 1$ and comment on the connection with Coulomb branch physics.  In Section \ref{sec:non-pert} we study the exponentially suppressed contributions to the strong coupling expansion of the HHLL giant gravitons correlators and show that the aforementioned universality is lifted non-perturbatively. Furthermore, we provide some arguments on the semi-classical nature of these non-perturbative effects from the dual holographic picture.
Lastly, in Section \ref{gluon} we comment on a different integrated correlator for the gluon-graviton scattering off D7-branes and show that it shares the same universal strong coupling perturbative expansion of the giant gravitons correlators. We conclude and discuss possible future directions in Section \ref{conclusion}. This work is accompanied by two appendices. Appendix \ref{app:exp} provides technical details concerning the derivation of the exponentially suppressed corrections discussed in Section \ref{sec:non-pert}, while Appendix \ref{app:cat} clarifies how certain non-perturbative terms can be recast in the language of Cheshire cat resurgence.

\section{Sphere and AdS giant gravitons}\label{BPS}

In this section, we discuss correlation functions of superconformal primary operators in $\cN=4$ SYM. Furthermore, we also introduce the main players of the paper: the sphere giant graviton and the AdS giant graviton operators. In particular, we present some results concerning the correlation functions of two giant gravitons with a single-trace operator, which will be an important input for determining the integrated four-point correlators of main interest for this work. 

\subsection{Multitrace operators}
The most common basis of half-BPS operators in $\mathcal{N}=4$ SYM is given by trace operators.
Amongst the many fields of $\mathcal{N}=4$ SYM we have six scalar fields $\Phi^I(x)$ with $I=1,...,6$ which all transform in the adjoint representation of the $SU(N)$ gauge group.
We can then define the multi-trace operators,
\begin{align}  \label{eq:multi-tr}
\cO_{\vec p}\,(x, Y)=\cO_{(p_1 \dots p_m)}\,(x, Y) \coloneqq \frac{p_1 \dots p_m}{p}\cO_{p_1}(x, Y) \dots \cO_{p_m}(x, Y) \, ,
\end{align} 
where each $\mathcal O_{p_i}(x, Y)$ represents the single-trace operator
\begin{align}  \label{eq:single-tr}
\cO_{q}(x, Y) \coloneqq \frac{1}{q} \Tr( Y \cdot\Phi(x)  )^q \, ,
\end{align} 
with $Y_{I}$ a null $SO(6)$ polarisation vector,~i.e. $Y\cdot Y= 0$. Here $p=\sum_{i=1}^m p_i$ is the protected conformal dimension of the operator $\cO_{\vec p}\,(x, Y)$.

Correlation functions of these operators satisfy several constraints. Importantly, the two- and three-point functions of multitrace operators are protected by supersymmetry and are tree-level exact \cite{Lee:1998bxa}. For example, the two-point function of identical operators is
\begin{align} \label{eq:OO_2pointFT}
    \vev{ \cO_{\vec p}(x_1, Y_1)\cO_{\vec p}(x_2, Y_2)} =\mathfrak R_{\vec p}(N) (d_{12})^p \,, \qquad {\rm with} \quad   d_{12}\coloneqq \frac{Y_1\cdot Y_2}{x_{12}^2} \ , 
\end{align}
where $\mathfrak R_{\vec p}(N)$ is a two-point coefficient that depends on the number of colours $N$ but is otherwise independent from the Yang-Mills coupling constant. 

Another result that will be needed in the following is the normalised three-point function between two identical operators $\cO_{\vec m}$ and one multitrace operator $\cO_{\vec p}$, which is given by
\begin{align}\label{eq:OO_3pointFT}
\frac{\vev{\mathcal{O}_{\vec m}(x_1,Y_1) \mathcal{O}_{\vec m}(x_2,Y_2) \cO_{\vec p}(x_3,Y_3)} }{\vev{\mathcal{O}_{\vec m}(x_1,Y_1) \mathcal{O}_{\vec m}(x_2,Y_2)}}= \mathfrak{C}_{\vec m \vec m \vec p}(N)\times \left(\frac{d_{23}d_{31}}{d_{12}}\right)^{\frac{p}{2}}\,,
 \end{align}
where $\mathfrak{C}_{\vec m \vec m \vec p}(N)$ is a three-point coefficient which can be computed in free theory via Wick contractions and is hence again coupling constant independent.

\subsection{Giant graviton operators}\label{sec:GG}

In this paper, we consider half-BPS operators with a conformal dimension that scales with the number of colours $N$, so that at large $N$ they become heavy. Such operators are particularly important from a holographic point of view since they are dual to extended objects such as D-branes in an AdS$_5\times S^5$ background. For such heavy operators, it is more convenient to introduce a different basis of gauge invariant objects. 

\subsubsection{Schur polynomial basis}

 Rather than the multi-trace operators defined in \eqref{eq:multi-tr}, to describe heavy operators it is more convenient to use the Schur polynomial basis, first introduced in \cite{Corley:2001zk}.
 This basis exploits the mapping between $U(N)$ representations and Young diagrams associated to representations of the symmetric group $S_n$. The same method holds for $SU(N)$ representations, up to some subtleties associated with the relative $U(1)$ factor between $U(N)$ and $SU(N)$ \cite{Brown:2007bb}. 
 
 In the Schur polynomial basis, given a representation $R$ described by a Young tableau with $m$ boxes, we define the dimension $m$ operator
\begin{equation}\label{eq:Schur_def}
    \Schur{R}(x,Y) \coloneqq \frac{1}{m!} \sum_{\sigma \in S_m}\chi_R(\sigma) \Tr (\sigma\varphi)(x,Y)~, \qquad {\rm with}\quad \varphi(x,Y)\coloneqq Y\cdot\Phi(x)
\end{equation}
where $\chi_R$ is the character of the representation $R$. In \eqref{eq:Schur_def} we use the trace notation introduced in \cite{Corley:2001zk} where
\begin{equation}
   \Tr (\sigma\varphi) \coloneqq \sum_{i_1,\dots i_m=1}^N \varphi^{i_1}_{i_{\sigma(1)}} \dots \varphi^{i_m}_{i_{\sigma(m)}}~,
\end{equation}
where $i_1,\dots i_m$ are fundamental indices of $\mathfrak{su}(N)$ and $\sigma \in S_m$ is a permutation of $m$ elements. The Schur basis is then obtained by considering the set of all operators \eqref{eq:Schur_def} where $R$ runs over all irreducible representations of $S_m$.

We now consider two special classes of operators $ \Schur{R}$ constructed from two particular representations. The first class is obtained by selecting the totally antisymmetric representation, described by a Young diagram with a single column with $m$ boxes. We denote such a representation as $R=(\vec 1)_m$ (to denote a Young diagram with $m$ rows with one box each). 

The associated operators \eqref{eq:Schur_def} with $R=(\vec 1)_m$ can be re-expressed in terms of subdeterminants over the gauge algebra as
\begin{equation}\label{eq:detPhi}
   \Schur{(\vec 1)_m}(x,Y) = \det_m \varphi(x,Y) \, ,
\end{equation}
where the $m$-subdeterminant of any $N\times N$ matrix $X$ is defined as
\begin{align}\label{eq:determinant_general}
   \det_m X \coloneqq \frac{1}{m!} \epsilon_{i_1 \dots i_m \ell_1\dots \ell_{N-m}}\epsilon^{j_1 \dots j_m \ell_1\dots \ell_{N-m}} X_{j_1}^{i_1}\cdots X_{j_m}^{i_m}\,,
\end{align}
where again $i_n,j_n,\ell_n=1,\dots, N$ are fundamental indices of $\mathfrak{su}(N)$. Note that for $m=N$ the subdeterminant \eqref{eq:determinant_general} reduces to the standard determinant of the $N\times N$ matrix $X$.
This class of gauge invariant observables \eqref{eq:detPhi} is useful to describe baryon-like operators in $\cN=4$ SYM. The subdeterminant operators can also be decomposed in terms of the different gauge invariant basis given by the multitrace operators \eqref{eq:multi-tr}. Furthermore, since the dimension of $\Schur{(\vec 1)_m}(x,Y)$ is given precisely by the number of boxes $m$ we must have that the dimension of such operators is bounded by the total number of colours $N$, given that when $m>N$ we have $\det_m X= 0$ for any $N\times N$ matrix $X$.

At large $N$ the class of operators \eqref{eq:detPhi} of dimension $m = \alpha N$, with  $0< \alpha \leq 1$ fixed, plays a very important role in the AdS/CFT correspondence. According to the holographic dictionary, such subdeterminant operators are dual to particular D3-brane solutions called \textit{sphere giant gravitons}, which wrap an $S^3$ inside the $S^5$ part of the background $AdS_5\times S^5$ geometry \cite{McGreevy:2000cw,Balasubramanian:2001nh,Berenstein:2002ke,deMelloKoch:2004crq,Jiang:2019zig,Jiang:2019xdz}. The class of operators $\Schur{(\vec 1)_m}$ are therefore also named \textit{sphere giant graviton} operators.  

In more detail, let us first express the $AdS_5\times S^5$ background metric in global coordinates,
\begin{equation}
ds^2 = ds_{AdS_5}^2+ ds^2_{S^5} \,,\label{eq:AdSSmet}
\end{equation}
where the $AdS_5$ and $S^5$ metrics are given by 
\begin{align}
   ds^2_{AdS_5} &\label{eq:AdS}= -\cosh^2\rho \, dt^2 +d\rho^2 +\sinh^2\rho \, d\tilde\Omega_3^2~,\\*
    ds^2_{S^5}&\label{eq:S}=d\theta^2+\sin^2\theta d\phi^2 +\cos^2\theta d\Omega_3^2~,
\end{align}
where $d\tilde\Omega_3^2$ and $d\Omega_3^2$ are the metric on $S^3$.
We then have that the operator $ \Schur{(\vec 1)_m}$ with  $m = \alpha N$, and  $0<\alpha \leq 1$ fixed, corresponds to a classical solution where the D3-brane is localised at the point $\theta=\theta_0$ and $\rho=0$ and extends along $\Omega_3$ while rotating in the $\phi$ direction at the speed of light. The dimension of the subdeterminant operator is related to the D3-brane solution via
\begin{equation}
    \cos^2\theta_0= \frac{m}{N}= \alpha\,.\label{eq:theta0}
\end{equation}
Consequently, the simple linear algebra bound $0\leq \alpha\leq 1$ has a neat holographic interpretation since the parameter $\alpha$ is interpreted as the cosine square of the angle $\theta_0$ which parametrises the D3-brane orientation inside the $S^5$ part of the background geometry.

Using the Schur polynomial basis, it is natural to introduce a class of operators dual to the subdeterminant operators, i.e. operators that are described by the totally symmetric Young tableau with $m$ boxes in a single row. We denote this representation as $R= (m,\vec 0)$ and define the symmetric Schur polynomials as $\Schur{(m,\vec 0)}(x,Y)$. Also in this case it is possible to decompose $\Schur{(m,\vec 0)}(x,Y)$ in terms of the basis of multitrace operators by using \eqref{eq:Schur_def}. We will not need such a decomposition for the present work, as it becomes very cumbersome to deal with at large $N$. 

In the large-$N$ limit, these operators have a similar holographic interpretation to the antisymmetric case, with the key difference that the dimensions have no upper bound. The operator $\Schur{(m,\vec 0)}$ with $m=\beta N$, and $\beta>0$ fixed, corresponds to a D3-brane solution called the \textit{AdS} (or \textit{dual}) \textit{giant graviton}, which extends along an $S^3$ inside the $AdS_5$ directions \cite{Hashimoto:2000zp, Grisaru:2000zn,Chen:2019gsb}. The class of operators $\Schur{(m,\vec 0)}$ are therefore called $AdS$ (or \textit{dual}) \textit{giant graviton} operators.

More precisely, using global coordinates for the $AdS_5$ part of the dual geometry as in \eqref{eq:AdS},
we have that the operator $\Schur{(m,\vec 0)}$ with $m=\beta N$ and fixed $\beta>0$ corresponds to a classical solution where the D3-brane is localised at $\theta=\pi/2$ and $\rho=\rho_0$ and extends along $\tilde\Omega_3$ while still rotating at the speed of light in the $\phi$ direction. In this case, the mapping of the parameters reads
\begin{equation}
    \sinh^2\rho_0=\beta~,\label{eq:rho0}
\end{equation}
which is consistent with the fact that $\beta$ is positive and unbounded.

 Additionally, thanks to the holographic interpretation in terms of D3-branes stretched along the $AdS_5$ factor, we see that at the planar level the insertion of `very heavy' symmetric Schur-polynomial operators ($\beta \gg 1$) effectively moves the $\cN=4$ SYM theory along its Coulomb branch \cite{Ivanovskiy:2024vel}. Therefore, by studying correlation functions of light operators in the presence of very heavy symmetric Schur-polynomial operators, we can shed light on properties of correlation functions on the moduli space of $\cN=4$ SYM. We will come back to this point later. 

\subsubsection{Generating functions of giant gravitons operators}

In the following, we will find it convenient to analyse properties of giant graviton operators from their associated generating functions.

Following the conventions of \cite{Yang:2021kot,Holguin:2022zii}, we first define the generating function of sphere giant gravitons:\footnote{We denote the operators $\cG_S$ to indicate the holographic interpretation in terms of sphere giant graviton solution of the subdeterminant operators, and similarly we will denote AdS giant gravitons as $\cG_{AdS}$.}
\begin{equation}\label{eq:D_generating}
    \cG_S(x,Y;z) \coloneqq \sum_{m=0}^N z^m\, \Schur{(\vec{1})_m}(x,Y) = \det \left( \mathbbm 1 +z\, Y\cdot \Phi\right)(x)~\,,\qquad {\rm with}\, z\in \mathbb{C}\,.
\end{equation}
Any subdeterminant operator with dimension $m$ is then obtained through a contour integral over the auxiliary variable $z$, 
\begin{equation}\label{eq:D_alpha_m}
    \Schur{(\vec{1})_m}(x,Y)= \oint \frac{dz}{2\pi \ii } \frac{1 }{ z^{m+1} }\cG_S(x,Y;z)~ , 
\end{equation}
where the integration contour is a counter-clockwise loop centred around $z=0$. 
In particular, as commented above, we are interested in the giant graviton operator with dimension $m=\alpha N$ where the parameter $\alpha$ lies in the range $0<\alpha \leq 1$.
From \eqref{eq:D_alpha_m} we see that this operator, which will be denoted by $\cG_{S;\, \alpha}$, is given by
\begin{equation} \label{eq:D_alpha_def}
\cG_{S;\, \alpha} (x,Y)\coloneqq \oint \frac{dz}{2\pi \ii } \frac{1 } {z^{\alpha N+1} }\cG_S(x,Y;z)\,.
\end{equation}

Similarly, we can write the dual generating function for the AdS giant gravitons by considering the inverse of the determinant \cite{Yang:2021kot},
\begin{equation}\label{eq:G_generating}
    \cG_{AdS}(x,Y;z) \coloneqq \sum_{m=0}^\infty z^m\,\Schur{(m,\vec{0})}(x,Y) =  \left[\det \left( \mathbbm 1 -z\, Y\cdot \Phi\right)\right]^{-1}(x)~,\qquad {\rm with}\,z\in \mathbb{C}\,,
\end{equation}
and the symmetric operator of dimension $m$ is then given by the contour integral 
\begin{equation}
    \Schur{(m,\vec{0})}(x,Y) = \oint \frac{dz}{2\pi \ii } 
    \frac{1}{z^{m+1}} \cG_{AdS}(x,Y;z)\,, \label{eq:G_alpha_m}
\end{equation}
where again we integrate over a counter-clockwise loop around $z=0$.
The AdS giant graviton operator with dimension  $m=\beta N$, where $\beta>0$, will be denoted by $\cG_{AdS;\, \beta}$, and from \eqref{eq:G_alpha_m} it can be expressed as
\begin{equation}\label{eq:G_alpha_def}
    \cG_{AdS;\, \beta} (x,Y) \coloneqq \oint \frac{dz}{2\pi \ii } \frac{1}{z^{\beta N+1}}\cG_{AdS}(x,Y;z) \, . 
\end{equation}
We stress again that in this case the parameter $\beta$ is not bounded from above, since the dimension of the AdS giant graviton can be larger than $N$.

Our ultimate goal is computing a class of integrated correlation functions involving two giant gravitons (both the sphere giant $\cG_{S;\, \alpha}$ and the AdS giant $\cG_{AdS;\, \beta}$) and two half-BPS operators $\cO_2$, as defined in \eqref{eq:single-tr}. 
Since these correlators involve two heavy and two light operators we will refer to them as \textit{Heavy-Heavy-Light-Light} (HHLL) correlators.

In the next section we show that the main ingredients for determining the large-$N$ limit of these integrated correlators are the three-point functions of single-trace operators \eqref{eq:single-tr} in the presence of a pair of identical giant gravitons, which we now briefly review.

\subsection{HHL three-point functions}

As previously mentioned, the three-point functions of half-BPS operators in $\cN=4$ SYM are tree-level exact and they are fixed by superconformal symmetry up to a coefficient depending only on the number of colours $N$ and the conformal dimensions of the operators, as in \eqref{eq:OO_3pointFT}. 

Since this will be one of the main ingredients for our calculation, we move to discuss more explicitly the normalised three-point functions of two identical giant gravitons with a generic multi-trace operator of dimension $p$.

The normalised Heavy-Heavy-Light (HHL) three-point correlator is given by
\begin{align}
\frac{\vev{\cH (x_1,Y_1) \cH (x_2,Y_2) \cO_{\vec p}(x_3,Y_3)} }{\vev{\cH (x_1,Y_1) \cH(x_2,Y_2)}}= \mathfrak{C}_{\cH,\cH, {\vec p}}(N) \times \left(\frac{d_{23}d_{31}}{d_{12}}\right)^{\frac{p}{2}}\,.
\end{align}
Importantly, when the heavy operator is either a sphere (i.e. $\mathcal{H}=\cG_{S;\alpha}$) or an AdS (i.e. $\mathcal{H}=\cG_{AdS;\beta}$) giant graviton, the three-point function coefficient $\mathfrak{C}_{\mathcal{H},\mathcal{H},\vec p}(N)$ admits a topological expansion at large $N$ and fixed ${ \vec p}$ of the form
\begin{equation} \label{eq:3pt_genus}
   \mathfrak{C}_{\cH,\cH,{\vec p}}(N) = \sum_{g=0}^\infty N^{\frac{p}{2}-g} \mathfrak{C}_{\cH,\cH,{\vec p}}^{(g)} \, .
\end{equation}

The leading large-$N$ contributions to the three-point functions were determined in \cite{Yang:2021kot} (see also \cite{Holguin:2022zii}) using semi-classical methods and were furthermore shown to match with the holographic picture of the sphere and AdS giant gravitons in terms of D3-branes. 
For later convenience, here we simply reproduce the results of the three-point functions computed in \cite{Yang:2021kot,Holguin:2022zii} (see also \cite{Bissi:2011dc} for earlier work). At leading order in the large-$N$ expansion, the three-point function of single trace operators in the presence of a pair of sphere giant gravitons reads
\begin{align}\label{eq:O2p_D}
\mathfrak{C}_{\cG_{S;\alpha},\cG_{S;\alpha},(2p)}^{(0)} &= - 2^{-p} \oint_{z=0} {dz \over 2\pi i} {1-z^2 \over z^{2p+1} \sqrt{1+2z^2(2\alpha-1) + z^4 } }  
 \cr 
 &=(-2)^{-p+1} \, _2F_1(-p,p;1;1-\alpha)\, . 
\end{align}
The light single-trace operator must have even dimension for the three-point function not to vanish. 
Analogously for the AdS giant graviton we have
\begin{align}\label{eq:O2p_G}
\mathfrak{C}_{\cG_{AdS;\beta},\cG_{AdS;\beta},(2p)}^{(0)} &=  2^{-p} \oint_{z=0} {dz \over 2\pi i } {1-z^2 \over z^{2p+1} \sqrt{1-2z^2(2\beta+1) + z^4 } }  
 \cr 
 &=-(-2)^{-p+1} \, _2F_1(-p,p;1;1+\beta)\, .
\end{align}
It is interesting to note that the sphere giant graviton three point function \eqref{eq:O2p_D} and the AdS one \eqref{eq:O2p_G} are simply related by the exchange $\alpha \leftrightarrow -\beta$ which is again linked to the map $N\leftrightarrow -N$.
The results presented in equations \eqref{eq:O2p_D}-\eqref{eq:O2p_G} will shortly be instrumental in computing the HHLL integrated correlators.

\section{Integrated correlators with heavy insertions} \label{integrated correlators}
In this section we are interested in studying four-point correlation functions of two (sphere or AdS) giant gravitons and two light operators $\cO_2$ as a special case of HHLL correlators. In particular, we consider particular averages of these HHLL correlators where the space-time insertion points of the operators are integrated over with respect to a certain measure, thus leading to HHLL \textit{integrated} correlators. 

We now proceed to review the derivation of \cite{Brown:2024tru}, where these integrated correlators were computed for the case of the maximal sphere giant graviton operator,~i.e.~for $\cG_{S;\alpha}$ with $\alpha=1$. By exploiting this derivation we will manage to compute HHLL integrated correlators for general giant graviton operators without the need for discussing the microscopic details of the heavy insertions. 

\subsection{Four-point functions of half-BPS operators}
Four-point correlation functions of half-BPS operators in $\cN=4$ SYM are strongly constrained by superconformal symmetry. 
In particular, correlators of two light operators $\cO_2$ and two generic heavy insertions $\cH$ take the following general form \cite{Eden:2000bk, Nirschl:2004pa}, 
\begin{align} \label{eq:22HH_general}
     \vev{\cO_2(x_1, Y_1) \cO_2(x_2, Y_2) \cH(x_3, Y_3)\cH(x_4, Y_4)} = \mathcal{C}_{\textrm{free}}(x_i, Y_i) + \mathcal{R}_4 (x_i, Y_i) \, d_{34}^{\Delta_{\cH}-2} \, \mathcal{C}_{\cH}(u,v; \tau) \, ,
\end{align}
where $\Delta_{\cH}$ is the dimension of the operator $\cH$. 
In this expression $\mathcal{C}_{\textrm{free}}(x_i, Y_i) $ denotes the free-theory part which can be computed via standard Wick contractions, whereas the second term, $\mathcal{C}_{\cH}(u,v; \tau)$, includes all quantum corrections. The complete R-symmetry dependence of this four-point function can be factorised and it is fully contained in the term $\cR_4 (x_i, Y_i)d_{34}^{\Delta_{\cH}-2}$ whose explicit expression is not needed here but which can be found for example in section 2 of \cite{Brown:2024yvt}. The so-called \textit{reduced correlator} $\mathcal{C}_{\cH} (u,v; \tau)$ contains all the dynamics of this observable, being a function of the complexified Yang-Mills coupling constant
\begin{equation}\label{eq:tau_def}
    \tau:= \tau_1 + \ii\, \tau_2 = \frac{\theta}{2\pi} +\ii \frac{4\pi }{\gym^2}\,,
\end{equation}
and of the spacetime cross-ratios
\begin{align}
u\coloneqq \frac{x_{12}^2 x_{34}^2 }{ x_{13}^2 x_{24}^2} \, , ~~~~~~ v\coloneqq \frac{x_{14}^2 x_{23}^2 }{ x_{13}^2 x_{24}^2} \, .
\end{align}

In order to study the properties of the reduced correlator $\mathcal{C}_{\cH} (u,v; \tau)$ in the planar limit, we consider the 't Hooft regime:
\begin{equation}\label{eq:thooft}
N\to\infty\,,\qquad{\rm with}\quad\lambda=\gym^2 N~~~\textrm{fixed}~.
\end{equation}
Computing the large-$N$ expansion of correlators of the form $\vev{\cO_2 \cO_2 \cH \cH}$ for operators $\cH$ with dimension $\Delta_\cH \sim N$ is generically a challenging task since the dimension of the heavy operator necessarily affects the large-$N$ expansion.

For this reason very limited results for these four-point functions are available in the literature. In the case of the sphere giant graviton of dimension $N$ (i.e. the determinant operator \eqref{eq:D_alpha_def} with $\alpha=1$), the perturbative expansion at small $\lambda$ for the corresponding reduced correlator $\cC_\cH$ has been determined up to two loops in \cite{Jiang:2019xdz} and  further progress has been made at three loops in \cite{Jiang:2023uut}. More recently, the leading strong coupling expression, namely the supergravity contribution in the dual holographic picture, has been obtained in \cite{Chen:2025yxg} by using a superconformal bootstrap approach.

In this paper we aim at obtaining information on the HHLL correlators to all orders both in the weak and the strong coupling regime by considering a space-time average of the four-point function, defined as the integral of $\mathcal{C}_{\cH}(u,v; \tau)$ over a specific measure of the cross-ratios \cite{Binder:2019jwn, Chester:2020dja}, 
\begin{align} \label{eq:measureFT}
\cI_{\cH}(\tau; N) \coloneqq -\frac{2}{\pi} \int_0^{\infty} dr \int^{\pi}_0 d \theta {r^3 \sin^2 \theta \over u^2} \mathcal{C}_{\cH} (u,v; \tau) \, , 
\end{align}
where $u=1-2r \cos\theta  +r^2$ and $v=r^2$. 

This seemingly ad-hoc choice of integration measure is dictated by supersymmetry.
The reason for this is rather indirect.
Firstly, in the $\cN=4$ theory we can turn on a mass parameter $m$ while preserving $\cN=2$ supersymmetry. In the resulting theory, known as $\cN=2^*$ SYM, we can evaluate the partition function on the four-sphere via supersymmetric localisation thus reducing the quantum field theory path integral to a matrix model \cite{Pestun:2007rz}. The particular integrated correlator defined in \eqref{eq:measureFT} can be obtained from the $\cN=2^*$ partition function on $S^4$ by taking a suitable combination of four derivatives with respect to the parameters of the theory.
From a path integral point of view the action of these derivatives on the partition function of $\cN=2^*$ produces the specific integration measure appearing in \eqref{eq:measureFT}.

However, the same quantity can also be computed in the localised matrix model for generic coupling constant $\tau$ and number of colours $N$, thus offering a pathway to probe the non-perturbative structure of the corresponding four-point correlators. As already shown in \cite{Brown:2024tru}, from this matrix model formulation we can obtain exact results for the integrated correlator of two sphere giant graviton operators of dimension $N$ in the leading and subleading orders in the large-$N$ expansion. 

We now proceed to present some more details regarding the matrix model calculation of the integrated correlator \eqref{eq:measureFT}.

\subsection{Matrix model details}

We now review some basic facts about the matrix model formulation for the partition function of $ \cN=2^*$ SYM on $S^4$, which we then use to compute the integrated correlators defined in \eqref{eq:measureFT}. 

A well-known result due to Pestun  \cite{Pestun:2007rz} shows that the partition function of  $\cN=2^*$ SYM with gauge group $SU(N)$ on $S^4$ can be expressed as a matrix model integral using supersymmetric localisation,
\begin{equation}\label{eq:Z_matrix_model}
    \begin{aligned}
\cZ(\tau, \tau_p; m) =\! \int_{\mathbb{R}^N} \! d\mu(a_i)  \left \vert \exp\left[ \ii\, \pi \tau \Big(\sum_{i=1}^N a_i^2\Big) +\ii \sum_{p=3}^\infty \pi^{\frac{p}{2}} \tau_p \Big(\sum_{i=1}^N a_i^p \Big)\right]\right\vert^2 \!   
Z_{1\textrm{-loop}}(a; m) \left\vert Z_{\rm inst}(\tau, \tau_p, a;m) \right\vert^2\phantom{\bigg|} \, . 
\end{aligned}
\end{equation}
The integration measure is given by
\begin{equation} \label{eq:measure_a}
  d\mu(a_i)  =  \prod_{i=1}^N da_i \,\prod_{i<j} a_{ij}^2~  \delta\bigg( \sum_{i=1}^N a_i\bigg) \, ,
\end{equation}
where $a_{ij} \coloneqq a_i -a_j$, while the integration variables $a_i$ are the eigenvalues of the $N\times N$ hermitian matrix $a$ and satisfy the $\mathfrak{su}(N)$ tracelessness condition $\sum_{i=1}^N a_i =0$. 

The exponential factor in \eqref{eq:Z_matrix_model} contains both a contribution from the classical action, proportional to the complexified Yang-Mills coupling $\tau$, as well as sources for chiral primary operators on the $S^4$, each multiplied by the higher-dimensional couplings $\tau_p$.
The integrand factors $Z_{1\textrm{-loop}}$ and $Z_{\rm inst}$ correspond respectively to the perturbative one-loop determinant and the non-perturbative instanton contributions \cite{Nekrasov:2002qd}. In particular, for the one-loop determinant we have
\begin{equation}\label{eq:Z_1loop_mm}
    Z_{1\textrm{-loop}}(a; m) = \frac{1}{H(m)^N} \prod_{i<j} \frac{H^2(a_{ij})}{H(a_{ij}+m)H(a_{ij}-m)}\,,~~~~H(x) \coloneqq \rme^{-(1+\gamma)x^2} G(1+\ii x) G(1-\ii x)\,,
\end{equation}
where $G(x)$ denotes Barnes $G$-function.
Since presently we are interested in computing the 't Hooft large-$N$ limit where instanton corrections are exponentially suppressed in $N$, we will not need an explicit expression for Nekrasov instanton partition function $Z_{\mathrm{inst}}$. 

For $m=\tau_p=0$, both $Z_{1\textrm{-loop}}$ and $ Z_{\rm inst}$ evaluate to $1$ and the matrix integral \eqref{eq:Z_matrix_model} reduces to a Gaussian matrix model. After a rescaling of the matrix $a$,
\begin{equation}\label{eq:rescaling_tau}
    a \rightarrow \frac{a}{\sqrt{2\pi \tau_2}}\,,
\end{equation} 
the expectation value of any observable $f(a_i)$ in the Gaussian matrix model is defined by
\begin{equation}\label{eq:GaussianVEV}
    \llangle f(a_i)\rrangle \coloneqq \frac{1}{\mathcal{N}} \int_{\mathbb{R}^N}   d\mu(a_i)  \exp\bigg(\!\! - \sum_i a_i^2\bigg) f(a_i) \, ,
\end{equation}
where $\mathcal{N}$ is a normalisation factor   
such that $\llangle 1 \rrangle = 1$.

A very important class of matrix model observables is obtained by taking derivatives of \eqref{eq:Z_matrix_model} with respect to the couplings $\tau,\tau_p$ and their complex conjugates.  If we consider a combination of derivatives $\partial_{\tau_{p_1}}\dots \partial_{\tau_{p_\ell}}$ acting on \eqref{eq:Z_matrix_model} we produce an observable which is associated with the insertion of chiral/antichiral operators located at the North/South poles on $S^4$, i.e. schematically we have
\begin{equation}\label{eq:multiTr_S4}
   \partial_{\tau_{p_1}}\dots \partial_{\tau_{p_\ell}} \cZ(\tau, \tau_p; m)\quad \longleftrightarrow \quad T_{\vec p}(a) \coloneqq \tr \big(a^{p_1}\big) \tr \big( a^{p_2} \big)\,\cdots \tr \big(a^{p_\ell}\big)\,,
\end{equation}
where we denote by $\vec{p}$ the vector $\vec{p} = (p_1,...,p_\ell)$ and for ease of notation we identify the complexified Yang-Mills couplding $\tau$ with $\tau_2$.

However, such operators defined on $S^4$ do not directly correspond to chiral primaries on $\mathbb R^4$ \cite{Gerchkovitz:2016gxx}. This is due to a non-trivial mixing amongst operators with different dimensions on the sphere. To derive a precise mapping between operators defined on $S^4$ and those defined on $\mathbb R^4$, we need to perform a Gram-Schmidt orthogonalisation procedure, as prescribed in \cite{Gerchkovitz:2016gxx} (see also \cite{Billo:2017glv}).
The $\mathbb{R}^4$ operator $\cO_{\vec p}(a)$ is then constructed via the normal-ordering procedure,
\begin{equation}
\label{eq:normalord_On}
   \cO_{\vec p}(a) \coloneqq 
    T_{\vec p}(a) +  \sum_{\vec q\,\vdash q< p} u_{\vec p, \vec q}(N)~T_{\vec q}(a) \, ,
\end{equation}
where the sum in \eqref{eq:normalord_On} runs over all partitions\footnote{Note that partitions where any $q_i=1$ are here excluded due to the $SU(N)$ tracelessness condition, $\tr(a)=0$.}  $\vec{q}$ of $q<p$, thus producing all multi-trace operators with dimension lower than $p$.
The coefficients $u_{\vec p, \vec q}(N)$ which appear in the definition  \eqref{eq:normalord_On} for the operator $\cO_{\vec p}(a)$ are determined by imposing the orthogonality conditions 
\begin{align}\label{eq:orthogonalization} 
    \llangle\cO_{\vec p}(a)\, T_{\vec r}(a)\rrangle =0\,,~~~ \forall \,0
    \leq r<p \quad {\rm{and}} \quad \forall \,
    {\vec r}\,~{\rm{partitions~of~}}r \,,
\end{align}
where $\llangle\phantom x\rrangle$ denotes the Gaussian matrix model expectation value \eqref{eq:GaussianVEV} and where $p=\sum_{i=1}^
\ell p_i$ with $\vec{p} = (p_1,...,p_\ell)$. Note that the above condition includes the case $T_{\vec r}(a) = 1$ thus necessarily implying that $ \llangle \cO_{\vec p}(a)\rrangle=0$ for $p > 0$.

An important property of the normal ordered operators \eqref{eq:normalord_On} is that their correlation functions in the gaussian matrix model reproduce the two- and three-point function coefficients of the $\cN=4$ SYM theory on $\mathbb{R}^4$. In particular, comparing with the field theory results \eqref{eq:OO_2pointFT} and \eqref{eq:OO_3pointFT}, one can prove that for $p,m>0$ we have
\begin{equation}\label{eq:MM_to_FT}
    \llangle \cO_{\vec p}(a) \cO_{\vec p}(a)  \rrangle = \mathfrak{R}_{\vec p} (N)~, ~~~~~ \frac{\llangle \cO_{\vec m}(a) \cO_{\vec m}(a) \cO_{\vec p}(a)  \rrangle}{\llangle \cO_{\vec m}(a) \cO_{\vec m}(a)  \rrangle}  =\mathfrak{C}_{\vec m\vec m\vec p} (N)~.
\end{equation}

The gaussian matrix model can therefore be used as a tool for computing the combinatorial coefficients of the field theory Wick contractions. Since we want to exploit the property \eqref{eq:MM_to_FT}, it is convenient to define the inverse of the dictionary \eqref{eq:normalord_On},
\begin{align}\label{eq:multiTrace_to_NO}
    T_{\vec q}(a) =\cO_{\vec q}(a) +  \sum_{\vec p \,\vdash p<q} v_{\vec q, \vec p}(N)~\cO_{\vec p}(a)\,,
\end{align}
which maps the multitrace operators $T_{\vec q}(a) $ on $S^4$ defined in \eqref{eq:multiTr_S4} to the normal ordered operators $\cO_{\vec q}(a) $ on $\mathbb{R}^4$. As shown in \cite{Brown:2024tru}, this change of basis is surprisingly symmetric so that the coefficients $u_{\vec q, \vec p}$ and $v_{\vec q, \vec p}$ are related by 
\begin{align}\label{eq:beta_to_alpha}
    v_{\vec q, \vec p}(N) = (-1)^{\frac{q-p}{2}} u_{\vec q, \vec p}(N)\,.
\end{align}
We refer to \cite{Brown:2024tru} for explicit examples of operators written in the two bases.

\subsection{A recipe for  HHLL integrated correlators}

We now return to discussing how to compute the integrated correlator $\cI_{\cH}(\tau; N)$ defined in \eqref{eq:measureFT} from the mass-deformed matrix model \eqref{eq:Z_matrix_model}. 

Firstly, from the matrix model point of view the insertion of an $\cO_2$ operator is realised by taking derivatives with respect to the mass parameter, $m$, following the idea that the mass deformation switches on a specific R-symmetry channel (the so-called moment map operator) of the $\mathbf{20}^\prime$ operator $\cO_2$ in $\cN=4$ SYM. The insertion of heavy operators $\cH$ is more involved. In principle, one should take the proper linear combination of $\tau_p$ derivatives \eqref{eq:multiTr_S4} to produce the normal ordered $\mathbb{R}^4$ operator $\cO_{\vec p}(a)$ as in \eqref{eq:normalord_On} that realises the linear combination of multitrace operators associated to the desired heavy operator $\cH$ on $\mathbb{R}^4$. We denote such a generic combination of $\tau_p$ derivatives and mixing coefficients by $\partial_\cH$. From the matrix model point of view we have that the integrated correlator is given by
\begin{equation}\label{eq:IH_mm_general}
     \mathcal{I}_{\cH} (\tau;N) = \frac{\partial_\cH \partial_\cH \partial_m^2 \log \mathcal{Z}(\tau, \tau_p; m) \, {\vert}_{\tau_p, m=0}}{\partial_\cH \partial_\cH \log \mathcal{Z}(\tau, \tau_p; m) \, {\vert}_{\tau_p, m=0}} \,.
\end{equation}

In order to bypass the difficulties for the actual computation of the particular linear combination of derivative $\partial_\cH$  which inserts the desired heavy operator in the matrix model we follow \cite{Brown:2024tru} and exploit the special properties \eqref{eq:MM_to_FT} of the normal ordered operators to devise a reverse engineering method.

We rewrite the partition function \eqref{eq:Z_matrix_model} for $\tau_p=0$ in the zero-instanton sector (i.e. $|Z_{\mathrm{inst}} |^2\to1$) as the following integral over the hermitian traceless matrix $a$ rescaled as in \eqref{eq:rescaling_tau},
\begin{align}\label{eq:Z2*}
   \cZ(\gym^{2};m)=\int \!da~\rme^{-\tr\,a^2}\,Z_{1-\textrm{loop}}\left(\frac{ a\,\gym}{2\sqrt{2} \pi} ;m \right)\,.
\end{align} 
By expanding the Barnes G-functions \eqref{eq:Z_1loop_mm} in the limit of small $m$, it is possible to rephrase the one-loop determinant factor, $Z_{1-\textrm{loop}}$, as an insertion in the Gaussian matrix model:
\begin{align}\label{eq:d2mZ_in_gaussMM}
  \partial_m^2\log \cZ(\gym^2;m)\Big|_{m=0}= \llangle\mathbf{M}_2(a,\gym^2)\rrangle =  
 \frac{1}{\mathcal{N}} \int \!da~\rme^{-\tr\,a^2}\, \mathbf{M}_2(a,\gym^2)\,,
\end{align} 
where $\mathbf{M}_2$ can be expanded perturbatively in $\gym$ as an infinite sum of operators on the sphere defined in \eqref{eq:multiTr_S4},
\begin{align}\label{eq:M2_doubleTraces}
\mathbf{M}_2(a,\gym^2)=-2\sum_{\ell=1}^\infty\sum_{j=0}^{2\ell}(-1)^{j+\ell} (2\ell+1) \binom{2\ell}{j}\,\zeta(2\ell+1)\Big(\frac{\gym^2}{8\pi^2}\Big)^\ell T_{(2\ell-j,j)}(a)\,.
\end{align}
Using the change of basis in~\eqref{eq:multiTrace_to_NO} we then rewrite~$\mathbf{M}_2$
in terms of normal-ordered operators~$\cO_{\vec p}(a)$,
\begin{align}\label{eq:M2_NO}
\mathbf{M}_2(a,\gym^2)=\!-2\sum_{\ell=1}^\infty\sum_{j=0}^{2\ell}(-1)^{j+\ell} (2\ell+1) \binom{2\ell}{j}\,\zeta(2\ell+1)\Big(\frac{\gym^2}{8\pi^2}\Big)^\ell \! \left[\cO_{(2\ell-j,j)}(a) +\!\! \sum_{\vec p\,\vdash p< 2\ell}v_{(2\ell-j,j), \vec p}~\cO_{\vec p}(a)\right]\,.
\end{align}

The integrated correlator in the presence of generic heavy operators \eqref{eq:IH_mm_general} may then be rewritten as \cite{Brown:2024tru} 
\begin{align}\label{eq:IH_as_vevM2}
    \cI_{\cH}  (\gym^2; N) = \frac{\llangle\mathcal{H}(a) \mathcal{H}(a) \mathbf{M}_2(a,\gym^2)\rrangle }{\llangle\mathcal{H}(a) \mathcal{H}(a)\rrangle}- \llangle\mathbf{M}_2(a,\gym^2)\rrangle\,.
\end{align}
By plugging in the above equation the expansion for $\mathbf{M}_2$ given in \eqref{eq:M2_NO} and then using the correspondence between three-point functions of normal ordered operators and field theory three-point coefficients \eqref{eq:MM_to_FT}, we can finally map the integrated correlators $ \cI_{\cH}$ in terms of an infinite sum of field theory three-point coefficients: 
\begin{equation}\label{eq:IH_3pt}
    \begin{split}
    \cI_{\cH}  (\gym^2; N) = \, -2\sum_{\ell=1}^\infty\sum_{j=0}^{2\ell}&(-1)^{j+\ell} (2\ell+1) \binom{2\ell}{j}\,\zeta(2\ell+1)\Big(\frac{g_{\rm YM}^2}{8\pi^2}\Big)^\ell \\ &
\times \left(\!\mathfrak{C}_{\cH,\cH,(2\ell-j,j)} +\! \sum_{\vec p \,\vdash p<2\ell}v_{(2\ell-j,j), \vec p}~\mathfrak{C}_{\cH,\cH,\vec p}\!\right) \, ,
\end{split}
\end{equation}
where the sum over partitions ${\vec p}$ does not include the case $p=0$ as this contribution cancels out in \eqref{eq:IH_as_vevM2} due to the second, disconnected factor.

From the above expression, it appears manifest that the calculation of the perturbative part of the integrated correlators $\cI_{\cH}$ relies on two simple inputs: the mixing coefficients $v_{(2\ell-j,j), \vec p}$, originating from the expansion \eqref{eq:M2_NO} of $Z_{\rm 1-loop}$ and which are completely independent from the heavy operators considered, and the field theory three-point function coefficients $\mathfrak{C}_{\cH,\cH,{\vec p}}$, which for special heavy insertions can be obtained in the regime of large conformal dimensions from semi-classical arguments.

\section{Exact results for giant graviton correlators} \label{exact}

In this section we use the matrix model result \eqref{eq:IH_3pt} to derive the HHLL integrated correlators \eqref{eq:measureFT} for two heavy giant gravitons and two superconformal primary operators $\mathcal{O}_2$ in the stress-tensor multiplet.

\subsection{Weak coupling perturbative expansions}

We specialise the formula \eqref{eq:IH_3pt} to the case of giant graviton operators, where the dimension of the heavy insertions is proportional to the number of colours $N$, and consider the large-$N$, fixed-$\lambda$ limit as in \eqref{eq:thooft}. 
In this limit, $\cI_\cH$ can be expanded in a genus expansion in powers of $1/N$,
\begin{align}\label{eq:I_H_genus}
       \mathcal{I}_{\cH}(\lambda; N) = \sum_{g=0}^{\infty} N^{1-g} \, \mathcal{I}^{(g)}_{\cH}(\lambda) \, . 
\end{align}
On the string theory side, this expansion is consistent with the holographic interpretation of these operators in terms of D3-branes. In this paper we focus our attention towards the leading order genus-zero term $\mathcal{I}^{(0)}_{\cH}(\lambda)$ in the genus expansion \eqref{eq:I_H_genus}, which on the dual string theory side corresponds to the disk string amplitude. 

As shown in \cite{Brown:2024tru}, the leading large-$N$ contribution is dominated by the single-trace operators. Expanding the second line of \eqref{eq:IH_3pt} at leading order in $N$ leads to
\begin{align}
    \mathcal{I}_{\cH}(\lambda; N) &\notag =  -2N \sum_{\ell=1}^\infty (-1)^{\ell} (2\ell+1)\,\zeta(2\ell+1)\Big(\frac{\lambda}{8\pi^2}\Big)^\ell  \\
    & \phantom{=} \label{eq:IH_lead} \times \Bigg[ 2\mathfrak{C}_{\cH,\cH,(2\ell)}^{(0)} \, + \, \sum_{j=0}^{\ell} \binom{2\ell}{2j} \sum_{p=1}^{\ell-1} v_{(2\ell-2j,2j),(2p)}^{(0)} \mathfrak{C}_{\cH,\cH,(2p)}^{(0)} \Bigg] + O(N^0)~, 
\end{align}
where at leading order the mixing coefficients $v_{(2\ell-2j,2j), (2p)}^{(0)}$ are given by,
\begin{align} \label{eq:beta0}
v_{(2\ell-2j,2j), (2p)}^{(0)} = 2^{p-\ell} \bigg[ \binom{2\ell-2j}{\ell-j+p}\,C_j+ \binom{2j}{j+p}\,C_{\ell-j} \bigg] \, ,
\end{align}
with $C_n$ denoting the $n^{th}$ Catalan number.

Substituting the above expression for $v_{(2\ell-2j,2j), (2p)}^{(0)}$ in \eqref{eq:IH_lead}, after some manipulations we can rewrite the leading contribution to $\mathcal{I}_{\cH}(\lambda; N)$ in the very simple form,
\begin{align}\label{eq:IH_3pt_lead}
    \mathcal{I}^{(0)}_{\cH}(\lambda) = -4 \sum_{\ell=1}^\infty (-1)^{\ell} \,\zeta(2\ell+1)\Big(\frac{\lambda}{16\pi^2}\Big)^\ell \sum_{p=1}^{\ell}  2^p \binom{2\ell+1}{\ell+p+1} \binom{2\ell+1}{\ell+p} \mathfrak{C}_{\cH,\cH,(2p)}^{(0)}\,.
\end{align}
We stress that this formula is general and applies to any heavy operator $\cH$ whose conformal dimension $\Delta_\cH\sim N$.

At this point we specialise the analysis of HHLL integrated correlators to the cases of the sphere giant gravitons with dimension $\alpha N$ and the AdS giant gravitons with dimension $\beta N$.
This can be done directly at the level of equation \eqref{eq:IH_3pt_lead} by simply using the field theory expressions for the three-point functions coefficients $\mathfrak{C}_{\cH,\cH,(2p)}^{(0)}$ given in \eqref{eq:O2p_D} and \eqref{eq:O2p_G} respectively for the sphere and AdS giant graviton.

To simplify the notation, from now on we drop the superscript $\,^{(0)}$ since in the rest of the paper we will only be discussing the leading large-$N$ planar order. Moreover, we now explicitly display the dependence from the physical parameters $\alpha$ and $\beta$ defining the dimensions of the heavy giant graviton operators $\mathcal{H} = \cG_{S;\alpha}$ and $\mathcal{H} = \cG_{AdS;\beta}$, hence we will be referring to the leading large-$N$ integrated correlators as 
\begin{equation}\label{eq:Iplan}
\mathcal{I}_{S}(\lambda;\alpha)\coloneqq \mathcal{I}^{(0)}_{\cG_{S;\alpha}}(\lambda)\,,\qquad\qquad \mathcal{I}_{AdS}(\lambda;\beta)\coloneqq \mathcal{I}^{(0)}_{\cG_{AdS;\beta}}(\lambda)\,.
\end{equation}

Starting with the sphere giant graviton case, we substitute the three-point function coefficient \eqref{eq:O2p_D} in \eqref{eq:IH_3pt_lead} to obtain the weak coupling perturbative expansion
\begin{align} 
    \mathcal{I}_{S}(\lambda;\alpha) &= 8 \sum_{\ell=1}^\infty (-1)^{\ell} \,\zeta(2\ell+1)\Big(\frac{\lambda}{16\pi^2}\Big)^\ell \sum_{p=1}^{\ell}   (-1)^p \binom{2\ell+1}{\ell+p+1} \binom{2\ell+1}{\ell+p}  \, _2F_1(-p,p;1;1-\alpha) \, ,
\end{align}
where $0\leq \alpha \leq 1$. The summation over $p$ can be  performed and the resulting expression can be conveniently separated into two terms, 
\begin{align} \label{eq:pert1}
    \mathcal{I}_{S}(\lambda;\alpha) = \mathcal{I}_{S; 0}(\lambda) + \mathcal{I}_{S; 1}(\lambda;\alpha) \, ,
\end{align}
where we defined
\begin{align} \label{eq:pert1I0}
\mathcal{I}_{S; 0}(\lambda) &\coloneqq 4 \sum_{\ell=1}^\infty (-1)^{\ell+1} \,\zeta(2\ell+1)\Big(\frac{\lambda}{16\pi^2}\Big)^\ell \binom{2\ell{+}1}{\ell}^2 \, , \\ 
\mathcal{I}_{S; 1}(\lambda;\alpha) &\coloneqq 4 \sum_{\ell=1}^\infty (-1)^{\ell} \,\zeta(2\ell+1)\Big(\frac{\lambda}{16\pi^2}\Big)^\ell  \alpha^{2(\ell+1)} \binom{2\ell {+} 1}{\ell}  \, _2 F_1 \big(\ell+1,\ell+2;1;1-\alpha\big) \, . \label{eq:pert1I1}
\end{align}
When $\alpha=1$ (i.e. for the maximal determinant operator), the hypergeometric function $_2 F_1$ reduces to one thus recovering the result of \cite{Brown:2024tru},
\begin{equation}
    \mathcal{I}_{S}(\lambda;1) = 4 \sum_{\ell=1}^\infty (-1)^{\ell+1} \,\zeta(2\ell+1)\Big(\frac{\lambda}{16\pi^2}\Big)^\ell \bigg[\binom{2\ell{+}1}{\ell}^2- \binom{2\ell{+}1}{\ell} \bigg] \, .
\end{equation}

An analogous procedure can be repeated for the dual giant graviton. We substitute the three-point function coefficient \eqref{eq:O2p_G} in \eqref{eq:IH_3pt_lead} to derive
\begin{align} \label{eq:pert2}
    \mathcal{I}_{AdS}(\lambda;\beta)  = \mathcal{I}_{AdS; 0}(\lambda) + \mathcal{I}_{AdS; 1}(\lambda;\beta) \, ,
\end{align}
where we defined
\begin{align} \label{eq:pert2I0}
\mathcal{I}_{AdS; 0}(\lambda) &\coloneqq 4 \sum_{\ell=1}^\infty (-1)^{\ell} \,\zeta(2\ell+1)\Big(\frac{\lambda}{16\pi^2}\Big)^\ell \binom{2\ell{+}1}{\ell}^2 = - \, \mathcal{I}^{(0)}_{S} (\lambda)\, , \\
\mathcal{I}_{AdS; 1}(\lambda;\beta) &\coloneqq 4 \sum_{\ell=1}^\infty (-1)^{\ell+1} \,\zeta(2\ell+1)\Big(\frac{\lambda}{16\pi^2}\Big)^\ell  (1{+}\beta)^{\ell} \binom{2\ell {+} 1}{\ell} (\ell+1) \, _2F_1 \big(\! -\ell, -\ell;2;\frac{1}{1+\beta}\big) \, , \label{eq:pert2I1}
\end{align}
and we stress that now we must have $\beta \geq 0$. 
We note that for integer $\ell \in \mathbb{N}$ the hypergometric function in the above expression satisfies the identity
\begin{equation}\label{eq:2F1Id}
(1+\beta)^\ell(\ell+1)\,_2F_1 \big(\! -\ell, -\ell;2;\frac{1}{1+\beta}\big) = \beta^{2(\ell+1)}\,_2 F_1 \big(\ell+1,\ell+2;1;1+\beta\big)\,,\qquad \forall \,\ell\in \mathbb{N}\,,
\end{equation}
so that the AdS giant graviton weak coupling expansion can be re-expressed as
\begin{align} \label{eq:pert2I1-2}
\mathcal{I}_{AdS; 1}(\lambda;\beta) =4 \sum_{\ell=1}^\infty (-1)^{\ell+1} \,\zeta(2\ell+1)\Big(\frac{\lambda}{16\pi^2}\Big)^\ell  \beta^{2(\ell+1)} \binom{2\ell {+} 1}{\ell}  \, _2 F_1 \big(\ell+1,\ell+2;1;1+\beta\big) \, . 
\end{align}

From equation \eqref{eq:pert2I1-2} it is immediate to see that the perturbative series of the two integrated correlators $ \mathcal{I}_{S}(\lambda;\alpha)$ and $\mathcal{I}_{AdS}(\lambda;\beta)$ are related to each other at every $\ell$-loop order by $\beta \leftrightarrow - \alpha$, i.e. we have $ \mathcal{I}_{S}(\lambda;\alpha) = - \mathcal{I}_{AdS}(\lambda;-\alpha)$.
But we stress that the physical parameters satisfy the constraints $0\leq \alpha \leq1 $ and $\beta\geq0$, which are clearly not kept invariant under the exchange $\alpha \leftrightarrow -\beta$. As a consequence, we will shortly see that for physical values of the parameters $\alpha$ and $\beta$ the strong coupling expansions of $ \mathcal{I}_{S}(\lambda;\alpha)$ and $\mathcal{I}_{AdS}(\lambda;\beta)$ will not be related by the above symmetry.

We remark that the formulae given in \eqref{eq:pert1} and \eqref{eq:pert2} provide the first all-loop results for the integrated four-point functions of two sphere or AdS giant gravitons of generic dimensions and two half-BPS operators of dimension two, even though very little is known for the unintegrated correlators in the literature. These integrated correlators results thus provide important constraints for the four-point correlators involving sphere giant gravitons or AdS giant gravitons. For example, at one-loop order these correlators should be expressed in terms of a single one-loop box integral hence the integrated correlator results fix the correlators completely. At two loops the conformal integrals are the double box and one-loop box squared; in this case the integrated correlators provide constraints for the coefficients of these two integrals.\footnote{Other approaches, for example OPE analysis, may provide additional constraints that can help with fixing the two-loop (and even higher-loop) correlators completely.}

We now proceed to use the weak-coupling expressions \eqref{eq:pert1} and \eqref{eq:pert2} to go beyond the small-$\lambda$ expansion and study the HHLL integrated correlators in the strong-coupling regime.

\subsection{From weak to strong coupling} \label{sec:ana-con}

In this subsection we use the weak coupling perturbative data to construct analytic continuations in the 't Hooft coupling $\lambda$, which give us access to the strong coupling regime.

We begin by analysing the analytic properties of the two expansions \eqref{eq:pert1} and \eqref{eq:pert2}.
Both weak coupling perturbative series are in fact convergent expansions for which it is straightforward to determine the radii of convergence. For the contributions  $\mathcal{I}_{S; 0}(\lambda)= - \mathcal{I}_{AdS; 0}(\lambda)$ it is  immediate to see that the radius of convergence is $|\lambda| < \pi^2$. 

Focusing now on $\mathcal{I}_{S; 1}(\lambda;\alpha)$ we use the integral representation of the hypergometric function to show that for $0\leq \alpha<1$ it satisfies the large-$\ell$ asymptotic behaviour 
\begin{align}
\alpha^{2(\ell+1)}\, _2 F_1 \big(\ell+1,\ell+2;1;1-\alpha\big) \sim    \frac{\left(1+ \sqrt{1-\alpha }\right)^{2 (\ell+1)} }{2\,  \sqrt{\pi \, \ell}\, \sqrt[4]{1-\alpha } } \, , \qquad \ell \gg 1
\end{align}
from which we easily deduce the radius of convergence for $\mathcal{I}_{S; 1}(\lambda;\alpha)$ and given by
\begin{align} \label{eq:converg1}
    |\lambda| < R_S(\alpha)\coloneqq \pi^2 \frac{4}{ (1+\sqrt{1-\alpha})^2} \, . 
\end{align}
Note that for the giant graviton correlator we have $\alpha\in (0,1]$ thus implying that $R_S(\alpha)> \pi^2$. We conclude that the convergence radius of $\mathcal{I}_{S}(\lambda;\alpha)$ is dictated by that of $\mathcal{I}_{S; 0}(\lambda)$, which is $|\lambda| < \pi^2$ and is thus in particular independent from $\alpha$. 

Similarly, we find the convergence radius for $\mathcal{I}_{AdS; 1}(\lambda;\beta)$ to be
\begin{align} \label{eq:converg2}
    |\lambda| < R_{AdS}(\beta) \coloneqq \pi^2 \frac{4}{ (1+\sqrt{1+\beta})^2} \, , 
\end{align}
which is related to \eqref{eq:converg1} through $\beta \to - \alpha$,~i.e. $R_{AdS}(\beta) = R_S(-\beta)$. Since for the AdS giant graviton $\beta> 0$ we have that $R_{AdS}(\beta) < \pi^2$ which in turn implies that the radius of convergence for $\mathcal{I}_{AdS}(\lambda;\beta)$ is also dictated by \eqref{eq:converg2}, rather than the previous bound $|\lambda| < \pi^2$. 

We stress that it is to be expected that the weak coupling expansion of such observables yields convergent series. Firstly, at large $N$ the Feynmann diagrams contributing to any physical quantity are the planar ones which only grow exponentially fast (rather than the standard factorial growth) at any given loop order. Secondly, in $\mathcal{N}=4$ SYM we do not expect any factorial growth of the perturbative coefficients due to renomalons-type of diagrams, thus yielding convergent weak coupling expansions at weak 't Hooft coupling order by order in the genus expansion.

However, we find it peculiar that while for the sphere giant graviton the radius of convergence at weak coupling is $\pi^2$, as already noted in many other quantities in  $\mathcal{N}=4$ SYM in the 't Hooft large-$N$ expansion \cite{Dorigoni:2021guq, Hatsuda:2022enx, Beisert:2006ez, Basso:2020xts, Dunne:2025wbq}, for the AdS giant graviton the radius of convergence is given by the unusual\footnote{Another example, which also exhibits a radius of convergence distinct from  $\pi^2$, is the HHLL integrated correlators where one considers the heavy operators $(\cO_2)^{p}$ with $p \propto  N^2$ \cite{Paul:2023rka}.}  $R_{AdS}(\beta) < \pi^2$. It would be interesting to understand from a Feynman diagram point of view why the dual giant graviton weak coupling expansion behaves so differently.

The convergence of the perturbative series implies that we can analytically continue the weak coupling expansions to generic values of $\lambda$, in particular to $\lambda \gg 1$. 
To this end we first notice that the convergent perturbative series \eqref{eq:pert1} and \eqref{eq:pert2} can be analytically continued via Mellin integral representations.\footnote{Analogous representations have appeared in the same context of integrated correlators using $SL(2, \mathbb{Z})$ spectral representation \cite{Paul:2022piq, Collier:2022emf}.} 
For the sphere giant graviton correlators these take the form
\begin{align} \label{eq:inte11}
 \mathcal{I}_{S; 0}(\lambda) &=  -  \int_{{\rm Re}\,s=\frac{1}{2} } {ds \over  2 \pi \ii}  \, {4\pi \, 
 \zeta(2s{+}1) \over \sin(\pi s)}   \binom{2s{+}1}{s}^2   \Big(\frac{\lambda}{16\pi^2}\Big)^s \, , \\
 \mathcal{I}_{S; 1}(\lambda;\alpha) &=    \int_{{\rm Re}\,s=\frac{1}{2} } {ds \over  2 \pi \ii}  \, {4\pi \, 
 \zeta(2s{+}1) \over \sin(\pi s)}   \alpha^{2(s+1)} \binom{2s {+} 1}{s}  \, _2 F_1 \big(s+1,s+2;1;1 {-} \alpha\big)  \Big(\frac{\lambda}{16\pi^2}\Big)^s \, , \label{eq:inte12}
\end{align}
while for the AdS giant graviton correlators we have
\begin{align} \label{eq:inte21}
 \mathcal{I}_{AdS; 0}(\lambda) &=  -  \mathcal{I}_{S; 0}(\lambda) \\
 \mathcal{I}_{AdS; 1}(\lambda;\beta) &= - \int_{{\rm Re}\,s=\frac{1}{2} } {ds \over  2 \pi \ii}  \, {4\pi \, 
 \zeta(2s{+}1) \over \sin(\pi s)}    (1{+}\beta)^{s} \binom{2s {+} 1}{s} (s+1) \, _2F_1 \big(\! -s, -s;2;\frac{1}{1{+}\beta}\big)   \Big(\frac{\lambda}{16\pi^2}\Big)^s \, .  \label{eq:inte22}
\end{align}
When $\lambda \to 0$ we can close the contour of integration to the right half-plane ${\rm Re}(s) >1/2$. Picking up the residues from the poles of $1/\sin(\pi s)$ located at $s\in \mathbb{N}^{\geq0}$ we reproduce the perturbative series \eqref{eq:pert1} and \eqref{eq:pert2}. 

However, it is important to stress that the above Mellin integral representations are valid for generic $\lambda \geq 0$. 
The reason for that is as follows. Firstly, since the perturbative weak coupling expansions  \eqref{eq:pert1} and \eqref{eq:pert2} are convergent series they do define analytic functions of $\lambda$ within the corresponding radii of convergence discussed previously. Furthermore, the Mellin integral representations above define analytic functions of $\lambda$ which coincide with their corresponding weak coupling expressions within the convergence disk and thus define the unique analytic continuations in $\lambda$ for the weak coupling expressions and for physical values of the parameters $0< \alpha \leq 1\,,\, \beta> 0$. In particular, we can use these Mellin integrals to extract the strong coupling expansion of the integrated correlators.

Before proceeding, we would like to comment on the expressions of the hypergeometric functions in the integrated correlators which is inherently tied up with the continuation $\alpha \to -\beta$. For any positive integer $\ell$ appearing in the perturbative expansions \eqref{eq:pert1} and \eqref{eq:pert2}, one may use alternative expressions for hypergeometric functions. For instance, as already mentioned both formulae \eqref{eq:pert2I1}  and \eqref{eq:pert2I1-2} give the correct results for the perturbative expansion of $\mathcal{I}_{AdS; 1}(\lambda;\beta)$. 
However, the form of the hypergeometric function that we used in \eqref{eq:pert2I1} is the most convenient for finding an analytic continuation of the integrated correlator in $\lambda$ from weak to strong coupling. Using the expression \eqref{eq:pert2I1-2} in \eqref{eq:inte22} would lead to an ill-defined integral representation when $\beta>0$ since had we used the analytic continuation 
\begin{align}
    _2 F_1 \big(\ell+1,\ell+2;1;1+\beta\big)\qquad{\rm with}\,\, \ell \in \mathbb{N}\quad  \to \quad \,_2 F_1 \big(s+1,s+2;1;1+\beta\big)\qquad {\rm with} \,\, s\in \mathbb{C}\,, 
    \end{align}
we would have found that for generic $s\in \mathbb{C}$ and for physical values of the parameter $\beta>0$ the hypergeometric function is evaluated on its branch cut.

A similar discussion also applies to $ \mathcal{I}_{S; 1}(\lambda;\alpha)$ where it appears manifest from the hypergeometric function at the integrand that the integral representation \eqref{eq:inte12} is only valid for $\alpha>0$. 
To discuss $ \mathcal{I}_{S; 1}(\lambda;\alpha)$ with $\alpha<0$ we must start from the weak coupling expression \eqref{eq:pert1I1} and use the hypergeometric functions identity \eqref{eq:2F1Id} in ``reverse'', thus showing that the correct integral representation for $ \mathcal{I}_{S; 1}(\lambda;\alpha)$ with $\alpha<0$ is given by \eqref{eq:inte22} with $\beta = -\alpha$ and an overall minus sign.
In conclusion, we see that the correct continuation of $ \mathcal{I}_{S; 1}(\lambda;\alpha)$ to $\alpha<0$ is exactly $-\mathcal{I}_{AdS; 1}(\lambda;-\alpha)$, thus yielding the previously stated symmetry for the full correlators
\begin{equation}
\mathcal{I}_{S}(\lambda;\alpha) = -\mathcal{I}_{AdS}(\lambda;-\alpha)\,,\qquad \alpha<0\,.\label{eq:Z2}
\end{equation}
We will come back to the geometric nature of the above identity when we discuss the non-perturbative effects which appear at strong coupling $\lambda\gg1$.

As a sanity check we can also show that both correlators vanish in the limit $\alpha,\beta\to0$, as they should given that the heavy operators trivialise in this limit.
Firstly, we notice that the integrands of \eqref{eq:inte12} and \eqref{eq:inte22} satisfy the limit
\begin{align}
& \lim_{\alpha\to 0} \left[\alpha^{2(s+1)}\,_2F_1(s+1,s+2;1;1{-}\alpha) \right] = \lim_{\beta\to 0} \left[(1+\beta)^s(s+1)\,_2F_1(-s,-s;2;\frac{1}{1{+}\beta}) \right]  = {2s{+}1 \choose s} \,,
\end{align}
valid for $s\in \mathbb{C}$.
Given \eqref{eq:inte11}-\eqref{eq:inte21} we then derive the limits
\begin{align}
\label{eq:abzero}
&\lim_{\alpha\to 0}\Big[ \mathcal{I}_{S; 1}(\lambda;\alpha)\Big] =  -\lim_{\beta\to 0}\Big[ \mathcal{I}_{AdS; 1}(\lambda;\beta)\Big] = - \mathcal{I}_{S; 0}(\lambda)\,,
\end{align}
thus implying that both integrated correlators $\mathcal{I}_S(\lambda;\alpha) =\mathcal{I}_{S; 0}(\lambda) + \mathcal{I}_{S; 1}(\lambda;\alpha)$ as well as $\mathcal{I}_{AdS}(\lambda;\beta) =\mathcal{I}_{AdS; 0}(\lambda) + \mathcal{I}_{AdS; 1}(\lambda;\beta)$ vanish in the limit $\alpha,\beta\to 0$.
Note that the same result could have been derived starting directly from the perturbative weak coupling expansions \eqref{eq:pert1I1} and \eqref{eq:pert2I1}.

\subsection{Universality at strong coupling} \label{subsec:universality}

With the integral representations \eqref{eq:inte11}-\eqref{eq:inte12} and  \eqref{eq:inte21}-\eqref{eq:inte22}, it is straightforward to obtain the strong coupling expansion of the integrated correlators.

To this end we find it useful to change integration variables to $s\to 1-s$ and rewrite the above integrals as
\begin{align}
 \mathcal{I}_{S; 0}(\lambda) &\label{eq:inte11SC}=-\mathcal{I}_{AdS; 0}(\lambda)  = \int_{{\rm Re}\,s=\frac{1}{2}} \frac{ds}{2\pi \ii} \frac{8 \pi ^{s-\frac{3}{2}} (2 s-3)  \xi (2 s-2) \tan (\pi  s) \Gamma (s-2)^2}{\Gamma \left(s-\frac{3}{2}\right)}\lambda ^{1-s}\,,\\
  \mathcal{I}_{S; 1}(\lambda;\alpha) &\label{eq:inte12C}=- \int_{{\rm Re}\,s=\frac{1}{2}} \frac{ds}{2\pi \ii} 4^s \pi ^{s-1} (2 s-3) \xi (2 s-2) \Gamma (s-2) \, _2F_1(s-2,s-1;1;1-\alpha ) \lambda ^{1-s}\,,
\end{align}
and  for the dual giant graviton correlators,
\begin{equation}
 \mathcal{I}_{AdS; 1}(\lambda;\beta) = - \int_{{\rm Re}\,s=\frac{1}{2}} \frac{ds}{2\pi \ii} 4^s \pi ^{s-1} (2 s-3)  \xi (2 s-2) \Gamma (s-1)  \, _2F_1\left(s{-}1,s{-}1;2;\frac{1}{\beta {+} 1}\right) (\beta {+} 1)^{1-s} \lambda ^{1-s}\,.
\label{eq:inte21SC}
\end{equation}
For convenience we introduced the completed Riemann zeta function $\xi(s) \coloneqq \pi^{-s/2}\Gamma(s/2) \zeta(s)$ which we remind the reader is a meromorphic function of $s\in \mathbb{C}$ with simple poles at $s=0$ and $s=1$.

Let us begin with the sphere giant graviton correlators $\mathcal{I}_{S}(\lambda;\alpha)$, again separating the two contributions $\mathcal{I}_{S; 0}(\lambda)$ and $\mathcal{I}_{S; 1}(\lambda;\alpha)$. 
Since we are interested in the limit $\lambda \gg1$ we can close the contour of integration to the right-half plane ${\rm Re}(s) >1/2$ and collect the residues from all the poles.
It is easy to see that $\mathcal{I}_{S; 0}(\lambda)$ has non-trivial residues from the poles at $s=1, s=2$ and $s=m+1/2$ with $m=2,3, \ldots$.
This leads to the strong coupling perturbative expansion\footnote{From here onward with the superscript ${}^{({\rm p})}$ we denotes all terms which are purely perturbative in the large-$\lambda$ expansion. We will consider `non-perturbative' contributions ${}^{({\rm np})}$ in Section \ref{sec:non-pert}.}
\begin{align} \label{eq:I0-strong}
    \mathcal{I}_{S; 0}(\lambda)\vert_{\rm strong} &\sim \mathcal{I}_{S; 0}^{({\rm p})}(\lambda) \\
    &= 2 [\log (\lambda )+2 \gamma +2-2 \log (4 \pi ) ] + \frac{4 \pi ^2}{3 \lambda } - \sum_{n=1}^{\infty}   \frac{16 n \zeta (2 n+1) \Gamma
   \left(n-\frac{1}{2}\right)^2 \Gamma \left(n+\frac{1}{2}\right)}{  \lambda ^{  n+\frac{1}{2}}\, \pi ^{3/2}\, \Gamma (n)} \, , \nonumber
\end{align}
where $\gamma$ is Euler-Mascheroni constant.
Here with $\sim$ we denote the fact that for $\lambda \gg 1$ the function $ \mathcal{I}_{S; 0}(\lambda)$ is asymptotic in the sense of Poincar\'e to the formal, factorially divergent power series expansion $\mathcal{I}_{S; 0}^{({\rm p})}(\lambda)$.

On the other hand, when closing the integration contour to the right half-plane ${\rm Re}\, s>1/2$ we note that $\mathcal{I}_{S; 1}(\lambda; \alpha)$ has only two non-trivial residues coming from the poles located at $s=1$ and $s=2$, thus leading to the terminating perturbative expansion
\begin{align}\label{eq:IS1P}
    \mathcal{I}_{S; 1}(\lambda;\alpha)\vert_{\rm strong} \sim  \mathcal{I}_{S; 1}^{({\rm p})}(\lambda;\alpha) = 2 \alpha - 2 [\log (\lambda )+2 \gamma +2-2 \log (4 \pi ) ]  - \frac{8 \pi ^2}{3 \lambda } \, . 
    \end{align}
We note that although $\mathcal{I}_{S; 1}(\lambda;\alpha)$ is definitely not a polynomial in $\log(\lambda)$ and $\lambda^{-1}$, we find nonetheless that $\mathcal{I}_{S; 1}(\lambda;\alpha)$ is asymptotic to the perturbative expansion $\mathcal{I}_{S; 1}^{({\rm p})}(\lambda;\alpha)$ which terminates after finitely many terms.

Altogether we have
\allowdisplaybreaks{
\begin{align}\label{eq:IS_strong_fin}
  \mathcal{I}_{S}(\lambda;\alpha)\vert_{\rm strong} &=   \,  \mathcal{I}_{S; 0}(\lambda) \vert_{\rm strong}+  \mathcal{I}_{S; 1}(\lambda;\alpha) \vert_{\rm strong} \\*
  & \sim \mathcal{I}_{S; 0}^{({\rm p})}(\lambda)+ \mathcal{I}_{S; 1}^{({\rm p})}(\lambda;\alpha) = 
  2 \alpha - \frac{4 \pi ^2}{3 \lambda }  - \sum_{n=1}^{\infty}   \frac{16 n \zeta (2 n+1) \Gamma
   \left(n-\frac{1}{2}\right)^2 \Gamma \left(n+\frac{1}{2}\right)}{  \lambda ^{  n+\frac{1}{2}}\, \pi ^{3/2}\, \Gamma (n)} \, . \nonumber
\end{align}}
We notice that for $\alpha=1$ this expression reduces precisely to the results of \cite{Brown:2024tru} for the maximal giant graviton.

The computation for $\cI_{AdS}(\lambda;\beta)$ is analogous. Firstly, thanks to the identity \eqref{eq:inte11SC} we immediately read the perturbative expansion of $\mathcal{I}^{({\rm p})}_{AdS; 0} (\lambda)$ from that of $\mathcal{I}^{({\rm p})}_{S; 0} (\lambda)$ given in \eqref{eq:I0-strong}.
To derive the asymptotic expansion of $\mathcal{I}_{AdS; 1}(\lambda;\beta)$ for $\lambda \gg 1$ we close the contour of integration in \eqref{eq:inte21SC} to the right half-plane ${\rm Re}(s)>1/2$ and notice that the integrand has a single non-trivial residue from the pole at $s=1$.
This yields the simple asymptotic perturbative expansion 
\begin{align}\label{eq:IAdS1P}
    \mathcal{I}_{AdS; 1}(\lambda;\beta)\vert_{\rm strong} \sim \mathcal{I}_{AdS; 1}^{({\rm p})}(\lambda;\beta) =  2 \log(1+\beta) + 2 [\log (\lambda )+2 \gamma +2-2 \log (4 \pi ) ] \, . 
    \end{align}
Exactly as noticed in \eqref{eq:IS1P} for $ \mathcal{I}_{S; 1}^{({\rm p})}(\lambda;\alpha)$, we stress that although the function $  \mathcal{I}_{AdS; 1}(\lambda;\beta)$ is not a polynomial in $\log(\lambda)$ and $\lambda^{-1}$, its asymptotic expansion gives rise to the simple perturbative expression $\mathcal{I}_{AdS; 1}^{({\rm p})}(\lambda;\beta)$ which terminates after finitely many terms.
This fact will shortly have important consequences for the large-$\lambda$ non-perturbative transseries completions of both   $  \mathcal{I}_{S; 1}(\lambda;\alpha)$ and  $  \mathcal{I}_{AdS; 1}(\lambda;\beta)$.

Combining with $\mathcal{I}_{AdS; 0}$ we obtain the complete perturbative expansion for the AdS giant graviton correlator,
\begin{align}
    \mathcal{I}_{AdS}(\lambda;\beta)\vert_{\rm strong}  &\notag \sim\mathcal{I}_{AdS; 0}^{({\rm p})}(\lambda)+\mathcal{I}_{AdS; 1}^{({\rm p})}(\lambda;\beta) \\*
    &\label{eq:IAdS_strong_fin} =2 \log(1+\beta)- \frac{4 \pi ^2}{3 \lambda }  +  \sum_{n=1}^{\infty}   \frac{16 n \zeta (2 n+1) \Gamma
   \left(n-\frac{1}{2}\right)^2 \Gamma \left(n+\frac{1}{2}\right)}{  \lambda ^{  n+\frac{1}{2}}\, \pi ^{3/2}\, \Gamma (n)} \, . 
\end{align}
We stress that after having moved the contour of integration past all the poles, as in the above analysis, we should in principle analyse the contribution coming from ${{\rm Re}}(s)\to \infty$ for each of the integrals \eqref{eq:inte11SC}, \eqref{eq:inte12C}, and \eqref{eq:inte21SC}. We anticipate that such limits do not vanishing and actually hide all the non-perturbative and exponentially suppressed terms at large $\lambda$, which will be discussed in Section \ref{sec:non-pert}.

A few comments are in order. Firstly, according to the AdS/CFT correspondence, the strong coupling regime of the correlators here derived is holographically dual to a series expansion in stringy corrections. More precisely, in the holographic dictionary we have that $1/\sqrt{\lambda} \sim \ell_s^2$, where $\ell_s$ is the string length. Secondly, we highlight that the strong coupling expansions for both $\mathcal{I}_{S}(\lambda;\alpha)$ and $ \mathcal{I}_{AdS}(\lambda;\beta)$ depend on the orientation parameters $\alpha$ and $\beta$ only through their leading $\lambda^0$ terms, which holographically correspond to the supergravity contributions to the HHLL correlators. Thirdly, unlike the perturbative series at weak $\lambda$ (whose convergence properties have been discussed in the previous subsection) the strong coupling expansions are asymptotic, factorially divergent and non-Borel summable power series. All higher order terms in these asymptotic series are fully controlled by the perturbative expansions $\mathcal{I}^{({\rm p})}_{S; 0}(\lambda) = - \mathcal{I}^{({\rm p})}_{AdS; 0}(\lambda)$. This leads to the fact that the integrated correlators $\mathcal{I}_{S}(\lambda;\alpha)$ and $ \mathcal{I}_{AdS}(\lambda;\beta)$ have the same asymptotic series at large $\lambda$ (up to an overall minus sign in the half-integer powers), despite their small-$\lambda$ expansions being different. 

At strong coupling, the remaining differences between $\mathcal{I}_{S}(\lambda;\alpha)$ and $ \mathcal{I}_{AdS}(\lambda;\beta)$ are then entirely due to non-perturbative effects. Therefore, the natural next step is to analyse the contributions that are missed by the perturbative asymptotic expansions \eqref{eq:IS_strong_fin} and \eqref{eq:IAdS_strong_fin},
namely the exponentially suppressed, non-perturbative terms at large $\lambda$, as we will discuss in section \ref{sec:non-pert}.

Before turning to the analysis of the non-perturbative terms, let us first comment on $ \mathcal{I}_{AdS}(\lambda;\beta)$ when the parameter $\beta$ of the AdS giant graviton is taken to be large. We will see that this interesting limit is associated with rich semi-classical physics. 

\subsection{Very heavy AdS giant gravitons and Coulomb-branch physics} \label{sec:Coulomb}

In this subsection we discuss some important properties of the planar HHLL integrated AdS giant gravitons correlator with dimension $\Delta= \beta N$ in the limit where $\beta \gg 1$, namely the limit where the AdS giant gravitons become `very heavy'. As briefly commented in subsection \ref{sec:GG} and extensively discussed in \cite{Ivanovskiy:2024vel}, the large-$\beta$ behaviour of the AdS giant gravitons maps $\cN=4$ SYM to its Coulomb branch physics.  Here we provide an exact result for the integrated correlator in the presence of such AdS giant gravitons and discuss its connection with the semi-classical analysis of Coulomb branch physics.

To understand the large-$\beta$ behaviour of the HHLL integrated correlators of AdS giant gravitons we begin by considering the perturbative expansion order by order. From the weak coupling expansions given in \eqref{eq:pert2}-\eqref{eq:pert2I1} it is straightforward to write an explicit form for $\cI_{AdS}(\lambda;\beta)$ at large $\beta$.  
As manifest from the first few perturbative terms, 
\begin{equation}
    \cI_{AdS}(\lambda;\beta) =   \frac{3 \beta  \lambda  \zeta (3)}{2 \pi ^2} -\frac{15 \beta  (\beta +4) \lambda ^2 \zeta (5)}{32 \pi ^4}+\frac{35 \beta  \left(2 \beta ^2+15 \beta +30\right) \lambda ^3 \zeta (7)}{512 \pi ^6}+\dots \, ,
\end{equation}
we see that at each order in the weak coupling expansion, one may introduce an `effective coupling' $\lambda_\beta\coloneqq \beta\lambda$, which corresponds to the so-called large-charge 't Hooft coupling \cite{Bourget:2018obm}, and consider the $1/\beta$ expansion as $\beta\gg 1$ with $\lambda_\beta$ being fixed. In this limit, it is easy to see from the perturbative series \eqref{eq:pert2}-\eqref{eq:pert2I1} that only $\mathcal{I}_{AdS; 1}(\lambda;\beta)$ contributes, for which we can also replace the hypergometric function with $1$.
Thus the leading large-$\beta$ contribution to $\mathcal{I}_{AdS; 1}(\lambda;\beta)$, here denoted by $ \cI_{AdS}(\lambda;\beta)\big|_{\mathrm{lead-}\beta}$, takes the very simple and compact form,
\begin{align}\label{eq:large_beta_weak}
    \cI_{AdS}(\lambda;\beta)\big|_{\mathrm{lead-}\beta} &= 4 \sum_{\ell=1}^\infty (-1)^{\ell+1}\binom{2\ell}{\ell} (\ell+1) \zeta(2\ell+1) \Big(\frac{\lambda_\beta}{16\pi^2}\Big)^\ell \cr 
    &= 4 \int_0^{\infty} dw {w   \over \sinh^2(w) }\left[1- J_0\left( { w \sqrt{\lambda_\beta}  \over \pi} \right) \right] \, , 
\end{align}
where in the second line we have resummed the convergent weak coupling expansion using the identity
\begin{equation}
2^{1-s} \zeta(s) \Gamma(s+1) =  \int_0^\infty dw \frac{w^s}{\sinh^2(w)} \,,
\end{equation}
valid for ${\rm Re}(s)>1$, and express the sum over $\ell$ in terms of the Bessel function $J_0(x)$.  
From this integral representation we can easily derive the large-$\lambda_\beta$ transseries representation, 
\begin{align} \label{eq:largeb}
        \cI_{AdS}(\lambda;\beta)\big|_{\mathrm{lead-}\beta} &\notag =  4+4\gamma +2 \log \left({\lambda_\beta \over 16\pi^2}  \right)  + 8 \sum_{n=1}^{\infty} \left( n \sqrt{\lambda_\beta} K_1(n \sqrt{\lambda_\beta})-  K_0(n \sqrt{\lambda_\beta}) \right)\\
        &\notag =  4+4\gamma +2 \log \left({\lambda_\beta \over 16\pi^2}  \right) \\
        &\phantom{=}- 8  \sum_{n=1}^\infty (2 \pi n \sqrt{\lambda_\beta})^\frac{1}{2} \, e^{-n \sqrt{\lambda_\beta} } \sum_{k=0}^\infty \frac{(4k^2-8k-1) \Gamma(k-\frac{1}{2})^2}{ \pi \,\Gamma(k+1)}  (-2 n \sqrt{\lambda_\beta})^{-k}\,,
\end{align}
 We note that $ \cI_{AdS}(\lambda;\beta)\big|_{\mathrm{lead-}\beta}$ has only a finite number of perturbative terms at large $\lambda_\beta$, while the modified Bessel functions of the second kind, $K_\nu(x)$,  lead to the exponentially suppressed terms of the form $\exp[-n \sqrt{\lambda_\beta}]$ presented in the second expression.

Importantly, we note that the expression \eqref{eq:large_beta_weak} for the large-$\beta$ limit of the integrated correlators has exactly the same form as the results for a different class of HHLL integrated correlators in the  large-charge 't Hooft limit discussed in \cite{Brown:2025cbz}  (see equation (5.19) in that reference), which was derived from a semi-classical analysis.  
To map the present discussion with the results of the above reference, the effective coupling $\lambda_\beta$ should be replaced by $\lambda_\beta/(16\pi^2) \to M_s^2$. In \cite{Brown:2025cbz} the parameters $M_s$ denote the masses (labelled by the index $s$) of particular $\mathcal{N}=4$ SYM states which have become massive after giving a vev to the adjoint scalars, thus bringing the theory to its Coulomb phase. We now make this point more precise. 

From the results of \cite{Brown:2025cbz, Ivanovskiy:2024vel}, the above discussion suggests that the insertion of two symmetric Schur polynomials can be described semi-classically at large $\beta$. Following the analysis carried out in those references, in the semi-classical regime the insertion of the two symmetric Schur polynomials located at $(x_1,Y_1)$ and $(x_2,Y_2)$ can be effectively described by assigning the following vev to the adjoint scalar fields,
\begin{equation} \label{eq:vev}
    \vev{\Phi_I (x)}_\theta =  \frac{ 2\sqrt{\lambda_\beta} }{ 4\pi \sqrt{d_{12}}} \left( e^{i\theta} \Omega^{(N)}_2 \frac{(Y_1)_I}{(x-x_1)^2} + e^{-i\theta} {\Omega}^{(N)}_2 \frac{(Y_2)_I}{(x-x_2)^2} \right)  \,, \qquad {\rm with} \quad I=1,\dots, 6\, , 
\end{equation}
where $\theta \in [0, 2\pi ]$ is the modulus of the classical solutions, and $\Omega^{(N)}_2$ is a $N \times N$ matrix of the form\footnote{This is exactly of the same form as the scalar vev generated by the so-called heavy `canonical operators' in \cite{Brown:2025cbz} by setting the parameter $K=2$ in equation (3.3) of the reference. However, one should note that the regime studied in \cite{Brown:2025cbz} is different from what we consider here; in that reference the dimension of the heavy operators is much larger than $N^2$. Furthermore, the value $K=2$ is precisely the case that cannot be included for the heavy `canonical operators' considered in \cite{Brown:2025cbz}.}
\begin{align}
    \label{eq:Zclass-intr}
\Omega^{(N)}_2 = 
\begin{pmatrix}  
1 & 0 & 0 & \cdots & 0  \\
0 & -1 & 0 &  \cdots & 0    \\
0 & 0  & 0 & \cdots &  0 \\
\vdots & \vdots & \vdots & \ddots  & \vdots \\
0 & 0 & 0 & \cdots & 0 
\end{pmatrix} \, . 
\end{align}
Note that the scalar vev \eqref{eq:vev} breaks the gauge group from $SU(N) \to U(N-2) \times U(1)$, and gives rise to $4(N-2)$ massive scalars each with mass $M_1=\sqrt{\lambda_\beta}/(4\pi)$ and two massive scalars of mass $M_2=2\sqrt{\lambda_\beta}/(4\pi)$, thus effectively moving the $\mathcal{N}=4$ SYM theory to its Coulomb branch.

The HHLL correlator considered here may then be interpreted in terms of two-point functions of light operators in the semi-classical background generated by two heavy Schur polynomial operators, and the correlator is determined in terms of propagators of massive (and massless) states. In particular, in this limit the resulting correlator is given by the square of an infinite sum of ladder integrals \cite{Giombi:2020enj} (see \cite{Caetano:2023zwe, Brown:2025cbz} for further details); and the integrated correlator turns out to be identical to equation (5.19) of \cite{Brown:2025cbz} as mentioned earlier.  By summing over the contributions from all the massive states, we note that the number of massive states with mass $M_1=\sqrt{\lambda_\beta}/(4\pi)$ dominates in the large-$N$ limit, and we see immediately that the integrated correlator is indeed given by \eqref{eq:large_beta_weak}.

\section{Non-perturbative effects}
\label{sec:non-pert}

We now turn to consider the exponentially suppressed terms that contribute to $\mathcal{I}_{S}(\lambda; \alpha)$ and $\mathcal{I}_{AdS}(\lambda; \beta)$ at strong coupling.  We will shortly see that these are precisely the contributions that account for the difference between these integrated correlators beyond their identical, universal power series expansion discussed in Section \ref{subsec:universality}. 

\subsection{Exponentially suppressed corrections}
\label{sec:Exp}

It is again convenient to consider $\mathcal{I}_{S; 0}(\lambda)$ and $\mathcal{I}_{S; 1}(\lambda; \alpha)$ (or $\mathcal{I}_{AdS; 0}(\lambda)$ and $\mathcal{I}_{AdS; 1}(\lambda; \beta)$) separately. As commented in the previous section, the strong coupling perturbative expansion of $\mathcal{I}_{S; 0}(\lambda)$ (and equivalently $\mathcal{I}_{AdS; 0}(\lambda)$) given in \eqref{eq:I0-strong} is an asymptotic and factorially divergent power series which is in fact not Borel summable. The resurgent properties of this asymptotic series have been analysed in \cite{Brown:2024tru} where it was shown that the required non-perturbative terms for completing the asymptotic series $\mathcal{I}_{S; 0}^{({\rm p})}(\lambda)$ are given by
\begin{align}
    \mathcal{I}_{S; 0}^{({\rm np})}(\lambda) &= \pm \ii \, \frac{ 64\sqrt{\lambda} }{\pi ^2} \sum_{n=1}^{\infty} e^{-2 n \sqrt{\lambda } } \int_0^{ \infty} dz \, e^{-2 n \sqrt{\lambda} \, z }   \left[ (z+1)^3 G_{3,3}^{2,3}\left((z+1)^2|
\begin{array}{c}
 -\frac{1}{2},\frac{1}{2},\frac{1}{2} \\
 0,0,-\frac{3}{2} \\
\end{array}
\right)-\frac{\pi^3 }{4 } \right] \cr 
&\label{eq:I0NP}= \pm \ii  \sum_{n=1}^{\infty} e^{-2 n \sqrt{\lambda } } \left[\frac{32}{\sqrt{\lambda } \, n^2} + \frac{24}{\lambda \,  n^3} + \frac{5}{\lambda ^{3/2}\, n^4} -\frac{9}{4 \lambda ^2\, n^5} + \frac{123}{64 \lambda ^{5/2} \, n^6} + O(\lambda ^{-2})\right] \, ,
\end{align}
where $G^{a,b}_{c,d}$ denotes a Meijer $G$-function which we have expanded around $z=0$ to obtain the expression in the second line. 
Recalling that $\mathcal{I}_{AdS; 0}(\lambda)=-\mathcal{I}_{S; 0}(\lambda)$ we easily deduce $\mathcal{I}_{AdS; 0}^{({\rm np})}(\lambda)$. 

The overall factor $ \pm \ii $, called the \textit{transseries parameter}, must be understood as a piecewise constant function of ${\rm arg}(\lambda)$,~i.e.~$\sigma(\lambda) = \pm \ii$ for ${\rm arg}(\lambda) \gl 0$. Since the perturbative expansion $\mathcal{I}^{(\rm{p})}_{S; 0}(\lambda)$ (and equivalently $\mathcal{I}^{(\rm{p})}_{AdS; 0}(\lambda)$) given in \eqref{eq:I0-strong} is non-Borel summable for $\lambda>0$, one must perform a lateral Borel resummation from either ${\rm arg}(\lambda) > 0$ or ${\rm arg}(\lambda) < 0$.
However, this process does produce two distinct analytic functions of $\lambda$, neither of which is real for $\lambda>0$ despite \eqref{eq:I0-strong} being manifestly so. Once we add the appropriate non-perturbative terms \eqref{eq:I0NP} to these two different analytic continuations we find a well-defined analytic function for ${\rm Re}(\lambda)>0$ which is real for $\lambda>0$. Since this analysis has already been carried out in \cite{Brown:2024tru} we will not discuss it any further here. However, in Appendix \ref{app:cat} we present some details regarding Borel resummation, its analytic properties and relations to non-perturbative effects.

Moving to the novel contributions $\mathcal{I}_{S; 1}(\lambda;\alpha)$ and $\mathcal{I}_{AdS; 1}(\lambda;\beta)$ we now present their associated non-perturbative sectors. To compute the large-$\lambda$ exponentially suppressed terms $\mathcal{I}_{S; 1}^{({\rm np})}(\lambda;\alpha)$ and $\mathcal{I}_{AdS; 1}^{({\rm np})}(\lambda;\beta)$, we use the integral representations \eqref{eq:inte12C} and \eqref{eq:inte21SC} respectively. 
Firstly, to isolate the purely non-perturbative effects we move the contour of integration from ${\rm Re}\,s=1/2$ to ${\rm Re}\,s=5/2$ by ignoring the poles at $s=1$ and $s=2$ whose residues produce the perturbative sectors $\mathcal{I}_{S; 1}^{({\rm p})}(\lambda;\alpha)$ in \eqref{eq:IS1P} and $\mathcal{I}_{AdS; 1}^{({\rm p})}(\lambda;\beta)$ in \eqref {eq:IAdS1P}. Then by changing variable $s \rightarrow s+1$, and using the Dirichlet series representation for the Riemann zeta function $\zeta(s) = \sum_{n=1}^\infty n^{-s}$, we can write the non-perturbative parts of $\mathcal{I}_{S; 1}(\lambda;\alpha)$ and $\mathcal{I}_{AdS; 1}(\lambda;\beta)$ 
as 
\begin{align}
  \mathcal{I}_{S; 1}^{({\rm np})}(\lambda;\alpha) &\label{eq:IS1npMell} = {-}\sum_{n=1}^{\infty} \int_{{\rm Re}\, s={\frac{3}{2}}} \frac{ds}{2 \pi \ii} \, 4(2s-1) \Gamma(s) \Gamma(s-1) \, _2F_1 \left( s-1,s; 1; 1-\alpha \right) \left(\frac{Y_n}{2} \right)^{-2s} \, ,\\
    \mathcal{I}_{AdS; 1}^{({\rm np})}(\lambda;\beta) &\label{nonpertAdS}= -\sum_{n=1}^{\infty} \int_{{\rm Re}\, s=\frac{3}{2}} \frac{ds}{2 \pi \ii} \, 4(2s-1) \Gamma(s)^2 \, _2F_1 \left( s,s; 2; \frac{1}{1+\beta} \right) \left(\frac{Y_n \sqrt{(\beta+1)}}{2} \right)^{-2s} \, ,
\end{align}
where $Y_n \coloneqq n \sqrt{\lambda}$ and having used in the process the Riemann zeta functional identity
\begin{equation}
\xi(s) = \xi(1-s)\,,
\end{equation}
with $\xi(s) = \pi^{-s/2}\Gamma(s/2)\zeta(s)$.

Leaving the technical details to Appendix \ref{app:exp}, from the above integral representation we find that the non-perturbative terms of  $\mathcal{I}_{S; 1}(\lambda;\alpha)$ for $0\leq \alpha<1$ take the form,\footnote{Note that \eqref{eq:IS1NPFin} becomes singular for $\alpha =1$. Hence, the non-perturbative corrections to the maximal giant graviton correlators, corresponding precisely to $\alpha=1$, cannot be obtained as a limit of \eqref{eq:IS1NPFin}. The reason for this phenomenon is explained in Appendix \ref{app:exp}.}
\allowdisplaybreaks{
\begin{align} 
    \mathcal{I}_{S; 1}^{({\rm np})}(\lambda;\alpha) = -16  \pi\, & (1-\alpha)^{\frac{1}{4}}  \sum_{n=1}^{\infty} \Big[ \notag   e^{-n \sqrt{\lambda}(1-\sqrt{1-\alpha})} \sum_{m=0}^\infty \frac{1}{m!} c_m(\sqrt{1-\alpha}) \left(2 n \sqrt{\lambda(1-\alpha)}\right)^{-m-1} \\*
    &\label{eq:IS1NPFin}\pm \ii \,e^{-n \sqrt{\lambda}(1+\sqrt{1-\alpha})} \sum_{m=0}^\infty \frac{(-1)^m}{m!} c_m(-\sqrt{1-\alpha})\left(2 n \sqrt{\lambda(1-\alpha)}\right)^{-m-1} \Big]\,, 
\end{align}}
where the explicit form of the perturbative coefficients $c_m(x)$ is given by  
\begin{align}\label{cm}
&c_m(x)  =\notag [m^2+m-1] x \, _3\tilde{F}_2\left(-\frac{1}{2},\frac{3}{2},-m;\frac{3}{2}-m,\frac{3}{2}-m;x\right)\\
&+\, _3\tilde{F}_2\left(-\frac{3}{2},\frac{1}{2},-m-1;\frac{1}{2}-m,\frac{1}{2}-m;x\right)-\frac{3}{4}  m (m+1) x^2 \, _3\tilde{F}_2\left(\frac{1}{2},\frac{5}{2},1-m;\frac{5}{2}-m,\frac{5}{2}-m;x\right)\,.
\end{align}
Similarly, the non-perturbative terms of  $\mathcal{I}_{AdS; 1}(\lambda;\beta)$ with $\beta \geq0$ are 
\begin{align}\label{eq:IadsFin}
     \mathcal{I}_{AdS; 1}^{({\rm np})} (\lambda;\beta)=&\, -16 \pi  (1+\beta)^{\frac{1}{4}}  \sum_{n=1}^\infty\Big[  \notag e^{-n \sqrt{\lambda} \left(\sqrt{\beta+1}-1 \right)} \sum_{m=0}^\infty \frac{1}{m!} {c}_m(\sqrt{1+\beta}) \left(-2n \sqrt{\lambda(1+\beta)}\right)^{-m-1} \\ 
     &\mp \ii \, e^{-n \sqrt{\lambda} \left(\sqrt{\beta+1}+1 \right)} \sum_{m=0}^\infty \frac{(-1)^{m}}{m!}\, c_m(-\sqrt{1+\beta})\left(2n \sqrt{\lambda(1+\beta)}\right)^{-m-1}\Big] \, .  
\end{align}
As mentioned in Section~\ref{subsec:universality}, we emphasise that while the perturbative large-$\lambda$ expansions~\eqref{eq:I0-strong}-\eqref{eq:IS1P}-\eqref{eq:IAdS1P} are obtained by summing over the residues from the poles in the half-plane ${\rm Re}(s) >1/2$ of the respective Mellin integrals \eqref{eq:inte11SC}-\eqref{eq:inte12C}-\eqref{eq:inte21SC}, the non-perturbative corrections \eqref{eq:I0NP}-\eqref{eq:IS1NPFin}-\eqref{eq:IadsFin} are entirely captured by the large-$s$ behaviour of the corresponding integrand. In particular, it is easy to see that the integrands of \eqref{eq:inte11SC}-\eqref{eq:inte12C}-\eqref{eq:inte21SC}  all have a saddle point for~${\rm Re}(s)\gg1$ when~$\lambda \gg1$. Rather than performing a saddle point expansion, we find it more systematic and instructive to compute the non-perturbative corrections as discussed above.

A few more comments are now in order. Firstly, 
it would naively seem that resurgence analysis does not play any role in connecting the perturbative expansions $ \mathcal{I}_{S; 1}^{({\rm p})} (\lambda;\alpha)$ and   $\mathcal{I}_{AdS; 1}^{({\rm p})} (\lambda;\beta)$ with their corresponding non-perturbative completions  $ \mathcal{I}_{S; 1}^{({\rm np})} (\lambda;\alpha)$ and   $\mathcal{I}_{AdS; 1}^{({\rm np})} (\lambda;\beta)$. The reason being that, unlike the factorially divergent non-Borel summable expansion $ \mathcal{I}_{S; 0}^{({\rm p})} (\lambda)$ given in \eqref{eq:I0-strong}, the perturbative sectors $ \mathcal{I}_{S; 1}^{({\rm p})} (\lambda;\alpha)$ and   $\mathcal{I}_{AdS; 1}^{({\rm p})} (\lambda;\beta)$ presented in \eqref{eq:IS1P} and \eqref{eq:IAdS1P} contain only finitely many terms. It appears impossible then to reconstruct any non-perturbative effect from these perturbative data.

However, this is not quite correct and resurgence analysis plays a more subtle role in here.
This is an example of a structure that has been referred to as {\em Cheshire cat resurgence}; a phenomenon first observed in quantum mechanical examples~\cite{Dunne:2016jsr,Kozcaz:2016wvy} but also in quantum field theory setups, see e.g.~\cite{Dorigoni:2017smz,Dorigoni:2019kux}.
It is possible to slightly modify the integral representations \eqref{eq:inte12C} and \eqref{eq:inte21SC} by introducing a deformation parameter, say $\epsilon$. 
When this deformation parameter is generic the strong coupling perturbative expansions of the corresponding deformed quantities
$ \mathcal{I}_{S; 1}^{({\rm p})} (\lambda;\alpha, \epsilon)$ and   $\mathcal{I}_{AdS; 1}^{({\rm p})} (\lambda;\beta, \epsilon)$ do give rise to asymptotic and factorially divergent power series expansions akin to what we see in \eqref{eq:I0-strong} for  $ \mathcal{I}_{S; 0}^{({\rm p})} (\lambda)$.

We can then use the whole machinery of resurgence analysis to extract the non-perturbative completions to the deformed models. As the deformation parameter is removed, i.e. as we send $\epsilon\to0$, while the perturbative sectors reduce to the terminating expressions \eqref{eq:IS1P} and \eqref{eq:IAdS1P}, crucially we have that the non-perturbative corrections do survive and reproduce \eqref{eq:IS1NPFin} and \eqref{eq:IadsFin}. Although in the undeformed case the full body of the resurgence structure has disappeared, the grin of the non-perturbative corrections still lingers on.
Since we have managed to compute the non-perturbative effects by other means, we present in Appendix \ref{app:cat} only the Cheshire cat resurgence analysis in the simpler case of the maximal giant graviton correlator, i.e. $ \mathcal{I}_{S; 1} (\lambda;\alpha)$ with  $\alpha =1$.

We note however that resurgence analysis plays a second, more standard role within the non-perturbative sectors alone.
From the expression for the perturbative coefficients $c_m(x)$ given in \eqref{cm}, it is easy to see that the power series in $\lambda^{-1}$ multiplying the leading exponential factors, respectively $\exp[-n\sqrt{\lambda}(1-\sqrt{1-\alpha})]$ in \eqref{eq:IS1NPFin} and  $\exp[-n\sqrt{\lambda}(\sqrt{1+\beta}-1)]$ in \eqref{eq:IadsFin}, is a factorially divergent power series which is non-Borel summable for $\lambda>0$. 
Very much like our previous discussion below \eqref{eq:I0NP}, the lateral Borel resummation of this power series for ${\rm arg}(\lambda) \gl 0$ produces two different analytic functions which differ by non-perturbative effects. 

This difference is precisely accounted for by the second, subleading exponential factor, respectively $\exp[-n\sqrt{\lambda}(1+\sqrt{1-\alpha})]$ in \eqref{eq:IS1NPFin} and  $\exp[-n\sqrt{\lambda}(\sqrt{1+\beta}+1)]$ in \eqref{eq:IadsFin}, which we note are both exponentially suppressed when compared to the associated leading exponentials. Just like before, the choice in sign for the transseries parameter $\sigma(\lambda) = \pm \ii$ multiplying the subleading exponential terms is correlated with the choice of  lateral Borel resummation of this power series for ${\rm arg}(\lambda) \gl 0$ giving rise to the transseries representations \eqref{eq:IS1NPFin}-\eqref{eq:IadsFin} for two unambiguous and real-analytic functions of $\lambda>0$.
For more details, we refer to Appendix \ref{app:exp} where we present the technical analysis, which lead to the expressions given in \eqref{eq:IS1NPFin}-\eqref{eq:IadsFin}.

It is also interesting to notice that exponentially suppressed factors (as summarised in \eqref{eq:IS1npSum}-\eqref{eq:IAdS1npSum}) take exactly the same form as the non-perturbative scales found  in \cite{Dunne:2025wbq} via a resurgence analysis approach to the so-called \textit{tilted cusp anomalous dimensions} in $\mathcal{N}=4$ SYM.
The tilted cusp anomalous dimension introduced in \cite{Basso:2020xts} arises in the study of a certain kinematical limit of the six-point maximally-helicity-violating amplitude in $\mathcal{N}=4$ SYM, and it is
a deformation by a parameter $a\in [0,\frac{1}{2}]$ of the conventional cusp anomalous dimension which corresponds to the case $a=\frac{1}{4}$. 
The resurgence structures of both the standard \cite{Aniceto:2015rua,Dorigoni:2015dha} and the tilted cusp anomalous dimension \cite{Dunne:2025wbq} are more complicated than those of the HHLL integrated correlators of giant gravitons discussed above. However, it is possible to map the tilt parameter $a$ to the giant graviton dimension parameters $\alpha$ or $\beta$ so that the non-perturbative scales found in  \cite{Dunne:2025wbq} coincide with what we found here.\footnote{For a precise match with the reference in the giant graviton case we simply need to identify the parameter $a$ which controls the tilt as $a = \frac{1}{2}\sqrt{1-\alpha}$. Since $0\leq \alpha \leq 1$ we clearly have $a\in[0,\frac{1}{2}]$. For the dual giant graviton we first need to introduce an effective coupling constant $\lambda_{\rm eff} \coloneqq \lambda(\beta+1)$ and then define the parameter $a = \frac{1}{2\sqrt{1+\beta}}$ which again lies in the range $a\in[0,\frac{1}{2}]$ since $\beta\geq0$.}

It would be very interesting to understand whether the semi-classical origin of these non-perturbative corrections is the same, especially in the holographic interpretation. Presently there is no direct semi-classical calculation for these observables from string theory; in the next subsection we will comment on the  holographic interpretation for the non-perturbative corrections to the integrated correlators of giant gravitons.

\subsection{Remarks on the holographic interpretation}

In the previous subsection, we found several different types of exponentially suppressed terms for $\mathcal{I}_{S}(\lambda;\alpha)$ and $\mathcal{I}_{AdS}(\lambda;\beta)$. 
Summarising here the results, we see that one type of non-perturbative corrections arises from $\mathcal{I}_{S; 0}(\lambda)$ (or equivalently $\mathcal{I}_{AdS; 0}(\lambda)$) due to the fact that the strong coupling expansion is asymptotic and non-Borel summable. These exponentially suppressed contributions are purely imaginary and carry the suppression factor 
\begin{align}
    \mathcal{I}_{S; 0}^{({\rm np})}(\lambda) = - \mathcal{I}_{AdS; 0}^{({\rm np})}(\lambda) \propto   \exp \left[-2n \sqrt{\lambda} \right]\, .  
\end{align}

The holographic interpretation of such contributions is easy to understand. Using the identification $\sqrt{\lambda} = L^2/\alpha'$ with $\alpha'=\ell_s^2$ the square of the string length scale and $L$ the scale of the $AdS_5\times S^5$ space, we can rewrite the exponential factor above as $\exp( - 4 \pi L^2 n T_F)$ with $T_F\coloneqq 1/(2\pi \alpha')$ the fundamental string tension.
As already argued in \cite{Dorigoni:2022cua}, such an exponential factor suggests that these non-perturbative contributions originate from $n$ coincident fundamental-string euclidean world-sheets wrapping a great two-sphere $S^2$ on the equator of the five-sphere $S^5$. The overall imaginary factor $\ii$ accompanying such corrections is suggestive of the fact that a string wrapping a two-sphere on the equator of the five-sphere would provide a saddle point which carries a negative fluctuation mode  (more generally, an odd number of negative modes).\footnote{It is worth noting that the exponential factor \eqref{eq:I0NP} coincides with that of the non-perturbative corrections found in the study of the strong coupling expansion for the dressing phase of the $AdS_5\times S^5$ superstring \cite{Arutyunov:2016etw}.} 
From a resurgence point of view this is reminiscent of uniton solutions in the principal chiral model \cite{Cherman:2013yfa,Cherman:2014ofa}.

In \cite{Dorigoni:2022cua} (see also \cite{Dorigoni:2024dhy}) it was also shown that the presence of such exponentially suppressed corrections combined with ${\rm SL}(2,\mathbb{Z})$ electro-magnetic duality of $\mathcal{N}=4$ SYM yields other types of non-perturbative effects at large $N$ and finite coupling $\tau$ corresponding to coincident $(p,q)$-strings. While these references pertained the study of particular LLLL integrated correlators, we believe a similar story should hold for the present class of correlators, although this deserves further study.

Importantly, for the HHLL integrated correlators we find novel types of exponentially suppressed terms originating from $\mathcal{I}_{S; 1}^{({\rm np})}(\lambda; \alpha)$ as given in \eqref{eq:IS1NPFin} and  $\mathcal{I}_{AdS; 1}^{({\rm np})}(\lambda; \beta)$ as given in \eqref{eq:IadsFin}. 
We divide such non-perturbative corrections into real and imaginary contributions, denoted respectively by $\vert_{\mathbb{R}}$ and $\vert_{\mathbb{I}}$,\footnote{As mentioned earlier, the maximal giant graviton which corresponds to $\alpha=1$ has to be treated separately. In this case the exponentially decayed terms simplify significantly; in fact, the imaginary contribution vanishes and $\mathcal{I}_{S; 1}^{({\rm np})}(\lambda;1) \propto \exp \left[ -n \sqrt{\lambda} \right]$. Interestingly, it has also been noted that the maximal giant graviton case is special for a different reason in that it is the only case which leads to integrable boundary states \cite{Berenstein:2005vf, deMelloKoch:2015uwu}. 
}
\begin{align} \label{eq:IS1npSum}
     \mathcal{I}_{S; 1}^{({\rm np})}(\lambda;\alpha)\vert_{ \mathbb{R}} \propto \exp \left[ -n \sqrt{\lambda}(1-\sqrt{1-\alpha}) \right] \, , \qquad  \mathcal{I}_{S; 1}^{({\rm np})}(\lambda;\alpha)\vert_{\mathbb{I}} \propto \exp \left[-n \sqrt{\lambda}(1+\sqrt{1-\alpha}) \right]  \, . 
\end{align}
\begin{align}\label{eq:IAdS1npSum}
\mathcal{I}_{AdS; 1}^{({\rm np})}(\lambda;\beta)\vert_{\mathbb{R}} \propto \exp \left[ -{n\sqrt{\lambda}(\sqrt{\beta+1} - 1)}  \right] \, , \qquad \mathcal{I}_{AdS; 1}^{({\rm np})}(\lambda;\beta)\vert_{\mathbb{I}} \propto \exp \left[ -{n\sqrt{\lambda}(\sqrt{\beta+1} + 1)}  \right] \, . 
\end{align}
As previously anticipated, we stress that even though $\mathcal{I}_{AdS}(\lambda;\beta)$ and $\mathcal{I}_{S}(\lambda;\alpha)$ share an identical perturbative asymptotic expansion at strong coupling, they crucially differ by the exponentially suppressed terms which are contained respectively in $\mathcal{I}_{AdS; 1}(\lambda;\beta)$ and $\mathcal{I}_{S; 1}(\lambda;\alpha)$. The nature of these different types of non-perturbative terms is particularly interesting. More specifically, the fact $\mathcal{I}_{S; 1}^{({\rm np})}(\lambda;\alpha)\vert_{\mathbb{R}}$ and $\mathcal{I}_{AdS; 1}^{({\rm np})}(\lambda;\beta)\vert_{\mathbb{R}}$ are real is due to the fact that the strong coupling expansions of $\mathcal{I}_{S; 1}(\lambda;\alpha)$ and $\mathcal{I}_{AdS; 1}(\lambda;\beta)$ contain only a finite number of terms. This is the typical behaviour of Cheshire cat resurgence \cite{Dunne:2016jsr, Kozcaz:2016wvy, Dorigoni:2017smz}, and such behaviour also appears in a similar context in the study of large-charge expansion of the integrated correlators \cite{Brown:2023why}. We explore more in detail this Cheshire cat resurgence structure in Appendix \ref{app:cat}. 

It is important to compare the scale of these novel non-perturbative corrections with \eqref{eq:I0NP} found previously. Starting with the large-$\lambda$ non-perturbative corrections to the sphere giant graviton integrated correlator, given the physical range for the parameter $0< \alpha <1$ we note the trivial fact $0<\sqrt{1-\alpha}<1$ and hence find this hierarchy of non-perturbative scales
\begin{align}
\mathcal{I}_{S; 1}^{({\rm np})}(\lambda;\alpha)\vert_{ \mathbb{R}} \gg  \mathcal{I}_{S; 1}^{({\rm np})}(\lambda)\vert_{\mathbb{I}}  \gg \mathcal{I}_{S; 0}^{({\rm np})}(\lambda)\,.
\end{align}
Similarly, for the AdS giant graviton we find that for $0<\beta<8$ we have the hierarchy
\begin{align}
\mathcal{I}_{AdS; 1}^{({\rm np})}(\lambda;\beta)\vert_{ \mathbb{R}} \gg  \mathcal{I}_{AdS;0}^{({\rm np})}(\lambda) \gg \mathcal{I}_{S; 1}^{({\rm np})}(\lambda;\beta)\vert_{\mathbb{I}} \,,
\end{align}
while for $\beta>8$ this becomes
\begin{align}
  \mathcal{I}_{AdS;0}^{({\rm np})}(\lambda) \gg \mathcal{I}_{AdS; 1}^{({\rm np})}(\lambda;\beta)\vert_{ \mathbb{R}} \gg \mathcal{I}_{S; 1}^{({\rm np})}(\lambda;\beta)\vert_{\mathbb{I}} \,.
\end{align}
Interestingly,  for $\beta=8$  we note that $\mathcal{I}_{AdS;0}^{({\rm np})}(\lambda)$ and $\mathcal{I}_{AdS; 1}^{({\rm np})}(\lambda;\beta)\vert_{ \mathbb{R}}$ have exactly the same exponentially suppressed factor. We do not know whether this is an accident or if this fact is hiding some interesting semi-classical structure.

To better understand the semi-classical origins of the non-perturbative effects we proceed as discussed above. From an holographic point of view $\sqrt{\lambda}$ is proportional to the string length squared $\ell_s^2$ and therefore it is natural to interpret the exponents of \eqref{eq:IS1npSum}-\eqref{eq:IAdS1npSum} as particular world-sheet areas. Just like $\mathcal{I}_{S; 0}^{({\rm np})}(\lambda) $, we believe that the purely imaginary terms, $\mathcal{I}_{S; 0}^{({\rm np})}(\lambda;\alpha)\vert_{\mathbb{I}}$ and  $\mathcal{I}_{AdS; 0}^{({\rm np})}(\lambda;\beta)\vert_{\mathbb{I}}$ should originate from semi-classical saddle points of the path-integral, whereas $\mathcal{I}_{S; 1}^{({\rm np})}(\lambda;\alpha)\vert_{\mathbb{R}}$ and $\mathcal{I}_{AdS; 1}^{({\rm np})}(\lambda;\beta)\vert_{\mathbb{R}}$ should correspond to path-integral contributions from local minima. 

Rewriting the exponential factors \eqref{eq:IS1npSum}-\eqref{eq:IAdS1npSum} in the holographic dictionary and making use of the mappings \eqref{eq:theta0}-\eqref{eq:rho0} between the parameters $\alpha,\beta$ and the geometry of the underlying D3-brane solutions, we find
\begin{align} \label{eq:IS1npSum2}
     \mathcal{I}_{S; 1}^{({\rm np})}(\lambda;\alpha)\vert_{ \mathbb{R}/\mathbb{I}} &\propto \exp \left[ -2\pi L^2 ( 1 \mp \sin(\theta_0)) n T_F\right]\,,\\
\mathcal{I}_{AdS; 1}^{({\rm np})}(\lambda;\beta)\vert_{\mathbb{R}/\mathbb{I}} &\propto \exp \left[ -2\pi L^2( \cosh(\rho_0)\mp1)n T_F \right]  \, , 
\end{align}
where the sign $\mp$ is correlated with the choice of real $\vert_{\mathbb{R}}$ and imaginary $\vert_{\mathbb{I}}$ contributions.

The string configurations which give rise to the leading exponential contributions, $\mathcal{I}_{S; 1}^{({\rm np})}(\lambda;\alpha)\vert_{\mathbb{R}}$ and $\mathcal{I}_{AdS; 1}^{({\rm np})}(\lambda;\beta)\vert_{\mathbb{R}}$, should correspond to minimal string worldsheets moving on the background geometry in the presence of a D3-brane either in the $S^5$ or $AdS_5$ factor. We have not succeeded in finding such configurations, however we can interpret the subleading contributions $\mathcal{I}_{S; 1}^{({\rm np})}(\lambda;\alpha)\vert_{\mathbb{I}}$ and $\mathcal{I}_{AdS; 1}^{({\rm np})}(\lambda;\beta)\vert_{\mathbb{I}}$ by considering the relative action with respect to the true minima above, i.e. we consider
\begin{align} \label{eq:IS1npSumRat}
     \frac{\mathcal{I}_{S; 1}^{({\rm np})}(\lambda;\alpha)\vert_{ \mathbb{I}}}{\mathcal{I}_{S; 1}^{({\rm np})}(\lambda;\alpha)\vert_{ \mathbb{R}}} &\propto \exp \left[ -4\pi L^2  \sin(\theta_0) n T_F\right] \, , \\
\frac{\mathcal{I}_{AdS; 1}^{({\rm np})}(\lambda;\beta)\vert_{\mathbb{I}}}{\mathcal{I}_{AdS; 1}^{({\rm np})}(\lambda;\beta)\vert_{\mathbb{R}}} &\propto \exp \left[ -4\pi L^2n T_F \right]  \, . \label{eq:IAdS1npSumRat}
\end{align}
These relations suggest that  semi-classically the purely imaginary contributions $\mathcal{I}_{S; 1}^{({\rm np})}(\lambda;\alpha)\vert_{\mathbb{I}}$ and $\mathcal{I}_{AdS; 1}^{({\rm np})}(\lambda;\beta)\vert_{\mathbb{I}}$ originate from superimposing to the local minima solutions two saddle-point solutions. In the case of the sphere giant graviton these saddle points are given by $n$-copies of the fundamental string wrapping a smaller $S^2$ inside of the $S^5$ which is stabilised by the presence of the D3-brane, while in the case of the AdS giant graviton, these saddle points are exactly the same semi-classical objects which produced the non-perturbative effects \eqref{eq:IAdS1npSum} and are given by $n$-copies of the fundamental string wrapping the equatorial $S^2$ inside of the $S^5$ factor. The fact that  \eqref{eq:IS1npSumRat} (unlike \eqref{eq:IAdS1npSumRat}) depends on $\theta_0$ (i.e. where we place the D3-branes in $S^5$) may be expected since these HHLL correlators can be viewed as a scattering process in the $AdS$ space. 
It is definitely worth investigating further the semi-classical holographic origin of the non-perturbative effects here presented.

Lastly, we note that the symmetry $\mathcal{I}_{S;1}(\lambda;\alpha) = - \mathcal{I}_{AdS;1}(\lambda;\beta)\vert_{\beta = -\alpha}$, an immediate consequence of the convergent weak coupling expansions \eqref{eq:pert1I1} and \eqref{eq:pert2I1-2}, is absolutely not obvious at strong coupling.
Both the asymptotic perturbative expansions  \eqref{eq:IS1P} and \eqref{eq:IAdS1P} as well as their non-perturbative completions \eqref{eq:IS1NPFin} and \eqref{eq:IadsFin} are not related by $\alpha \leftrightarrow -\beta$. Importantly, while the perturbative coefficients in the non-perturbative sectors \eqref{eq:IS1NPFin} and \eqref{eq:IadsFin} are indeed related by this symmetry, the exponential factors are not.
For example, we see that under $\alpha \to -\beta$ the exponentially suppressed term $\exp[-n (1-\sqrt{1-\alpha})]$ in  \eqref{eq:IS1NPFin} goes into $\exp[-n (1-\sqrt{1+\beta})]$, which blows up in the limit $\lambda \gg1$ for $\beta>0$ and is thus clearly absent in \eqref{eq:IadsFin}.

The reason for this discrepancy between weak coupling and strong coupling for the $\alpha \leftrightarrow -\beta$ symmetry is that the strong coupling expansion is only an asymptotic expansion. Hence, we need to be extremely careful in how we perform the large-$\lambda$ expansion. As already noted, the weak coupling symmetry $\mathcal{I}_{S;1}(\lambda;\alpha) = - \mathcal{I}_{AdS;1}(\lambda;\beta)\vert_{\beta = -\alpha}$ does not preserve the physical conditions $0\leq \alpha \leq 1$ and $\beta\geq 0$ for the giant gravitons parameters. Importantly, the strong coupling perturbative and non-perturbative sectors have been derived here only for physical values of the parameters, i.e. only for $0\leq \alpha\leq1$ and $\beta\geq0$, starting from the well-defined integral representations \eqref{eq:inte12C} and \eqref{eq:inte21SC}.\footnote{It is intriguing to note the striking similarity between the present discussion with the non-perturbative scales found in \cite{Dunne:2022esi}. This reference analyses the non-perturbative effects responsible for electron-positron pair-production from the vacuum via the Euler-Heisenberg effective action, which encodes the effective dynamics of the electromagnetic field after having integrated out the massive charged particles. In this context, the symmetry $\alpha \leftrightarrow -\beta$ can be seen as exchanging a spatially dependent magnetic field with a spatially dependent electric field with the same profile. We thank Gerald Dunne for bringing this fact to our attention.} 

From an holographic point of view we can understand the symmetry \eqref{eq:Z2} as an analytic continuation in the geometry. This symmetry may be viewed as a particle-hole duality in the LLM geometry~\cite{Lin:2004nb}, where sphere giant gravitons are interpreted as quasi-holes and AdS giant gravitons as quasi-particles. In our setup, starting with the sphere giant graviton we see from the geometric picture discussed below \eqref{eq:AdSSmet} that as $\alpha$ approaches $0$, the D3 solution reaches the point $\theta_0 \to \pi/2$ on the $S^5$ while still being located at $\rho=0$ in the radial direction.
From the metrics \eqref{eq:S}-\eqref{eq:AdS} we see that at $\theta_0=\pi/2$ and $\rho=0$ both the $S^3$ in the $S^5$, denoted by $\Omega_3$, and the $S^3$ in $AdS_5$, denoted by $\tilde{\Omega}_3$, have shrunk to zero volume so that at $\rho =0$ and $\theta=\pi/2$ it 
does not make any difference whether the brane is wrapping an $S^3$ in the $S^5$ or $AdS_5$ factor of the geometry. As we push further to negative $\alpha<0$ the brane now starts moving in the radial direction located at $\rho_0 = {\rm arcsinh} \sqrt{\beta}$ with $\beta = -\alpha$ while still keeping $\theta = \pi/2$.  The form of the exponentially suppressed terms \eqref{eq:IS1NPFin} and \eqref{eq:IadsFin} is a reminder that  the non-perturbative effects do feel the change in geometry as we continue the parameter $0 \leq \alpha \leq 1 $ to $\alpha = -\beta\leq 0$.

 We note that since the scaling dimensions of the sphere giant graviton and the AdS dual giant graviton are given respectively by $\Delta = \alpha N$ and $\Delta = \beta N$, the exchange $\alpha \leftrightarrow -\beta$ is essentially equivalent to $N \leftrightarrow -N$. This implies 
\begin{equation}
\left[ N \, \mathcal{I}^{(0)}_{\cG_{S;\alpha}}(\lambda) \right]_{N\to -N} = -N\, \mathcal{I}^{(0)}_{\cG_{S;-\alpha}}(\lambda) = + N \, \mathcal{I}^{(0)}_{\cG_{AdS;\beta}}\Big\vert_{\beta = -\alpha}\,,
\end{equation}
where we have included the $N$-dependence for the planar contribution \eqref{eq:Iplan} according to the genus expansion \eqref{eq:I_H_genus}. 
We emphasise that this symmetry between sphere giant and AdS giant gravitons under $N\leftrightarrow -N$ has been  verified in \cite{Abajian:2023jye} directly at the level of exact three-point functions, where the operators considered are either three generic symmetric Schur polynomials operators or three generic anti-symmetric Schur polynomials operators.\footnote{We thank Francesco Aprile for bringing these arguments to our attention.}  It would be interesting to study the fate of this symmetry for HHLL integrated correlators beyond the planar limit.

\section{Comments on gluon-graviton scatterings off D7-branes} \label{gluon}

In this section, we briefly comment on a different class of observables considered in the recent work \cite{Chester:2025ssu} and which shows very similar properties to the HHLL integrated correlators of giant gravitons studied in this paper.

More specifically, the authors of \cite{Chester:2025ssu} studied a certain integrated correlator of moment map operators in an $\cN=2$ SCFT with $USp(2N)$ gauge group and a matter content of four fundamental and one antisymmetric hypermultiplets, transforming respectively with an $SO(8)$ and an $SU(2)$ flavour symmetry. Denoting by $\mu$  the mass associated with the $SO(8)$ moment map operator and by $m$ the mass associated with the $SU(2)$ moment map operator, the mixed-flavor integrated correlator studied in \cite{Chester:2025ssu} is expressed as $\partial_{\mu}^2 \partial_{m}^2 \log Z(\tau;m,\mu) \vert_{m=\mu=0}$, where $Z(\tau;m,\mu)$ is the $\cN=2$ partition function deformed by the mass parameters $\mu$ and $m$. We refer to \cite{Chester:2025ssu} for the details of this computation. Here we remark that this integrated correlator has a well-defined holographic dual interpretation in terms of a four-point scattering amplitude of two gluons (transforming under $SO(8)$) and two gravitons on an  $AdS_5 \times S^5/\mathbb Z_2$ background in the presence of D7-branes.

At fixed 't Hooft coupling the large-$N$ expansion of this integrated correlator takes the following form, 
\begin{align}\label{eq:gl_grav_thooft}
    \partial_{\mu}^2 \partial_{m}^2 \log Z(\tau;m,\mu) \vert_{m=\mu=0} \sim N F_{1}(\lambda) + F_{2}(\lambda) + N^{-1} F_{3}(\lambda) + \ldots \, , 
\end{align}
where integral representations for the first few $F_i(\lambda)$ can be found in \cite{Chester:2025ssu}.\footnote{In \cite{Chester:2025ssu} the 't Hooft coupling is denoted as $\lambda_{\rm UV}$, and the notation $\lambda$ is for the so-called IR ’t Hooft coupling, which is related to $\lambda_{\rm UV}$ by a simple transformation (see (4.13) of \cite{Chester:2025ssu}). For our purposes we can simply replace $\lambda_{\rm UV}$ with $\lambda$, since we concentrate on the leading contribution $F_1(\lambda)$ in \eqref{eq:gl_grav_thooft}, whereas the difference between UV and IR ’t Hooft coupling arises in the subleading terms (i.e. $F_2(\lambda)$, $F_3(\lambda)$ etc.), which are not discussed here. } Focusing here only on the leading term $F_{1}(\lambda)$, we write it via the integral representation
\begin{align} \label{eq:F1-int}
    F_{1}(\lambda)  =2^6 \int_0^{\infty} {w_1^2\, dw_1 \over \sinh(w_1)^2}{w_2\, dw_2 \over \sinh(w_2)^2} J_1(x_2) 
\frac{w_2 J_0(x_2)J_1(x_1)- w_1 J_0(x_1)J_1(x_2)}{w_1^2 - w_2^2} \, , 
\end{align}
with  $x_i \coloneqq w_i \sqrt{\lambda}/\pi$. It is straightforward to compute the integral perturbatively in the small-$\lambda$ expansion. In particular, at each loop order we see that the perturbative coefficients contain products of two odd Riemann zeta values, e.g. 
\begin{align}
&F_{1}(\lambda) =   \frac{9  \zeta (3)^2}{2 \pi ^4} \lambda ^2- \frac{105  \zeta (3) \zeta (5)}{16 \pi ^6}\lambda ^3 + \frac{105 \left(5 \zeta (5)^2+12 \zeta (3) \zeta (7)\right)}{256 \pi ^8}\lambda ^4 \\ 
& -\frac{21  (65 \zeta (5) \zeta (7)+99 \zeta (3) \zeta (9))}{512 \pi ^{10}} \lambda ^5 + \frac{105  \left(427 \zeta (7)^2+1284 \zeta (5) \zeta (9)+2211 \zeta (3) \zeta
   (11)\right)}{65536 \pi ^{12}} \lambda ^6 + \ldots \, . \nonumber
\end{align}
The above weak coupling expansion is significantly more complicated than those of the giant gravitons integrated correlators, given in \eqref{eq:pert1I0}-\eqref{eq:pert1I1} and \eqref{eq:pert2I0}-\eqref{eq:pert2I1}. However, as shown in  \cite{Chester:2025ssu} the strong coupling expansion of $ F_{1}(\lambda)$ is remarkably of a much simpler form,
\begin{align} \label{eq:F1-strong}
        F_{1}(\lambda) \vert_{\rm strong} \sim 8 \left( 2 - \frac{4 \pi ^2}{ \lambda }  -  \sum_{n=1}^{\infty}   \frac{16 n \zeta (2 n+1) \Gamma
   \left(n-\frac{1}{2}\right)^2 \Gamma \left(n+\frac{1}{2}\right)}{  \lambda ^{  n+\frac{1}{2}}\, \pi ^{3/2}\, \Gamma (n)}  \right) \, . 
\end{align}
Even more interestingly, we note that $F_{1}(\lambda) \vert_{\rm strong}$ is in fact identical to $\mathcal{I}_{S}(\lambda; \alpha)\vert_{\rm strong}$ in \eqref{eq:IS_strong_fin} and to $\mathcal{I}_{AdS}(\lambda; \beta)\vert_{\rm strong}$ in \eqref{eq:IAdS_strong_fin}, except for the coefficients of the constant term and the $1/\lambda$ term and an overall factor of $8$. 

We deduce that the leading planar contribution, $F_{1}(\lambda)$, to the gluon-graviton scattering off D7-branes has the same strong-coupling asymptotic series as that of the giant graviton correlators $\mathcal{I}_{S}(\lambda; \alpha)$ and $\mathcal{I}_{AdS}(\lambda; \beta)$. We find this to be a remarkable universal property of brane scattering.

We note that the $\cN=2$ SCFT here considered belongs to a class of $\cN=2$ theories that are planar equivalent to $\cN=4$ \cite{Billo:2019fbi,Beccaria:2020hgy, Beccaria:2022kxy}. Therefore, at the planar level several observables in all of these planar-equivalent $\cN=2$ SCFTs, such as the free energy and Wilson loops expectation values, have the same expression as functions of $\lambda$ as their $\cN=4$ SYM counterparts  \cite{Beccaria:2021ism,Beccaria:2021vuc}.  What we find here may be even more striking: despite considering seemingly unrelated quantities (namely HHLL integrated correlators of giant gravitons in $\mathcal{N}=4$ SYM and integrated correlators for moment map operators in a $\mathcal{N}=2$ SCFT) at strong coupling all these observables share the same universal perturbative expansion. Moreover, this matching appears to extend beyond the leading large-$N$ planar limit \cite{DeLillo:2025stg} suggesting a deeper connection that warrants further investigation.

As already seen in Section \ref{sec:non-pert} for $\mathcal{I}_{S}(\lambda; \alpha)\vert_{\rm strong}$ and $\mathcal{I}_{AdS}(\lambda; \beta)\vert_{\rm strong}$, the difference amongst all these quantities should again be found in the non-perturbative effects. To extract the large-$\lambda$ exponentially suppressed corrections to \eqref{eq:F1-int}, we proceed as in \cite{Chester:2025ssu, Alday:2023pet} and use the identity 
\begin{align}
\frac{x_2 J_0(x_2)J_1(x_1)- x_1 J_0(x_1)J_1(x_2)}{x_1^2 - x_2^2}  =4 \sum_{\ell=1}^{\infty} {\ell\, J_{2\ell}(x_1)J_{2\ell}(x_2) \over x_1 \, x_2} \, , 
\end{align}
 to decouple the $w_1$ and $w_2$ integrals in \eqref{eq:F1-int} which can then be performed using the Mellin-Barnes representations 
 \begin{align}
 J_\mu(x) &= \int_{-\ii \infty}^{+\ii\infty} \frac{ds}{2\pi i} \frac{\Gamma(-s)}{\Gamma(\mu+s+1)} \left(\frac{x}{2}\right)^{\mu +2s}\,,\\
 J_\mu(x)J_\nu(x) &=\int_{-\ii \infty}^{+\ii\infty}  \frac{ds}{2\pi i}\frac{\Gamma(-s)\Gamma(2s+\mu+\nu+1)}{\Gamma(s+\mu+1)\Gamma(s+\nu+1)\Gamma(s+\mu+\nu+1)} \left(\frac{x}{2}\right)^{\mu+\nu+2s}\,.
 \end{align}
 This leads to the following expression for $F_1(\lambda)$: 
\begin{equation}\label{eq:F1IntRep}
F_1(\lambda) = \int_{c -\ii \infty}^{c+\ii\infty} \frac{ds\, dt}{(2 \pi \ii)^2}\, 2^{9-2s} \pi^{\frac{5}{2}} \frac{s(2 s{-}1) t (2 t{-}1)  \xi(2 s)\xi(2 t)  \, _4\tilde{F}_3(2,s,s{+}1,t{+}1;2{-}s,3{-}s,2{-}t;{-}1)}{ \sin(2 \pi s)  \sin (\pi  t) \Gamma ( s-\frac{1}{2}) } \left( \frac{\lambda}{4 \pi}\right)^{-s-t} \, ,
\end{equation}
where $_p\tilde{F}_q(a_1, \dots a_p; b_1, \dots b_q;z)$ denotes a regularised hypergeometric function while the completed Riemann zeta function is given by $\xi(s) = \pi^{-s/2}\Gamma(s/2)\zeta(s)$. The contours of integration must be chosen such that $-1< c <0$ for both the $s$ and $t$ variables.
In the limit $\lambda \gg1$ we can close the contour of integration to the right half-planes ${\rm Re}(s)>0$ and ${\rm Re}(t)>0$ collecting residues from the poles located at $t=0$ and $s=0, 1, 3/2, 5/2, \ldots$ as well as the residue from the pole at $t=1, s=0$.
This residue calculation immediately reproduces the asymptotic strong coupling perturbative expansion \eqref{eq:F1-strong}.  It is important to note that the $t$-integral has only a finite number of poles contributing to the strong coupling expansion, whereas the $s$-integral contributes with an infinite number of residues. Similar to the giant graviton integrated correlators $\mathcal{I}_{S;0}(\lambda)$ and $\mathcal{I}_{S;1}(\lambda; \alpha)$ (or $\mathcal{I}_{AdS;1}(\lambda)$), this fact has significant effects in the study of non-perturbative corrections as we now discuss. 

To extract the strong coupling non-perturbative contributions to $F_1(\lambda)$ we start from the integral representation \eqref{eq:F1IntRep} and firstly expand out the completed Riemann zeta functions using their Dirichlet series representation, as well as rewrite the hypergometric function in terms of its Gauss series, thus yielding
\begin{align} \label{eq:F1-InvMellin}
F_1(\lambda) = -\sum_{n_1,n_2=1}^{\infty} \sum_{\ell=1}^{\infty }  \frac{2^{6} \, \ell }{\pi} \int_{c' -\ii \infty}^{c' +\ii \infty}  ds dt  \,&\notag \left[ \frac{(2s-1)\Gamma (s-\ell-1) \Gamma (s-\ell) \Gamma (s+\ell-1) \Gamma (s+\ell)}{\Gamma (2 s-1)\, \cot (\pi  s)}  \right. \\
&\left. \,\, \times \, (2t-1) \Gamma (t-\ell) \Gamma (t+\ell)
  \, n_1^{-2s} n_2^{-2t}\left( \frac{\sqrt{\lambda} }{2} \right)^{-2s-2t}  \right] \, . 
\end{align}
The non-perturbative corrections are computed by performing the inverse Mellin integral in \eqref{eq:F1-InvMellin} at fixed $\ell$, and then evaluating the summation over  $\ell$ using $\zeta$-function regularisation. 
From an analysis similar to the one discussed in Appendix \ref{app:exp}, we find that the non-perturbative terms associated with $F_{1}(\lambda)$ are of the following types, 
\begin{align} \label{eq:n1n2}
   \ii \exp \left[ - 2n_1 \sqrt{\lambda} \right] \, ,  \quad   \ii  \exp \left[ - (2n_1 + n_2) \sqrt{\lambda} \right]\, ,  \quad  \exp \left[ - n_2 \sqrt{\lambda} \right] \, , 
\end{align}
where $n_1, n_2$ are positive integers.
It is important to notice that the first two non-perturbative scales are purely imaginary while the last one is real thus suggesting that semi-classically they correspond respectively to saddle points and true local minima to the path-integral. 
A proper semi-classical, holographic interpretation of these non-perturbative effects certainly deserves further investigation.

The first class of exponentially suppressed terms, of the form $ \ii \exp \left[ - 2n_1 \sqrt{\lambda} \right]$, originates from the non Borel-summable nature of $F_{1}(\lambda) \vert_{\rm strong}$ given in \eqref{eq:F1-strong}. This is analogous to what occurs for the integrated correlators of giant gravitons in \eqref{eq:I0NP}. The purely real exponential terms, i.e. the ones of the form $\exp \left[ - n_2 \sqrt{\lambda} \right]$ in \eqref{eq:n1n2}, arise from the $t$-integral which, as noted earlier, has only a finite number of poles. Such non-perturbative effects are naturally interpreted in terms of Cheshire cat resurgence as discussed previously in Section \ref{sec:Exp}.  Lastly, the second type of exponential term appearing in \eqref{eq:n1n2}, $\ii  \exp \left[ - (2n_1 + n_2) \sqrt{\lambda} \right]$, can be understood as a combination of the first and last contributions in \eqref{eq:n1n2}. Overall, we note that the structure of the non-perturbative sectors of  $F_{1}(\lambda)$, although similar to that of the giant graviton correlators \eqref{eq:IS1npSum}-\eqref{eq:IAdS1npSum}, is more involved, and importantly we discover that once again the universal nature of brane scattering, seen at the perturbative level, is lifted by non-perturbative effects.

\section{Conclusion and outlook} \label{conclusion}

In this paper we derive exact expressions for the integrated four-point correlators of two giant graviton operators and two half-BPS single trace operators of dimension two in the planar limit, providing remarkable progress with respect to the available results in the literature for the correlators themselves. Due to the holographic interpretation of the giant gravitons in terms of D3-branes, the class of correlators considered in this paper is dual to scattering processes of two gravitons off D3-branes.  From these exact expressions we extract the strong coupling expansions to all orders in the 't Hooft coupling $\lambda$ for both the sphere and the AdS giant gravitons with generic conformal dimensions. 

Remarkably, we find that the strong coupling asymptotic series for both classes of giant gravitons are identical, even though the weak coupling expansions depend crucially on the details of the giant graviton conformal dimensions. Moreover, we show that such a universal expansion is also shared by the strong coupling perturbative expansion of a seemingly unrelated integrated correlator in an $\cN=2$ SCFT with $USp(2N)$ gauge group, which is holographically dual to a mixed scattering amplitude of two gluons and two gravitons in the presence of D7-branes on a $AdS_5 \times S^5/\mathbb{Z}_2$ background. 

To understand how these distinct integrated correlators differ at strong coupling we analyse as well the exponentially suppressed, non-perturbative contributions. Crucially, we find that at strong coupling all these integrated correlators are distinguished by their non-perturbative sectors. Importantly, we show that the different types of exponentially suppressed corrections have very different origins both from the point of view of resurgence analysis and from a semi-classical perspective.  

This work opens up several future research directions worth investigating. A first natural question regards the origin of the universal behaviour of the strong coupling perturbative expansion of these different integrated correlators. We do note that these observables all share some common features: for example they are all holographically dual to scattering amplitudes of gravitons off D-branes. 
Furthermore, we believe it possible to promote the large-$\lambda$ expansions of all these observables to genuine $SL(2 ,\mathbb{Z})$ invariant functions of the complexified coupling $\tau$ via non-holomorphic Eisenstein series \cite{Brown:2024tru, Chester:2025ssu}. Of course, these common features of such integrated correlators are not enough to explain the surprising universal strong-coupling expansion discovered.

Furthermore, in this paper we have considered only the leading order in the large-$N$ 't Hooft limit. In the case of the sphere giant graviton of dimension exactly $N$, i.e. for $\alpha=1$, the integrated correlator has been determined in \cite{Brown:2024tru} also at the subleading order in $1/N$. However, no result is presently known for subleading corrections to the general giant graviton and dual giant graviton correlators. It is extremely important to explore whether the universal strong coupling perturbative behaviour here discussed persists beyond the leading planar order.
As commented previously, the presence at fixed $\tau$ of electro-magnetic $SL(2 ,\mathbb{Z})$ invariance relates perturbative terms in the leading planar correction to other perturbative terms at the non-planar level. Therefore, it is tantalising to conjecture that thanks to $SL(2 ,\mathbb{Z})$ electro-magnetic invariance such a universality may very well survive beyond the planar limit.  

It is also particularly interesting to extend such considerations to other integrated correlators, for example those considered in \cite{Billo:2024ftq} and which are closely related to the particular $\mathcal{N}=2$ SCFT considered in \cite{Chester:2025ssu} that we have commented on in this paper. Preliminary results in \cite{DeLillo:2025stg} show that remarkably all of these observables exhibit exactly the same asymptotic strong coupling expansion also at the next subleading order in the large-$N$ expansion.

Finally, it is of great interest to investigate whether the universality we observe for the integrated correlators also persists to the level of the full (un-integrated) correlators. Naively, we do not expect the correlators to take a universal form at strong coupling. Unfortunately, as already discussed, there are no known results for the four-point correlators involving AdS giant gravitons while for the sphere giant graviton correlators only the leading order at strong coupling has been determined in \cite{Chen:2025yxg} for the full determinant operator. A crucial, albeit difficult, task is computing all these correlators to higher orders in the strong coupling expansion. Holographically this process corresponds to computing stringy corrections to scattering amplitudes in $AdS_5 \times S^5$. This calculation will yield important CFT data for the study of $\mathcal{N}=4$ SYM in the presence of heavy operators, and may help with understanding the universality of the integrated correlators.

\vskip 1cm
\noindent {\large {\bf Acknowledgments}}
\vskip 0.2cm
We would like to thank Francesco Aprile, Marco Bill\`o, Shai Chester, Frank Coronado, Lorenzo De Lillo,  Zhihao Duan, Gerald Dunne, Marialuisa Frau, Alba Grassi, Cristoforo Iossa, Yunfeng Jiang, Shota Komatsu, Alberto Lerda, Paolo Vallarino, and Xinan Zhou for many useful discussions.   
A.B. is supported by a Royal Society funding RF$\backslash$ERE$\backslash$210067. C.W. is supported by a Royal Society University Research Fellowship,  URF$\backslash$R$\backslash$221015. F.G. and C.W. are supported by a STFC Consolidated Grant, ST$\backslash$T000686$\backslash$1 ``Amplitudes, strings \& duality".  D.D. is supported by
the Royal Society grants ICA$\backslash$R2$\backslash$242058 and IEC$\backslash$R3$\backslash$243103.

\appendix 

\section{Analysis of the non-perturbative terms} \label{app:exp}

In this appendix we derive equation \eqref{eq:IS1NPFin} for the sphere giant graviton non-perturbative terms $\mathcal{I}_{S; 1}^{({\rm np})}(\lambda;\alpha)$ and equation \eqref{eq:IadsFin} for the AdS giant graviton $\mathcal{I}_{AdS; 1}^{({\rm np})} (\lambda;\beta)$.
For both cases, the starting point of our analysis is the corresponding Mellin integral representation \eqref{eq:IS1npMell} and \eqref{nonpertAdS}.

Let us start by discussing the sphere giant graviton case.
Firstly, in the integral representation \eqref{eq:IS1npMell} we replace the hypergeometric function by its Gauss series,
\begin{equation}
\,_2F_1(a,b;c;z) = \sum_{n=0}^\infty \frac{(a)_n(b)_n}{(c)_n} \frac{z^n}{n!}\,,
\end{equation}
to arrive at
\begin{align}
    \mathcal{I}_{S; 1}^{({\rm np})}(\lambda;\alpha)&= -\sum_{n=1}^{\infty} \sum_{{m=0}}^{\infty} \frac{4(1{-}\alpha)^m}{\Gamma(m+1)^2} \int_{{\rm Re}\, s=\frac{3}{2}} \frac{ds}{2 \pi i} \,(2s-1) \Gamma(m+s) \Gamma(m+s-1) \left(\frac{Y_n}{2} \right)^{-2s}\,,
\end{align}
 and then perform the Mellin integral thus yielding
\begin{equation}\label{eq:IS1app1}
\mathcal{I}_{S; 1}^{({\rm np})}(\lambda;\alpha)= \sum_{n=1}^{\infty} \sum_{{m=0}}^{\infty} \frac{8(1{-}\alpha)^m}{\Gamma(m+1)^2} \left[ \left(\frac{Y_n}{2} \right)^{2m-1} \!\!\! K_1(Y_n) + G^{2,1}_{2,3} \left(0;m{-}1,m,1;\frac{Y_n^2}{4} \right)\right]\,,
\end{equation}
where~$K_\nu(x)$ is the Bessel function of the second kind while~$G^{a,b}_{c,d}$ denotes the Meijer $G$-function. 

We now use the asymptotic expansion at large argument for the Bessel function and the Meijer $G$-function,
\begin{align}
K_1(Y) &\sim -\sqrt{\frac{\pi }{2Y}}e^{-Y} \sum_{k=0}^\infty \frac{\Gamma \left(k-\frac{1}{2}\right) \Gamma \left(k+\frac{3}{2}\right)}{ \pi  \Gamma (k+1)} (-2 Y)^{-k} \,,\\
G^{2,1}_{2,3} \left(0;m{-}1,m,1;\frac{Y^2}{4} \right)&\notag \sim - \sqrt{\frac{\pi }{2Y}}  \left(\frac{Y}{2}\right)^{2 m}e^{-Y} \sum_{k=0}^\infty \frac{\Gamma \left(k-\frac{3}{2}\right) \Gamma \left(k+\frac{1}{2}\right)}{2\pi \Gamma (k+1)}\\
& \phantom{\sim} \times\big[ (2 k+1) (2 k+3)-16 k m\big](-2 Y)^{-k} \, ,
\end{align}
to expand \eqref{eq:IS1app1} at large $Y_n$ leading to the formal series expansion
\begin{align}
    \mathcal{I}_{S; 1}^{({\rm np})}(\lambda;\alpha) =-\sum_{n=1}^{\infty} \sum_{{m=0}}^{\infty} &\notag \frac{(1{-}\alpha)^m}{\Gamma(m {+} 1)^2} \sqrt{\frac{8 \pi}{ Y_n}} \, e^{-Y_n} \left(\frac{Y_n}{2} \right)^{2m} \\
    &\times \left[ \sum_{k=0}^\infty (-2Y_n)^{-k} \frac{\Gamma(k {+} 1/2)\Gamma(k {-} 3/2)}{\pi \, \Gamma(k{+}1)} (4k^2+3-16km)\right] \, . \label{eq:InpAppm}
\end{align}
The sum over $m$ is easily performed and it produces modified Bessel functions of the first kind,\allowdisplaybreaks{
\begin{align}\label{eq:InpApp2}
  &\notag  \mathcal{I}_{S; 1}^{({\rm np})}(\lambda;\alpha) =-\sum_{n=1}^{\infty}\sqrt{\frac{8\pi }{Y_n}}e^{-Y_n} \sum_{k=0}^\infty \frac{\Gamma \left(k-\frac{3}{2}\right) \Gamma \left(k+\frac{1}{2}\right)}{\pi  \Gamma (k+1)} (-2Y_n)^{-k}\\*
    & \times \left[ (4k^2+3)I_0\left(\sqrt{1-\alpha}Y_n\right) - 8 k Y_n \sqrt{1-\alpha}\, I_1\left(\sqrt{1-\alpha}Y_n \right)\right]\, .
\end{align}}
To proceed further we need to be careful.
We note that the argument of the Bessel functions is given by $\sqrt{1-\alpha}\,Y_n$ with $Y_n =n \sqrt{\lambda}$. Since we are interested in the large-$\lambda$ expansion of the non-perturbative terms, we should substitute the resurgent asymptotic expansion for the Bessel function at large values of the argument. However, for $\alpha=1$ the argument of the Bessel functions vanishes and we obtain the simpler expression
\begin{align}
{\mathcal{I}}_{S; 1}^{(\rm{np})}(\lambda;1) =-\sum_{n=1}^\infty e^{-Y_n} \sqrt{\frac{8 \pi}{Y_n}} \sum_{k=0}^\infty \frac{\left(4 k^2+3\right) \Gamma \left(k-\frac{3}{2}\right) \Gamma \left(k+\frac{1}{2}\right)}{\pi  \Gamma (k+1)} (-2Y_n)^{-k}\,,\label{eq:IS1npAppalpha1}
\end{align}
which arises from the $m=0$ summand of \eqref{eq:InpAppm}.
We highlight that the non-perturbative corrections to ${\mathcal{I}}_{S; 1}(\lambda;\alpha)$ with $\alpha=1$ are exponentially suppressed by a factor $e^{-Y_n} = e^{-n \sqrt{\lambda}}$.

However, when $0\leq \alpha<1$ we find that \eqref{eq:InpApp2} contains two different types of non-perturbative effects thanks to the known transseries expansion of the Bessel functions (see \cite[\href{https://dlmf.nist.gov/10.40}{(10.40.5)}]{NIST:DLMF})
\allowdisplaybreaks{
\begin{align}
I_\nu(z) &=\notag \frac{e^z}{(2\pi z)^{\frac{1}{2}}} \sum_{j=0}^\infty \frac{\left( \tfrac{1}{2} +\nu\right)_j\left( \tfrac{1}{2} -\nu\right)_j}{j!} (2z)^{-k} \\*
&\label{eq:TSI}\phantom{=}\pm \ii e^{\pm \nu \pi \ii} \frac{e^{-z}}{(2\pi z)^{\frac{1}{2}}} \sum_{j=0}^\infty \frac{\left( \tfrac{1}{2} +\nu\right)_j\left( \tfrac{1}{2} -\nu\right)_j}{j!}(-2z)^{-k} \,.
\end{align}}
Similar to our discussion below \eqref{eq:I0NP}, we have that the overall factor $\sigma_\nu(z) \coloneqq \pm \ii e^{\pm \nu \pi \ii} $ corresponds to the transseries parameter which must be understood as a piecewise constant function of ${\rm arg}(z)$,~i.e.~$\sigma_\nu(z) = \pm \ii e^{\pm \nu \pi \ii} $ for ${\rm arg}(z) \gl 0$. It is easy to see that for ${\rm arg}(z) = 0$, the formal power series multiplying the leading exponential factor $e^z$ in \eqref{eq:TSI} is not Borel summable. If we perform a lateral Borel resummation (as discussed in Appendix \ref{app:cat}) from either ${\rm arg}(z) > 0$ or ${\rm arg}(z) < 0$ we produce two distinct analytic functions of $z$. Once we add the appropriate non-perturbative terms, i.e. the terms proportional to $\sigma_\nu(z) e^{-z}$ in  \eqref{eq:TSI}, we find a well-defined analytic function for ${\rm Re}(z)>0$, i.e. the Bessel function $I_\nu(z)$.

Substituting the above transseries in \eqref{eq:InpApp2} we finally arrive at 
\begin{align}
  &  \mathcal{I}_{S; 1}^{({\rm np})} (\lambda;\alpha)=-2\sum_{n=1}^{\infty} \sum_{j=0}^\infty \sum_{k=0}^\infty (-1)^k \left(\frac{1}{2n \sqrt{\lambda}} \right)^{j+k} \frac{\Gamma(j+1/2)^2\Gamma(k+1/2)\Gamma(k-3/2)}{\pi^2 \, \Gamma(j+1) \, \Gamma(k+1)  } \frac{1}{(1-\alpha)^{1/4+j/2}} \\
&\times \Bigg[e^{-n \sqrt{\lambda}(1-\sqrt{1-\alpha})}\left(\frac{4k^2+3}{n \sqrt{\lambda}}+8k\frac{2j+1}{2j-1}\sqrt{1{-}\alpha}\right)  \pm \ii\, (-1)^j e^{-n \sqrt{\lambda}(1+\sqrt{1-\alpha})} \left(\frac{4k^2+3}{n \sqrt{\lambda}}-8k\frac{2j+1}{2j-1}\sqrt{1{-}\alpha}\right)   \Bigg]\,. \nonumber
\end{align}
We can rearrange the above sums and define the new summation variable $m=j+k$. Performing the sum over $j$, with $0\leq j\leq m$, we arrive at the large-$\lambda$ expansion of the non-perturbative terms
\begin{align} \label{eq:IS1NPFinApp} 
    \mathcal{I}_{S; 1}^{({\rm np})}(\lambda;\alpha) = -16  \pi (1-\alpha)^{\frac{1}{4}} & \sum_{n=1}^{\infty} \Big[ \notag   e^{-n \sqrt{\lambda}(1-\sqrt{1-\alpha})} \sum_{m=0}^\infty \frac{1}{m!} c_m(\sqrt{1-\alpha}) \,(2 n \sqrt{\lambda(1-\alpha)})^{-m-1} \\
    &\pm \ii \,e^{-n \sqrt{\lambda}(1+\sqrt{1-\alpha})} \sum_{m=0}^\infty \frac{(-1)^m}{m!} c_m(-\sqrt{1-\alpha})\,(2 n \sqrt{\lambda(1-\alpha)})^{-m-1}\Big]\,, 
\end{align}
where the perturbative coefficients $c_m(x)$ have been presented in \eqref{cm}.

We notice that ${\mathcal{I}}_{S; 1}(\lambda;\alpha)$ with $0<\alpha<1$ now contains two types of non-perturbative effects, one purely real (denoted by $\vert_\mathbb{R}$) and one purely imaginary (denoted by $\vert_\mathbb{I}$)  whose exponential suppression factors are given by
\begin{align} \label{eq:IS1npSumApp}
     \mathcal{I}_{S; 1}^{({\rm np})}(\lambda;\alpha)\vert_{ \mathbb{R}} \sim \exp \left[ -n \sqrt{\lambda}(1-\sqrt{1-\alpha}) \right] \, , \qquad  \mathcal{I}_{S; 1}^{({\rm np})}(\lambda)\vert_{\mathbb{I}} \sim \ii\,\exp \left[-n \sqrt{\lambda}(1+\sqrt{1-\alpha}) \right]  \, .
\end{align}
In the limit $\alpha\to 1$ we see that both $\mathcal{I}_{S; 1}^{({\rm np})}(\lambda;\alpha)\vert_{ \mathbb{R}}$ and $\mathcal{I}_{S; 1}^{({\rm np})}(\lambda;\alpha)\vert_{ \mathbb{I}}$ reduce to the single exponential suppression factor $e^{-n \sqrt{\lambda}}$ found in \eqref{eq:IS1npAppalpha1}.
However, we stress that we cannot simply take the limit for $\alpha \to 1$ of \eqref{eq:IS1npSumApp} to retrieve \eqref{eq:IS1npAppalpha1} since to derive \eqref{eq:IS1npSumApp} we crucially used in \eqref{eq:InpApp2} the asymptotic expansion for the Bessel functions at large argument which is not valid when the argument vanishes at $\alpha=1$.

In a very similar fashion, we can discuss the non-perturbative terms $\mathcal{I}_{AdS; 1}^{({\rm np})}(\lambda;\beta) $ for the dual giant graviton.
Again, we start from the corresponding Mellin integral representation \eqref{nonpertAdS} and rewrite the hypergeometric function in terms of its Gauss series,
\begin{align}
    \mathcal{I}_{AdS; 1}^{({\rm np})}(\lambda;\beta)  = -\sum_{n=1}^{\infty} \sum_{m=0}^{\infty} \frac{4}{(1{+}\beta)^m \, \Gamma(m{+}1) \Gamma(m{+}2)} \int_{{\rm Re}\, s=\frac{3}{2}} \frac{ds}{2 \pi i} \,(2s-1) \Gamma(m+s)^2 \left(\frac{Y_n\sqrt{\beta+1}}{2} \right)^{-2s} \, .
\end{align}
As before, this allows us to evaluate the Mellin integral, 
\begin{align}
    \mathcal{I}_{AdS; 1}^{({\rm np})}(\lambda;\beta) = \sum_{n=1}^{\infty} \sum_{m=0}^{\infty} \frac{8\, (1{+}\beta)^{-m}}{ \Gamma(m{+}1) \Gamma(m{+}2)} \left[\left(\frac{Y_n\sqrt{\beta {+} 1}}{2}\right)^{2m} \! K_0(Y_n\sqrt{\beta {+} 1}) \, + G^{2,1}_{2,3}\left(0;m,m,1;\frac{Y_n^2(\beta {+} 1)}{4} \right) \right] \, .
\end{align}
Since we are interested in the large-$\lambda$ regime, we simply need to expand the above expression at large $Y_n$,
\begin{align}
    \mathcal{I}_{AdS; 1}^{({\rm np})}(\lambda;\beta) =&\,  \sum_{n=1}^{\infty}  e^{-Y_n\sqrt{\beta+1}} \sum_{m=0}^{\infty} \frac{\sqrt{2\pi Y_n} }{ \Gamma(m+1) \Gamma(m+2)}  \left(\frac{Y_n}{2} \right)^{2m} \cr 
    & \times  \sum_{k=0}^{\infty} (1+\beta)^{1/4-k/2}\frac{   \Gamma(k-1/2)^2 }{ \pi \Gamma(k+1)} \left(-2Y_n \right)^{-k} \left[(4k^2-8k-1)-16km \right] \, . 
\end{align}
We then perform the summation over $m$ to obtain
\begin{align}
    \mathcal{I}_{AdS; 1}^{({\rm np})}(\lambda;\beta) =&\label{eq:IAdS1np} \,  \sum_{n=1}^{\infty} \sqrt{\frac{8\pi}{ Y_n}} e^{-Y_n \sqrt{\beta+1}} \sum_{k=0}^{\infty} \frac{\Gamma(k-1/2)^2}{\pi (1+\beta)^{k/2-1/4} \Gamma(k+1)} (-2Y_n)^k \cr & \times \left[(4k^2+8k-1)I_1(Y_n)-8kY_nI_0(Y_n)\right] \, .
\end{align}
 Using again the large-$x$ transseries expansion \eqref{eq:TSI} for the modified Bessel function $I_\nu(x)$,  we then deduce the exponentially suppressed terms for $\mathcal{I}_{AdS; 1}(\lambda;\beta)$,
\begin{align} \label{eq:Iads}
     \mathcal{I}_{AdS; 1}^{({\rm np})}&(\lambda;\beta) = \sum_{n=1}^{\infty} \sum_{k=0}^\infty \sum_{j=0}^{\infty} (-1)^k  \left(\frac{1}{2n \sqrt{\lambda}}\right)^{j+k} \frac{\Gamma(k-1/2)^2 \Gamma(j+1/2)\Gamma(j-1/2)}{\pi^2 \,\Gamma(j+1) \, \Gamma(k+1) } \frac{1}{(\beta + 1)^{\frac{k}{2}-\frac{1}{4}}} \\ 
     & \left[ e^{-n\sqrt{\lambda}(\sqrt{\beta+1}-1)} \left(-\frac{(4k^2+8k-1)(2j+1)}{n\sqrt{\lambda}}-8k(2j-1) \right) \right.\cr
     &\left.\mp \ii (-1)^j \,  e^{-n\sqrt{\lambda}(\sqrt{\beta+1}+1)} \left(\frac{(4k^2+8k-1)(2j+1)}{n\sqrt{\lambda}}-8k(2j-1) \right)   \right] \, , \nonumber
\end{align}
where we expressed the results back in terms of the 't Hooft coupling $\lambda$ using $Y_n = n\sqrt{\lambda}$.
Lastly, by changing summation variables to $m=j+k$ and then summing over $j$ with $0\leq j \leq m$ we finally arrive 
at
\begin{align}\label{eq:IadsFinApp}
     \mathcal{I}_{AdS; 1}^{({\rm np})} (\lambda;\beta)=&\, -16 \pi  (1+\beta)^{\frac{1}{4}}  \sum_{n=1}^\infty\Big[  \notag e^{-n \sqrt{\lambda} \left(\sqrt{\beta+1}-1 \right)} \sum_{m=0}^\infty \frac{1}{m!} {c}_m(\sqrt{1+\beta}) \left(-2n \sqrt{\lambda(1+\beta)}\right)^{-m-1} \\ 
     &\mp \ii \, e^{-n \sqrt{\lambda} \left(\sqrt{\beta+1}+1 \right)} \sum_{m=0}^\infty \frac{(-1)^{m}}{m!}\, c_m(-\sqrt{1+\beta})\left(2n \sqrt{\lambda(1+\beta)}\right)^{-m-1}\Big] \, ,  
\end{align}
with the same perturbative coefficients $c_m(x)$ as above.

It is interesting to specialise the non-perturbative sectors \eqref{eq:IS1NPFinApp} and \eqref{eq:IadsFinApp} to the special case $\alpha =\beta =0$. We note that in this limit, the leading exponential factors $e^{-n\sqrt{\lambda}(1-\sqrt{1-\alpha})}$ and $e^{-n\sqrt{\lambda}(\sqrt{1+\beta}-1)}$ both reduce to $1$, thus becoming perturbative contributions whose coefficients can be easily derived from the formula \eqref{cm} for $c_m(1)$. Once we add to the original perturbative series $\mathcal{I}_{S; 1}^{({\rm p})}(\lambda;0)$ in \eqref{eq:IS1P} and $\mathcal{I}_{AdS; 1}^{({\rm p})}(\lambda;0)$  in \eqref{eq:IAdS1P} the perturbative series accompanying the corresponding leading exponential factors (now reduced to unity), we find respectively $\mp \mathcal{I}_{S; 0}^{({\rm p})}(\lambda)$ as given in \eqref{eq:I0-strong}. 
Similarly, we see that when $\alpha=\beta=0$ the subleading exponential factors in \eqref{eq:IS1NPFinApp} and \eqref{eq:IadsFinApp} both reduce to $e^{-2n \sqrt{\lambda}}$. Again using \eqref{cm} to determine $c_m(-1)$, we realise that the power series corrections accompanying the subleading exponential respectively reduce to $\mp\mathcal{I}_{S; 0}^{({\rm np})}(\lambda)$ given in  \eqref{eq:I0NP}. 
We conclude that at the level of transseries we must have $\mathcal{I}_{S; 1}(\lambda;0) = -\mathcal{I}_{AdS; 1}(\lambda;0) = -\mathcal{I}_{S; 0}(\lambda)$, as previously obtained in \eqref{eq:abzero} from an integral representation analysis.

Furthermore, as also mentioned in Section \ref{sec:Exp} we see that the resurgence structure which connects the non-Borel summable asymptotic power series multiplying the leading exponential factor in both \eqref{eq:IS1NPFinApp} and \eqref{eq:Iads} to the subleading exponential, multiplied by the transseries parameter $\pm i$, is ultimately linked to the transseries representation \eqref{eq:TSI} for the Bessel function $I_\nu(z)$, which is precisely the cause for this standard resurgence behaviour within the purely non-perturbative sectors $\mathcal{I}_{S; 1}^{({\rm np})}(\lambda;\alpha)$ and $\mathcal{I}_{S; 1}^{({\rm np})}(\lambda;\beta)$.
The resurgence structure which connects these non-perturbative sectors to their respective terminating perturbative parts, i.e. $\mathcal{I}_{S; 1}^{({\rm p})}(\lambda;\alpha)$ and $\mathcal{I}_{S; 1}^{({\rm p})}(\lambda;\beta)$, is instead more subtle as discussed in the next appendix.

\section{Cheshire cat resurgence} \label{app:cat}

In this appendix, we show how to retrieve the large-$\lambda$ exponentially suppressed terms contained in 
$\mathcal{I}_{S; 1}(\lambda;\alpha)$ and $\mathcal{I}_{AdS; 1}(\lambda;\beta)$ via a special version of resurgence analysis dubbed \textit{Cheshire cat resurgence}.

Firstly as already noticed, the large-$\lambda$ perturbative expansion of 
$\mathcal{I}_{S; 0}(\lambda) = -\mathcal{I}_{AdS; 0}(\lambda)$ presented in \eqref{eq:I0-strong} is an asymptotic and factorially divergent power series.  
Using standard resurgence analysis (see e.g. the introductory lectures \cite{Dorigoni:2014hea}), it is possible to extract from the purely perturbative data \eqref{eq:I0-strong} the complete transseries expansion \cite{Brown:2024tru}, i.e. the exponentially suppressed non-perturbative terms presented in \eqref{eq:I0NP}. 
Conversely, the large-$\lambda$ perturbative expansion of $\mathcal{I}_{S; 1}(\lambda;\alpha)$ and $\mathcal{I}_{AdS; 1}(\lambda;\beta)$, given respectively in \eqref{eq:IS1P} and \eqref{eq:IAdS1P}, both contain only finitely many terms, thus making it impossible to retrieve directly any non-perturbative data from such simple expressions.

A more subtle resurgence structure is at play here.
We already know that the complete transseries expansions for $\mathcal{I}_{S; 1}(\lambda;\alpha)$ and $\mathcal{I}_{AdS; 1}(\lambda;\beta)$ do contain the infinite towers of exponentially suppressed terms presented in \eqref{eq:IS1NPFin} and \eqref{eq:IadsFin}. Very much like the lingering grin of Lewis Carrol's Alice in Wonderland's cat, the survival of exponentially suppressed corrections when the corresponding perturbative series does terminate after finitely many terms, has been referred to as {\em Cheshire cat resurgence}.

In this appendix we wish to clarify how this mechanism works. For simplicity we now present details only for $\mathcal{I}_{S; 1}(\lambda;1)$ while similar, albeit slightly more technical arguments can be presented for $\mathcal{I}_{S; 1}(\lambda;\alpha)$ with generic $0\leq \alpha\leq 1$ and $\mathcal{I}_{AdS; 1}(\lambda;\beta)$ with $\beta\geq0$.

Our starting point is the contour representation \eqref{eq:inte12}. Firstly, we notice that for $\alpha=1$ the hypergometric function reduces simply to $1$, hence we wish to study the large-$\lambda$ expansion of
\begin{align} 
 \mathcal{I}_{S; 1}(\lambda;1) &=    \int_{{\rm Re}\,s=1/2 } {ds \over  2 \pi \ii}  \, {4\pi \, 
 \zeta(2s{+}1) \over \sin(\pi s)}   \binom{2s {+} 1}{s}   \Big(\frac{\lambda}{16\pi^2}\Big)^s \, . \label{eq:inte12App}
\end{align}
Secondly, we rewrite the binomial coefficient in terms of gamma functions finding the expression
\begin{equation}
{2s+1 \choose s} = \frac{2^{2s+1} \Gamma \left(s+\frac{3}{2}\right)}{\sqrt{\pi } \Gamma (s+2)}\,.
\end{equation}
We claim that the origin behind the Cheshire cat resurgence behaviour stems from the particular shift in the arguments of the two gamma functions in the fraction above, namely a half-integer shift for the gamma function in the numerator and an integer shift for the gamma function in the denominator.

To reinstate the full resurgence analysis body we then deform the quantity under consideration and introduce the auxiliary function,
\begin{align} 
 \tilde{\mathcal{I}}_{S; 1}(\lambda;1; a,b) &\coloneqq    \int_{{\rm Re}\,s=1/2 } {ds \over  2 \pi \ii}  \, {4\pi \, 
 \zeta(2s{+}1) \over \sin(\pi s)}   \frac{2^{2s+1} \Gamma \left(s+a+\frac{1}{2}\right)}{\sqrt{\pi } \Gamma (s+b)}  \Big(\frac{\lambda}{16\pi^2}\Big)^s {ds \over  2 \pi \ii}\,. \label{eq:inteAux}
\end{align}
where $a,b\in \mathbb{C}$ play the role of deformation parameters.
By specialising the above equation to $a=1$ and $ b=2$ we retrieve the function of interest, i.e. $ \tilde{\mathcal{I}}_{S; 1}(\lambda;1; a,b)\vert_{a=1,b=2} =  \mathcal{I}_{S; 1}(\lambda;1)$.

We proceed to show that for general deformation parameters $a,b\in \mathbb{C}$ the function defined in \eqref{eq:inteAux} presents an asymptotic and factorially divergent power series expansion as $\lambda\to\infty$. 
For convenience we change the integration variable $s\to 1-s$ and use the functional equations for both the gamma function and Riemann zeta function to rewrite \eqref{eq:inteAux} as
\begin{align} 
 \tilde{\mathcal{I}}_{S; 1}(\lambda;1; a,b) &=   16 \sqrt{\pi} \int_{{\rm Re}\,s=1/2 } {ds \over  2 \pi \ii}  \, \zeta (2 s -2) \frac{  \Gamma (2 s-2)  \Gamma (s-b)}{\Gamma \left(s-a-\frac{1}{2}\right)}  \frac{\cos(\pi s) \sin(\pi (s-b)) }{\cos(\pi(s-a)) \sin(\pi s)} \,  \lambda ^{1-s}\, . \label{eq:inteAux2}
\end{align}
It now appears manifest that for the special case where the deformation parameters $a,b\in \mathbb{Z}$, the large-$\lambda$ expansion of $\tilde{\mathcal{I}}_{S; 1}(\lambda;1; a,b)$ does produce only finitely many terms given that the integrand is holomorphic for ${\rm Re}(s)$ large enough.
However, for generic values of $a,b \in \mathbb{C}$ the large-$\lambda$ expansion produces a factorially divergent asymptotic power series.

For simplicity, let us specialise $a=1$ and keep $b\in \mathbb{C}$ generic, this is enough to keep the body of the Cheshire cat resurgence visible.
This simplifies \eqref{eq:inteAux2} to
\begin{equation}
 \tilde{\mathcal{I}}_{S; 1}(\lambda;1; 1,b)  =- \int_{{\rm Re}\,s=1/2 } {ds \over  2 \pi \ii}  4^s (2 s-3) \zeta (2 s-2)  \Gamma (s-1)\Gamma (s-b) \frac{\sin (\pi  (s-b))}{\sin (\pi s)} \lambda ^{1-s} \,.\label{eq:IntAuxIR}
\end{equation}

The asymptotic large-$\lambda$ expansion is obtained from \eqref{eq:IntAuxIR} by closing the contour of integration to the right half-plane ${\rm Re}(s)>1/2$ and collecting the residue of the double pole at $s=1$ coming from $\Gamma(s-1) /\sin (\pi  s)$ and the simple poles residues at $s\in \mathbb{N}^{\geq 2}$ coming from $1/\sin (\pi  s)$.
This yields the asymptotic expansion, 
\begin{align}
\tilde{\mathcal{I}}_{S; 1}(\lambda;1; 1,b) \sim \tilde{\mathcal{I}}_{S; 1}^{({\rm p})}(\lambda;1; 1,b)  = A_1(\lambda;b)+ \frac{\sin (\pi  b)}{\pi} \sum_{n=1}^\infty 4^{n+1} (2 n-1)  \zeta (2 n) \Gamma (n) \Gamma (n+1-b) \lambda ^{-n}\,,\label{eq:IappAsy}
\end{align}
where $A_1(\lambda;b)$ denotes the contribution coming from the double pole at $s=1$,
\begin{equation}
A_1(\lambda;b) \coloneqq -\frac{2 \left[\log \left(\frac{\lambda }{16 \pi ^2}\right)+\gamma +2 -\psi(b)\right]}{\Gamma (b)}\,,
\end{equation}
with $\psi(x) \coloneqq \Gamma'(x) / \Gamma(x)$ denoting the digamma function.

Importantly we have that
\begin{equation}
\lim_{b\to 2} A_1(\lambda;b) = -2 \left[\log \left(\frac{\lambda }{16 \pi ^2}\right)+2 \gamma +1\right]\,,
\end{equation}
thus reproducing the constant and logarithmic term found in \eqref{eq:IS1P} when specialised to $\alpha=1$.
Furthermore, we see that as $b\to m$, with $m\in \mathbb{Z}$, the factorially divergent formal power series of perturbative corrections reduces to finitely many terms. In particular for $b\to 2$ only the summand with $n=1$ produces a non-vanishing contribution, i.e.
\begin{equation}
\lim_{b\to 2}\left[  \frac{\sin (\pi  b)}{\pi} \sum_{n=1}^\infty 4^{n+1} (2 n-1)  \zeta (2 n) \Gamma (n) \Gamma (n+1-b) \lambda ^{-n} \right] = \begin{cases}
&-\frac{8 \pi^2}{3 \lambda} \,,\qquad n=1\,,\\
&\,\,\phantom{-} 0 \,,\qquad\,\,\,\, n\geq 2\,.
\end{cases}
\end{equation}
We then conclude that the asymptotic expansion \eqref{eq:IappAsy} in the limit $b\to 2$ does reduce precisely to the terminating perturbative expansion \eqref{eq:IS1P} when specialised to $\alpha=1$.

We now proceed to define an unambiguous analytic continuation for the asymptotic expansion \eqref{eq:IappAsy} and derive from there the non-perturbative completion.
Firstly, as in \cite{Arutyunov:2016etw} we define the modified Borel transform
\begin{equation}
\mathcal{B}: \quad \sum_{n=1}^{\infty} c_n \lambda^{-n} \rightarrow  \hat{\phi}(w) \coloneqq \sum_{n=1}^{\infty} {2 \,c_n \over \zeta(2n) \Gamma(2n+1)} (2 w)^{2n}  \, .
\end{equation}
If the perturbative coefficients $c_n\sim (2n)!$ for $n\gg1$, as for the relevant case of \eqref{eq:IappAsy}, we have that the corresponding modified Borel transform $\hat{\phi}(w) $ has finite radius of convergence and thus defines an analytic function in the complex $w$-plane.
In particular, for the perturbative coefficients $c_n$ in \eqref{eq:IappAsy} we find 
\begin{align}
 \hat{\phi}(w;b) =&\label{eq:Borel}\frac{64 w^2}{\Gamma (b-1)} \left[\, _3F_2\left(1,1,2-b;\frac{3}{2},2;4 w^2\right)-2 \, _2F_1\left(1,2-b;\frac{3}{2};4 w^2\right)\right]\,. 
\end{align}

We can then define the directional Borel resummation of \eqref{eq:IappAsy} as
\begin{equation} \label{inverse-Borel}
\mathcal{S}_\theta\left[\tilde{\mathcal{I}}_{S; 1}^{(\rm p)}\right](\lambda;1; 1,b) = A_1(\lambda;b)+ \sqrt{\lambda} \int_0^{e^{i\theta} \infty} {dw  \over 4 \sinh^2( \sqrt{\lambda} w ) }  \hat{\phi}( w) \, , 
\end{equation}
which defines an analytic function for $\lambda>0$ when $\theta \in (-\pi/2,0) \cup (0,\pi/2)$.
Using the integral representation 
\begin{equation} \label{zeta-integral}
{2^{2s-1} \over  \Gamma(2s+1)} \int_0^{\infty} dw {w^{2s} \over  \sinh^2( w )} = \zeta(2s) \, ,
\end{equation}
valid for ${\rm Re}(s)> 1/2$, it is easy to see that the large-$\lambda$ asymptotic expansion of \eqref{inverse-Borel} does in fact coincide with \eqref{eq:IappAsy}.

Crucially, although \eqref{inverse-Borel} does define an analytic continuation of the formal power series \eqref{eq:IappAsy}, it suffers from two major problems:
it is neither unique nor is it real for $\lambda>0$ and $b\in \mathbb{R}$ for any value of $\theta$ despite \eqref{eq:IappAsy} being manifestly so. 
Both problems are related to the fact that in \eqref{inverse-Borel} we cannot integrate along the real line $\theta=0$ due to the fact that $\hat{\phi}(w;b) $ in \eqref{eq:Borel} has a branch cut along the positive real line for $w\in[1/2, \infty)$. 

This is a signal that our starting asymptotic series \eqref{eq:IappAsy} is not Borel summable. As anticipated, the large-$\lambda$ expansion requires the addition of exponentially small non-perturbative terms. These terms are encoded by the discontinuity in the resummation, related to what is usually called the Stokes automorphism factor, which is given by:
\begin{equation} \label{exp-decay}
(\mathcal{S}_+-\mathcal{S}_-)\left[\tilde{\mathcal{I}}_{S; 1}^{({\rm p})}\right](\lambda;1; 1,b) = \Delta \left[\tilde{\mathcal{I}}_{S; 1}^{({\rm p})}\right](\lambda;1; 1,b)  = \lambda \int_0^{\infty} dw \,   {1 \over 4 \sinh^2(  x w ) }  \, {\rm Disc}_0\, \hat{\phi}(w;b)   \, ,
\end{equation}
where $\mathcal{S}_\pm$ define the lateral Borel resummations
\begin{align}
\mathcal{S}_\pm \left[\tilde{\mathcal{I}}_{S; 1}^{({\rm p})}\right](\lambda;1; 1,b) = \lim_{\theta\to 0^{\pm}} \mathcal{S}_\theta \left[\tilde{\mathcal{I}}_{S; 1}^{({\rm p})}\right](\lambda;1; 1,b)\,.
\label{resubor}
\end{align}

The discontinuity of the Borel transform can be easily computed starting from the known discontinuity of the $_2F_1$ given by
\begin{align}
{\rm Disc}\,\, _2F_1\left(a,b;c ; z   \right)  &= \frac{2\pi \ii\, \Gamma(c)}{\Gamma(a) \Gamma(b) \Gamma(c-a-b+1) }  z^{1-c}  (z-1)^{c-a-b}  \cr
&\,\,\, \times\,  _2F_1\left(1-a,1-b;c-a-b+1;  1-z \right) \, , \label{eq:disc2F1}
\end{align}
and, using its integral representation, that of the generalised hypergeometric function
\begin{align}
{\rm Disc}\,\,_3F_2\left(1,1,2-b;\frac{3}{2},2;z\right) = -\ii \sin(\pi b) \sqrt{\pi } z^{-\frac{1}{2}} (z-1)^{b-\frac{1}{2}}  \Gamma (b-1) \, _2\tilde{F}_1\left(1,b;b+\frac{1}{2};1-z\right)\,,\label{eq:disc3F2}
\end{align}
both of which are valid for $z>1$.

Using the above expressions, we then find the discontinuity of the Borel transform \eqref{eq:Borel} along $w\in [1/2,\infty)$,
\begin{align}
{\rm Disc}_0\, \hat{\phi}(w;b) = -2\ii \sin(\pi b) \,{\rm disc}(w;b)\,,\label{eq:Disc}
\end{align}
where we defined for $w>1/2$
\begin{equation}
{\rm disc}(w;b) \coloneqq \frac{ 16\sqrt{\pi } w \left(4 w^2-1\right)^{b-\frac{3}{2}}\left[\left(4 w^2-1\right) \, _2F_1\left(1,b;b+\frac{1}{2};1-4 w^2\right)+1-2b\right]}{\Gamma \left(b+\frac{1}{2}\right)} \, . \label{eq:disc}
\end{equation}
We notice that the factor $\sin(\pi b)$ makes it so that ${\rm Disc}_0\, \hat{\phi}(w;b)=0$ for $b\in \mathbb{Z}$, a defining feature of Cheshire cat resurgence.

To obtain a resummation that is real, for real values of $\lambda$ and $b$, and unambiguous, we take an average of the two lateral resummations, referred to as a {\em median} resummation \cite{delabaere1999resurgent},
namely
\begin{align}
\mathcal{S}_{\rm med}\left[\tilde{\mathcal{I}}_{S; 1}^{({\rm p})}\right](\lambda;1; 1,b) &\notag =\mathcal{S}_\pm\left[\tilde{\mathcal{I}}_{S; 1}^{({\rm p})}\right](\lambda;1; 1,b) \mp \frac{1}{2} \Delta \left[\tilde{\mathcal{I}}_{S; 1}^{({\rm p})}\right](\lambda;1; 1,b) \\
&=\notag \mathcal{S}_\pm\left[\tilde{\mathcal{I}}_{S; 1}^{({\rm p})}\right](\lambda;1; 1,b) \pm \ii\, \sin(\pi b) \,\tilde{\mathcal{I}}_{S; 1}^{(\rm{np})}(\lambda;1; 1,b) \\
& = \mathcal{S}_\pm\left[\tilde{\mathcal{I}}_{S; 1}^{({\rm p})}\right](\lambda;1; 1,b) \pm \ii\, \sin(\pi b)\lambda \int_{\frac{1}{2}}^{\infty} dw \,   {1 \over 4 \sinh^2(  \sqrt{\lambda} w ) }  \, {\rm disc}(w;b) \,.\label{eq:Smed1}
\end{align}

As already stated, the resummation of the perturbative series jumps when we cross the Stokes direction $\theta=0$.
Furthermore, we know that the imaginary part of this jump depends on the fixed parameter $b$.
The particular coefficient, which we denote by $\sigma(b)$, multiplying the non-perturbative contributions $\tilde{\mathcal{I}}_{S; 1}^{(\rm{np})}(\lambda;1; 1,b)$ is known as \textit{transseries parameter}.
Although we do not write its explicit dependence, the transseries parameter $\sigma(b)$ is also a piecewise constant function of ${\rm arg}(\lambda)$ and it jumps when we cross the Stokes direction $\theta=0$. We take a normalisation of $\sigma(b)$ from \eqref{eq:Smed1}, obtaining
\begin{equation}
\mbox{Im} \,\sigma (b) 
= \pm \ii\,  \sin(\pi b) = \begin{cases} {\rm arg}(\lambda)>0\,,\\
 {\rm arg}(\lambda)<0\,,
\end{cases}
\label{eq:TSparamIm}
\end{equation}
above and below the Stokes direction $\theta=0$. 

Assuming an analytic dependence on the parameter $b$, it  seems then natural to {\em conjecture} that the full transseries parameter is given by
\begin{equation}
\sigma (b)= \e^{\pm \,\ii\, \pi b}\,,
\label{eq:TSexp}
\end{equation}
where again the choice in sign is dictated by the particular lateral Borel resummation employed to resum the perturbative series as in \eqref{eq:Smed1} which is furthermore correlated with ${\rm arg}(\lambda) \gl 0$.
Although we do not have a proof for \eqref{eq:TSexp}, we notice that this phenomenon where the full transseries parameter becomes an exponential function of the deformation parameters (in this case $b$) seems to be ubiquitous when discussing functions with a Cheshire cat structure, see e.g.~\cite{Dorigoni:2020oon, Broadhurst:2025iab}.

Under the above assumptions, we claim that the complete transseries representation of \eqref{eq:inteAux} is given by
\begin{align}
\tilde{\mathcal{I}}_{S; 1}(\lambda;1; 1,b) &= \mathcal{S}_\pm\left[\tilde{\mathcal{I}}_{S; 1}^{({\rm p})}\right](\lambda;1; 1,b) + \sigma(b) \,\tilde{\mathcal{I}}_{S; 1}^{(\rm{np})}(\lambda;1; 1,b)\,.
\end{align}
It is finally the time to specialise the above transseries to the physical value $b=2$ and arrive at the Cheshire cat resurgent transseries representation for $\mathcal{I}_{S;1}(\lambda;1) = \tilde{\mathcal{I}}_{S; 1}(\lambda;1, 1,2)$, 
\begin{align}
\tilde{\mathcal{I}}_{S; 1}(\lambda;1) &= \mathcal{I}_{S; 1}^{({\rm p})}(\lambda;1) + \tilde{\mathcal{I}}_{S; 1}^{(\rm{np})}(\lambda;1;1,2)\,,\label{eq:IS1ExApp}
\end{align}
where we dropped the resummation symbol $\mathcal{S}_\pm$ since, as previously discussed, the perturbative sector $\tilde{\mathcal{I}}_{S; 1}^{({\rm p})}(\lambda;1; 1,2) = \mathcal{I}_{S; 1}^{({\rm p})}(\lambda;1)$ reduces to the terminating asymptotic expansion \eqref{eq:IS1P}.

Crucially, although the full body of the resurgence structure has now disappeared by working at integer $b$, the grin of the non-perturbative effects still lingers on.
We read the non-perturbative completion $ {\mathcal{I}}_{S; 1}^{(\rm{np})}(\lambda;1) = \tilde{\mathcal{I}}_{S; 1}^{(\rm{np})}(\lambda;1;1,2)$ from \eqref{eq:Smed1},
\allowdisplaybreaks{
\begin{align}
{\mathcal{I}}_{S; 1}^{(\rm{np})}(\lambda;1) = \tilde{\mathcal{I}}_{S; 1}^{(\rm{np})}(\lambda;1;1,2) &\notag = \lambda \int_{\frac{1}{2}}^\infty dw \frac{1}{4\sinh^2(\sqrt{\lambda} w )} {\rm disc}(w;2)\\*
&= -\lambda \int_{\frac{1}{2}}^\infty dw \frac{1}{4\sinh^2(\sqrt{\lambda} w )} 16 \left(2 w \sqrt{4 w^2-1}+{\rm arccosh}(2 w)\right)\,.
\end{align}}
To evaluate this integral we first use the identity
\begin{equation}
\frac{1}{4\sinh^2(\sqrt{\lambda} w )} = \sum_{n=1}^\infty n e^{-2 n w \sqrt{\lambda}}\,,
\end{equation}
and then shift the contour of integration $w\to w+\frac{1}{2}$ as to factorise out the non-perturbative factor $e^{-n \sqrt{\lambda}}$. 
Lastly, we Taylor expand the integrand ${\rm disc}(w+\tfrac{1}{2};2)$ and integrate term by term to arrive at the formal power series expansion
\begin{align}
{\mathcal{I}}_{S; 1}^{(\rm{np})}(\lambda;1) =-\sum_{n=1}^\infty e^{-Y_n} \sqrt{\frac{8 \pi}{Y_n}} \sum_{k=0}^\infty \frac{\left(4 k^2+3\right) \Gamma \left(k-\frac{3}{2}\right) \Gamma \left(k+\frac{1}{2}\right)}{\pi  \Gamma (k+1)} (-2Y_n)^{-k} \, ,  \label{eq:IS1npApp}
\end{align}
where $Y_n = n \sqrt{\lambda}$, which manifests the non-perturbative nature of ${\mathcal{I}}_{S; 1}^{(\rm{np})}(\lambda;1)$.
We stress that the complete transseries expansion \eqref{eq:IS1ExApp} and in particular the non-perturbative corrections \eqref{eq:IS1npApp} have been derived here via Cheshire cat resurgence and thus crucially rely on the conjectural form \eqref{eq:TSexp} for the transseries parameter.
This conjecture is strongly validated by noticing that the above expression coincides with the result derived in \eqref{eq:IS1npAppalpha1} directly from the Mellin representation \eqref{eq:IS1npMell} without use of resurgence analysis.


\providecommand{\href}[2]{#2}\begingroup\raggedright\endgroup

\end{document}